\documentclass[11pt]{article}
\usepackage[utf8]{inputenc}
\pdfoutput = 1

\usepackage{amsmath}
\usepackage{amssymb}
\usepackage{bbm}
\usepackage{bbold}
\usepackage{dsfont}
\usepackage{braket}
\usepackage{caption}
\usepackage{cite}
\usepackage{color}
    \definecolor{darkgreen}{rgb}{0,0.5,0}
    \definecolor{darkred}{rgb}{0.5,0,0}
    \definecolor{darkblue}{rgb}{0,0,0.6}
    \definecolor{purple}{rgb}{0.4,.2,0.7}
    \definecolor{orange}{rgb}{0.7,0.3,0}
\usepackage[dvipsnames]{xcolor}
\definecolor{amethyst}{rgb}{0.54, 0.17, 0.89}
\definecolor{coral}{rgb}{1.0, 0.3, 0.4}

\usepackage{float}
\usepackage{graphicx}
\usepackage{fancyhdr}
\usepackage{tikz}
\usetikzlibrary{decorations.pathmorphing}
\usepackage{subcaption}
\usepackage[hyperfootnotes = true, colorlinks = true, linkcolor = blue, citecolor = red]{hyperref}
\usepackage{mathabx}
\usepackage[mathcal]{eucal}
\usepackage{pdfsync}
\usepackage{url}
\usepackage[normalem]{ulem}
\usepackage{comment}
\usepackage[english]{babel}
\usepackage{mathtools,slashed}
\usepackage{csquotes}
\usepackage{enumitem}
\usepackage[percent]{overpic}
\usepackage{epigraph} 
\usepackage{breqn}

\numberwithin{equation}{section}

\usepackage[margin = 2.2cm]{geometry}
\usepackage[ragged]{footmisc}
\def\beq{\begin{equation}\begin{aligned}}
\def\eeq{\end{aligned}\end{equation}}

\usepackage{color}
    \definecolor{darkgreen}{rgb}{0,0.5,0}
    \definecolor{darkred}{rgb}{0.5,0,0}
    \definecolor{darkblue}{rgb}{0,0,0.6}
    \definecolor{purple}{rgb}{0.4,.2,0.7}
    \definecolor{orange}{rgb}{0.7,0.3,0}
\usepackage{braket}
\usepackage{tikz}

\begin{document}

\thispagestyle{empty}
\begin{center}
    ~\vspace{5mm}
    
    {\LARGE \bf 
    Observing Spacetime
    }
    \noindent\rule{\textwidth}{1pt}
    \vspace{0.4in}
    
    {\bf Vijay Balasubramanian$^{1,2,3}$ and Tom Yildirim$^4$}

    \vspace{0.4in}

    $^1$ Department of Physics and Astronomy, University of Pennsylvania, Philadelphia, PA 19104, USA \vskip 1ex
    $^2$ Theoretische Natuurkunde, Vrije Universiteit Brussel (VUB) and The International Solvay Institutes, Pleinlaan 2, B-1050 Brussels, Belgium
    \vskip 1ex
    $^3$  Rudolf Peierls Centre for Theoretical Physics, University of Oxford, Beecroft Building, Parks Road Oxford OX1 3PU, UK\vskip 1ex
    $^4$ {Department of Physics, Keble Road, University of Oxford, Oxford, OX1 3RH, UK} \vspace{0.1in}
   
    {\tt vijay@physics.upenn.edu, tom.yildirim@physics.ox.ac.uk}
\end{center}

\vspace{0.4in}

\begin{abstract}
Complex states of quantum gravity in  flat and AdS gravity can have features that are inaccessible to classical asymptotic observers. The missing information appears to such observers to be hidden behind a horizon or in a baby universe.   Here we use the gravitational path integral to ask whether quantum observables can access the hidden data.  We show that generic probes give a universal result and contain no information about the state.  However, a probe appropriately fine-tuned to the state can give a large signal because of novel wormhole saddles in the path integral. Thus, in these settings, asymptotic observers cannot easily determine the state of the universe, but can check a proposal for it.  Using these fine-tuned probes we show that an asymptotic observer can detect information hidden in a disconnected baby universe. Furthermore we show that the state of a two-boundary black hole can be detected using Lorentzian operators localised on just one of the boundaries. 

\end{abstract}

\newpage
\tableofcontents

\section{Introduction}

The Bekenstein-Hawking entropy formula  $S = Ac^3/4G \hbar$ \cite{Bekenstein_1973, Hawking_1975}  implies that  microstates of a black hole span a Hilbert space of dimension $e^S$, where $A$ is the horizon area.  However, the asymptotic observer cannot easily detect the associated microstates.  In the classical limit, this difficulty manifests itself as a horizon which causally separates the interior and exterior of a black hole.   Quantum mechanically, however, several lines of argument suggest that the microstates can be detected by a distant observer who is sufficiently patient, or has access to sufficiently complicated probes.

For example, precise mass measurement will work, but takes a non-perturbatively long time \cite{Balasubramanian:2006iw}.  Indeed, as $\hbar \to 0$ the timescale for detection goes to infinity so that in the classical limit the microstates are not detectable by this mechanism in any finite time, making the classical description consistent with a causally disconnected interior and information loss. Similarly, Planck scale precision is required to resolve data separating different black hole microstates even if this information is available in charges that can be measured at asymptotic infinity \cite{Balasubramanian:2006jt}.  Recent work from a quantum information perspective has also suggested that degrees of freedom in the black hole interior are computationally complex to reconstruct by a boundary observer \cite{Penington:2019npb,Brown:2019rox,Engelhardt:2021mue,Engelhardt:2021qjs,Akers:2022qdl,Engelhardt:2024hpe,Balasubramanian:2022fiy,Balasubramanian:2023xdp}. Perhaps this is because low-order correlation functions are simple probes and the microstates are very complex.
The authors of \cite{Balasubramanian:2005mg}  argued that the microstates of black holes in AdS$_{5}$ are difficult to detect because they are structurally complex in terms of the underlying fundamental degrees of freedom -- i.e., statistical random combinations of the fields of the dual Yang-Mills theory. The result is that almost all probes, both simple and complex will give universal responses up to non-perturbatively small corrections.  These papers also argued that there should be special probes, tuned to the actual microstate, that can give a large response. Given such a probe, an observer can easily test whether it identifies the actual microstate, provided one is given multiple identically prepared copies of the black hole with which to do quantum experiments.. 

In this sense, the problem of detecting a black hole microstate is computationally difficult, but a proposed answer is easy to check -- like problems in the classical computational class NP (Non-deterministic Polynomial) or, rather, QMA (Quantum Merlin-Arthur) which is the quantum analog of NP. Here, we use gravitational path integral methods to explicitly show this for  microstates of black holes in General Relativity, and extend the conclusion to  detecting the state of a baby universe. In detail, we construct probes to detect superpositions of the single- and two- boundary shell states introduced in  \cite{Chandra:2022fwi,Balasubramanian:2025zey,Balasubramanian:2025hns} and \cite{Sasieta:2022ksu,Balasubramanian:2022gmo,Balasubramanian:2022lnw,Antonini:2023hdh,Balasubramanian:2024rek,Balasubramanian:2025jeu} respectively. The shell states are constructed by inserting heavy dust shell operators into the Euclidean gravitational path integral and subsequently evolving  in Euclidean boundary time. Depending on the length of this time evolution, these states correspond, after continuation to Lorentzian signature,  either to a black hole with a heavy shell propagating in a very large interior (\textit{Type A}), or (thermal) empty space entangled with a compact big-crunch baby universe containing the shell (\textit{Type B}).  

Interestingly, sufficiently large sets of either Type A or B states form complete bases of the non-perturbative gravity Hilbert space \cite{Balasubramanian:2025jeu,Balasubramanian:2025zey} with appropriate asymptotic boundary conditions. Such a basis was used to resolve the gravity Hilbert space factorisation problem \cite{Balasubramanian:2024yxk,Balasubramanian:2025zey}, account for the Beckenstein-Hawking entropy  \cite{Balasubramanian:2022gmo,Balasubramanian:2022lnw,Balasubramanian:2024rek,Climent:2024trz}, and  to establish the validity of the Gibbons-Hawking prescription for the gravitational partition function \cite{Balasubramanian:2025hns}. However, in the classical limit of either type of basis, the information specifying the state is hidden away in a region causally disconnected from the  asymptotic observer. Moreover, for a very heavy shell, the semiclassical spacetime geometry and state  of the low-energy effective fields accessible to the asymptotic observer are very close to those of the thermal state, displaying only fine quantum hair \cite{Balasubramanian:2022gmo}.  Consistently with this,  we show the gravitational path integral computes a universal response for the  expectation value of a generic Lorentzian boundary probe.  However, we also find a larger response for a probe fine-tuned to match the operator that created the microstate.  This larger response occurs because of novel wormhole  contributions to the gravity path integral, which can be thought of as detecting the quantum hair sourced by the shell.  In short, an observer can {\it check} a proposal for the microstate of the universe, but determining the state from scratch is exponentially hard. We show this is the case for both black holes with horizons and spacetimes with disconnected baby universe components, suggesting that causal disconnectedness of the latter from the asymptotic boundary is semiclassical artifact of the underlying complexity of the gravity state.  

These results hold for universes with one or two asymptotic boundaries. Interestingly the state of a two-boundary black hole can be checked using probes localized on just one  boundary, even though the interior is not in the entanglement wedge of that boundary.  

Our results hold for any gravity theory described by General Relativity at low energies, in any dimension.

Four sections follow. In Sec.~\ref{sec:onebdry} we construct probes that can be used to detect single-boundary gravity states. In Sec.~\ref{sec:two_bdry} we extend this to two-boundary gravity states and show that such states can be detected either through coordinated probes on both asymptotic boundaries or using probes localised to just one boundary. We then argue in Sec.~\ref{sec:QMA} that while these probes can be easily used to check a proposal for the gravity state, finding the state from scratch is hard.
We conclude with a summary and discussion in Sec.~\ref{sec:diss}.

\section{Observing a Single-Boundary Universe} \label{sec:onebdry}
In this section we establish that there are fine-tuned probes that a (non-local) Lorentzian boundary observer can use to detect the state of a single-boundary universe, and that generic probes yield a universal response, containing no information. This detection relies on the existence of novel non-perturbative wormhole saddles that contribute only when the probe is fine-tuned to the states in question. In particular, we will show that if the state is probed with the same operator as the one used to create the state, the response is an $\mathcal{O}(1)$ fraction larger than the universal contribution.

\begin{figure}[h]
    \centering
    \begin{subfigure}{0.45\linewidth}
    \centering
    \includegraphics[width=0.3\linewidth]{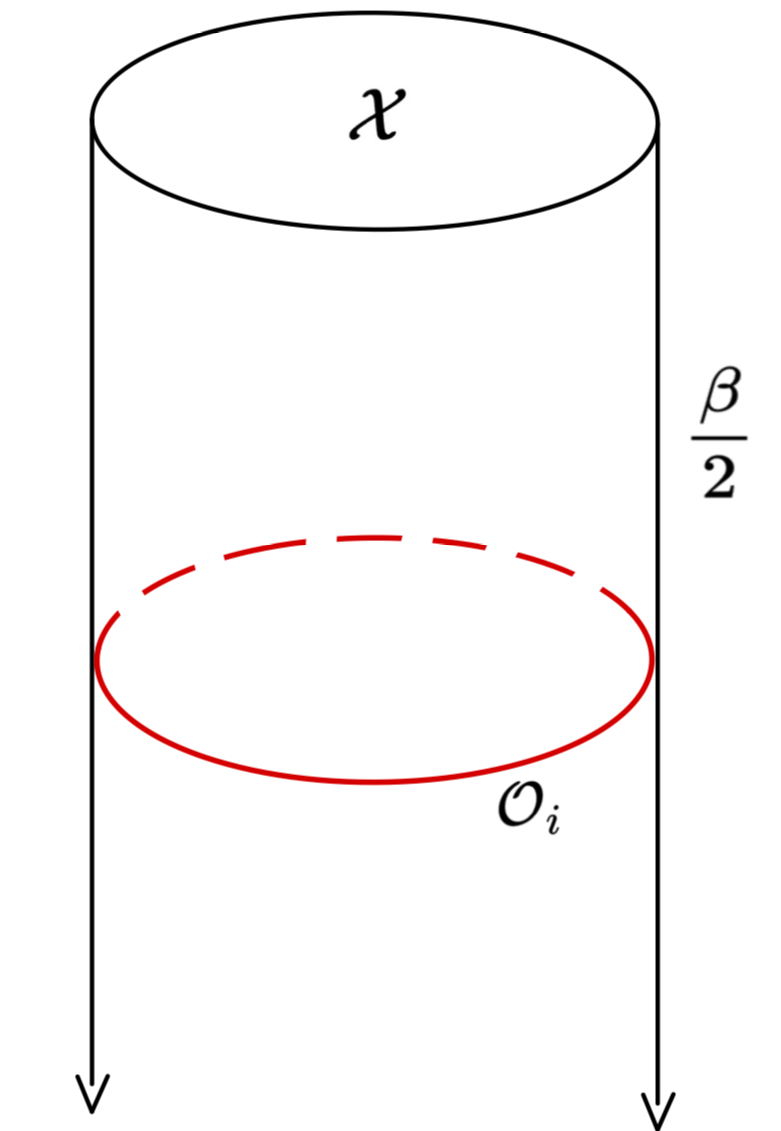}
    \caption{}
    \end{subfigure}
     \centering
    \begin{subfigure}{0.45\linewidth}
    \centering
    \includegraphics[width=0.3\linewidth]{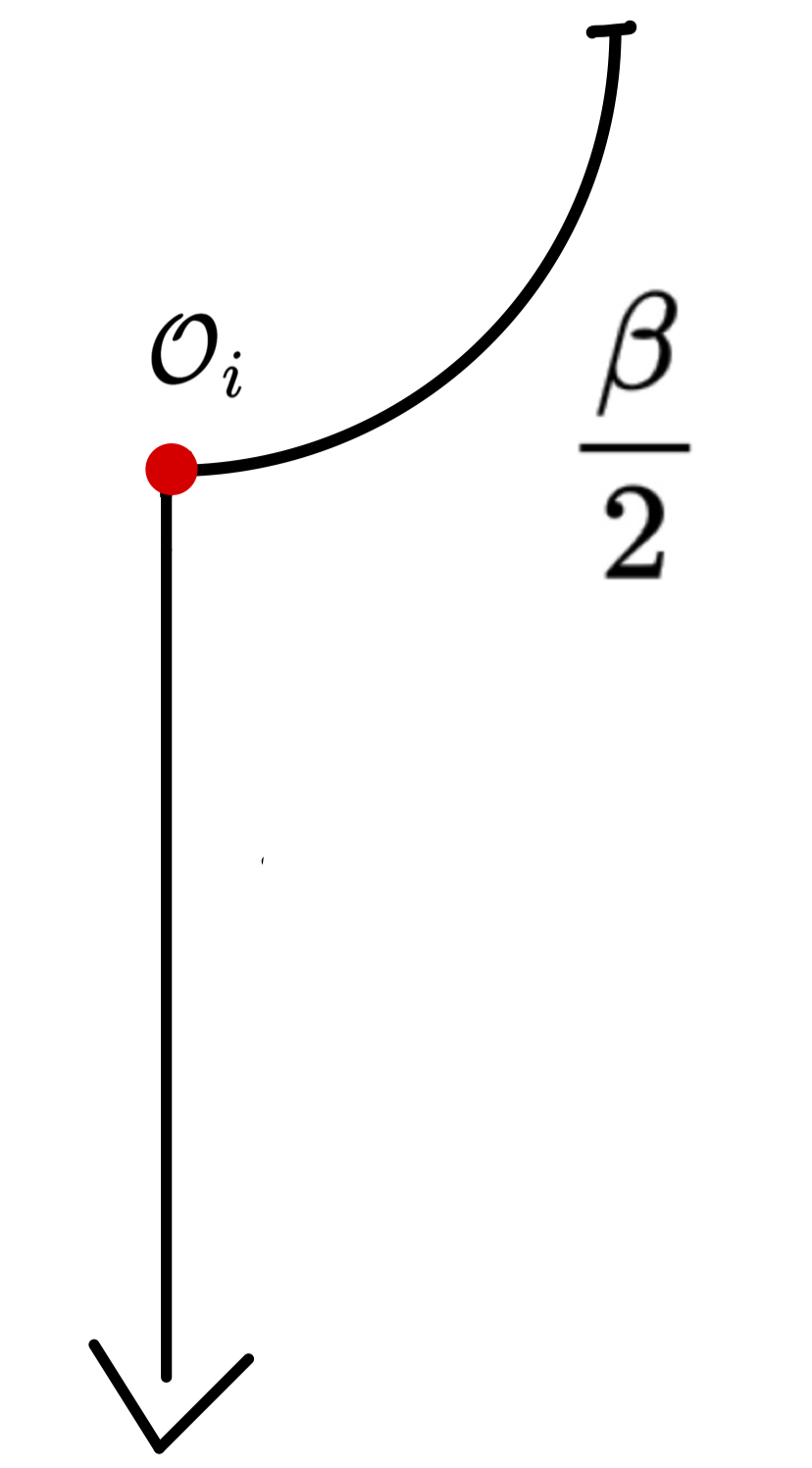}
    
    \caption{}
    \end{subfigure}

    \caption{Path integral boundary condition defining the single-sided shell states.  ({\bf a}) Euclidean boundary with topology $\mathbb{R}^{<0}\times\mathbb{S}^{d-1}$ for preparation of the shell states.  The arrows indicate a half-infinite line. The shell operator $\mathcal{O}_{i}$ is pictured in red and $\beta_{}/2$ is the Euclidean ``preparation time''. ({\bf b}) Euclidean boundary with the $\mathbb{S}^{d-1}$ suppressed. We adopt this convention, and sometimes depict this boundary with a curve or a kink to clarify diagrams (images adapted from \cite{Balasubramanian:2025zey}).}

     \label{fig:sinlge_def}
\end{figure}

 It was shown in \cite{Balasubramanian:2025zey} that the single-boundary, non-perturbative, gravity Hilbert space is spanned by the single-sided {\it shell states},\footnote{Also see \cite{Chandra:2022fwi} and \cite{Balasubramanian:2025hns}.} and we briefly review key aspects of their construction here. The single-sided shell states are defined by slicing open the Euclidean gravitational path integral with  a heavy dust shell operator inserted on the asymptotic boundary.\footnote{See \cite{Balasubramanian:2025jeu} for details on what we mean by cutting open a gravity path integral. } The cut-open boundary has topology $\mathbb{R}^{<0}\times \mathbb{S}^{d-1}$ (where $\mathbb{R}^{<0}$ is the half line) and  includes a $\mathbb{S}^{d-1}$ symmetric heavy ($m \sim \mathit{O}(1/G_{N})$) dust shell operator  $\mathcal{O}_{S}$ separated by a boundary length $\frac{\beta}{2}$ from the cut $\mathcal{X}$. In short, the state is defined by half infinite boundary time extending from $-\infty$ to  $\tau = 0$, followed by a $\mathcal{O}_S$ insertion, and finally   time evolution up to $ \tau =\beta/2$ (Fig.~\ref{fig:sinlge_def}). By varying the inertial mass $m_i$ of the shell operators $\mathcal{O}_{i}$ we can obtain an infinite family of shell states $|i\rangle$.  Throughout this paper we will  rely on the methods developed in \cite{Balasubramanian:2025jeu,Balasubramanian:2025zey} for manipulating shell states. We will  review some of the key aspects, but  refer the reader to these works for details. While we will focus here  on asymptotically AdS quantum gravity for concreteness, the discussion carries over straightforwardly to asymptotically flat gravity. See, for example, \cite{Balasubramanian:2022lnw,Balasubramanian:2025zey} for the construction of shell states in flat space.

\begin{figure}
    \centering
    \includegraphics[width=0.3\linewidth]{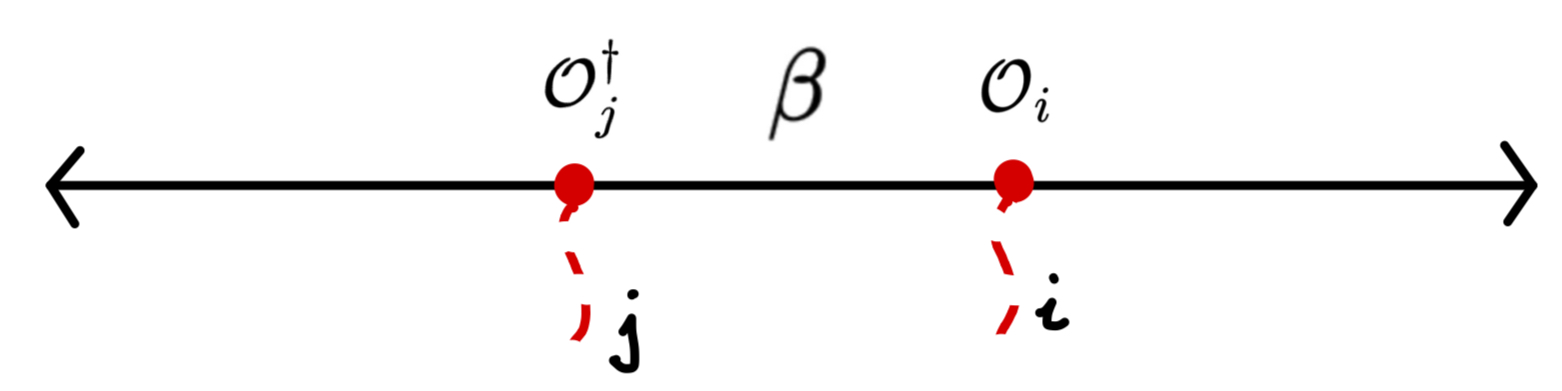}
    \caption{Shell-strip asymptotic boundary condition for the overlap $\braket{j|i}$ consisting of the line $\lim_{\alpha \to \infty} [-\alpha, \beta + \alpha]$ on which  $\mathcal{O}_{i}$ and  $\mathcal{O}^{\dagger}_{j}$ are inserted at $\tau =0 $ and  $\tau=\beta$ respectively (images adapted from \cite{Balasubramanian:2025zey}).}
    \label{fig:1s_shellbdry}
\end{figure}

We can compute the Gram matrix elements $G_{ij}\equiv \langle i|j\rangle$ by sewing the boundary condition defining the bra and ket along the cut $\mathcal{X}$. This results in an interval $[-\infty,0]$ followed by $\mathcal{O}_j$ insertion, boundary time evolution by an amount  $\beta$, $\mathcal{O}^{\dagger}_i$ insertion, and another  half line $[\beta,\infty]$ (Fig.~\ref{fig:1s_shellbdry}). We will call $\beta$  the ``length of the strip'' and denote the path integral with such a boundary condition as $S(\beta)$.\footnote{See \cite{Balasubramanian:2025zey,Balasubramanian:2025hns} and Fig.~\ref{fig:1s_shellbdry} for a specification of the limit procedure in which these strip boundary conditions are defined.} Following~\cite{Balasubramanian:2022gmo},  we  make different shell states  orthogonal at tree level in the gravitational path integral, $\overline{\langle i|j \rangle} =\delta_{ij} Z_{1} $, by taking their inertial mass differences to be arbitrarily large, since  it will take  $|m_{i}-m_{j}|$ bulk interactions in Planck units for such shells to annihilate.  However, the boundary condition for the magnitude squared of this overlap consists of two shell strip boundaries, and the gravity path integral for this quantity contains higher topology \textit{wormhole} contributions stabilized by the shell matter,  modifying the overlap to
\begin{equation} \label{eq:WHoverlap}
\overline{|\langle i|j \rangle|^2}=\overline{\langle i|j \rangle\langle j|i \rangle} = Z_{2} + \delta_{ij}Z_{1}^2 \, ,
\end{equation}
where $Z_{2}$ is a wormhole saddlepoint contribution, see \cite{Balasubramanian:2025zey,Balasubramanian:2025hns}. These wormhole contributions imply that the gravity path integral computes coarse-grained or alternatively ensemble averaged data of an underlying fine-grained theory (see \cite{Balasubramanian:2025jeu} and references therein). Understood this way the  overlap $\overline{\langle i|j \rangle}$ can vanish after coarse-graining due to erratically oscillating phases, while the coarse-grained magnitude $|\overline{\langle i|j \rangle}|^2=\overline{\langle i|j \rangle\langle j|i \rangle}$ does not. To reflect this we write quantities computed with the gravity path integral with an overline as in (\ref{eq:WHoverlap}). These non-perturbative effects imply that the heavy shell states are not truly orthogonal, and in fact have universal overlaps. Non-perturbative wormhole saddles of this kind were crucial for obtaining an microscopic account the Bekenstein-Hawking entropy \cite{Balasubramanian:2022gmo,Balasubramanian:2022lnw,Climent:2024trz}, resolving the Hilbert space factorisation puzzle, recovering unitarity in Hawking radiation \cite{Penington:2019kki,Almheiri:2019qdq}, and establishing the validity of the Gibbons-Hawking proposal for the gravity thermal partition function \cite{Balasubramanian:2025zey}. Indeed, the probe used to detect single-boundary Lorentzian state  discussed in this section will rely precisely on such a wormhole contribution. 

\begin{figure}
    \centering
    \includegraphics[width=0.5\linewidth]{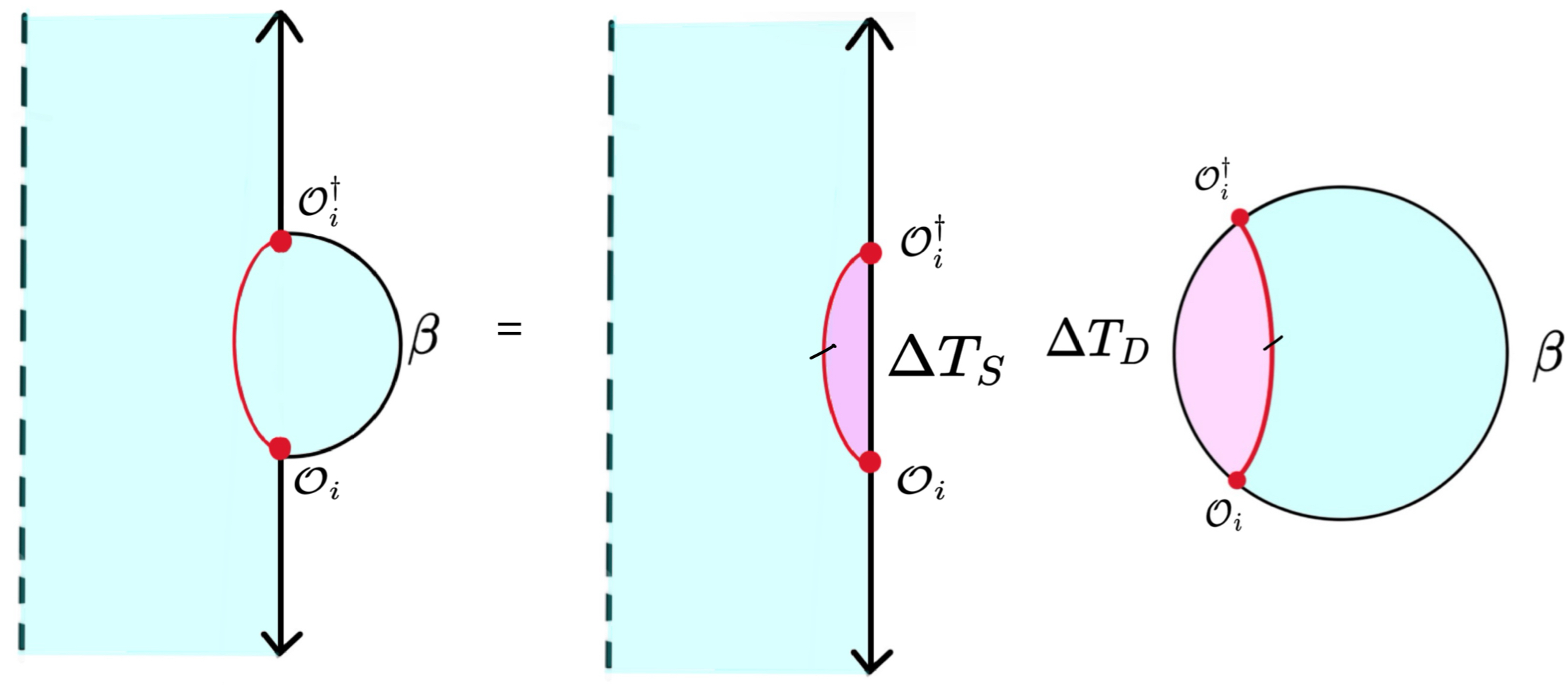}
    \caption{The saddle for the norm $\overline{\braket{i|i}}$ is constructed by considering the shell propagating on a disk and strip separately for some propagation times $\Delta T_{S,D}$ and then gluing them together along the $i$-shell worldvolume by discarding the shell homology region (purple). The junction conditions dynamically determine $\Delta T_{S,D}$ to yield an on shell glued geometry. We have suppressed the angular directions in these diagrams, and represent the radius-time plane with the dashed line representing the origin (images adapted from \cite{Balasubramanian:2025zey}).}
    \label{fig:singe_norm}
\end{figure}

\paragraph{Norm.}\label{sec:appStripNorm}  
To see how Euclidean saddle geometries subject to these \textit{shell strip} boundary conditions are constructed we evaluate the gravity path integral computing the norm $\overline{\langle i|i \rangle}$ within the saddlepoint approximation. In the saddle geometries the shells propagate into the bulk before being re-absorbed at the boundary a time $\beta$ later. Following \cite{Balasubramanian:2025zey}, we can construct these saddlepoints by filling in one side of the shell worldvolume with any of the saddle geometries for the Euclidean path integral with a periodic boundary condition of length $\beta$ (the ``disk''), we denote this path integral as $\overline{Z(\beta)}$, \footnote{It was shown in \cite{Balasubramanian:2025hns} that this path integral computes the thermal trace over the non-perturbative single-boundary gravity Hilbert space. Hence $\overline{Z(\beta)}$ is the canonical partition function of the gravity theory.}   and the other side with the Euclidean vacuum geometry with non-compact time (the ``strip" $\overline{S}$). This can be done by using the Israel junction conditions \cite{Israel:1966rt} to glue the geometries at the shell world-volume  (Fig.~\ref{fig:singe_norm}). For example, we can take a Euclidean AdS strip with a shell operator  $\mathcal{O}_{i}$ on the boundary at $\tau=0$ and the conjugate operator $\mathcal{O}^{\dagger}_{i}$  at time $\tau=\Delta T_{S}$ in addition to a Euclidean AdS disk with similar boundary insertions separated by  $\Delta T_{D}$ such that the boundary circumference of the disk is $\beta +\Delta T_{D}$. The saddlepoint geometry is constructed by  discarding the shell homology regions (purple  in Fig.~\ref{fig:singe_norm}) from the strip and disk and using the junction conditions to glue the remainder along the shell world-volumes. The junction conditions dynamically determine $\Delta T_{S}$ and $\Delta T_{D}$ such that the resulting geometry satisfies the equations of motion.

\paragraph{Large shell mass limit.} 
As discussed in \cite{Balasubramanian:2022gmo,Balasubramanian:2025zey}, in the  $m_i \to \infty$  limit the turning point of the shell in both the strip and disk bulk approaches the asymptotic boundary and the propagation times go to zero $\Delta T_{S},\Delta T_{D} \to 0$. Hence in this limit the shell homology regions pinch off and the remainder grows to cover the entire strip and disk saddle geometry. Furthermore, in this limit the shell contribution to the saddle action is a simple function it its mass, contributing a factor $Z_{m_i}\sim e^{-2(d-1)log(\mathrm{G_N} m_i)}$. In the above construction we could have glued any of disk saddles to the strip. For asymptotically AdS boundary conditions there are three disk saddles: the large/small AdS black hole (BH$_{L,S}$) and thermal AdS (TAdS), and there are therefore also three saddles to $\overline{\langle i|i \rangle}$. In the saddlepoint approximation we have 
\beq \label{eq:partitionfunc}
\overline{Z(\beta)}\approx e^{-I^{(BH_L)}(\beta)} +e^{-I^{(BH_S)}(\beta)} + e^{-I^{(TAdS)}(\beta)}.
\eeq
In the large shell mass limit we can therefore economically write this sum over saddles as:
\beq \label{eq:overlap}
\overline{\langle i|j \rangle} = \delta_{ij} \, Z_{m_i} \times 
\overline{Z}(\beta)\times \overline{S(}0)\, .
\eeq

\paragraph{Saddle geometries.} 
Following \cite{Balasubramanian:2025zey}, we associate a geometry to a shell state by considering the leading saddle geometry for the norm
$\overline{\langle i|i\rangle}$. In the large shell mass limit the norm is given by (\ref{eq:overlap}) and as there is only one strip saddle  the dominance is determined by the leading saddle to $\overline{Z(\beta)}$. In the case of AdS asymptotics the leading saddle to $\overline{Z(\beta)}$ is thermal Euclidean AdS for $\beta > \beta _{HP}$  and the large Euclidean black hole for $\beta < \beta _{HP}$, where $\beta_{HP}$ marks the Hawking-Page transition \cite{Hawking:1982dh}. In the case of flat asymptotics the thermal flat space  saddle is always leading, although the black hole can be made leading by a micro-canonical projection.\footnote{See, for example, Sec.~3.3 in \cite{Balasubramanian:2025hns}, and see \cite{Barbon:2025bbh} for discussion about the stability of the canonical saddles.}  The time reflection symmetric slice of the leading saddle can be continued to Lorentzian signature, giving two classes of single-sided shell states:
\begin{itemize}
\item {\bf Type A}: The leading saddle is the Euclidean  black hole, so the Lorentzian geometry is a black hole with a shell  in the interior, capped off by the vacuum behind the shell. In AdS space we find these saddles when  $\beta < \beta _{HP}$ and in asymptotically flat space in the finite energy micro-canonical ensemble. These are known as \textit{type A} shell states.
\item  {\bf Type B}: The leading saddle has a non-contractible thermal circle  and the Lorentzian geometry is the thermal space with a disconnected  closed ``Big Crunch" universe in which the shell propagates. In AdS space we find these saddles for  $\beta >\beta _{HP}$ and in asymptotically flat space for any $\beta $ in the canonical ensemble. These are known  as \textit{type B}  shell states.
\end{itemize}

Interestingly, in both the type A and type B shell bases the region in which the shell propagates (which therefore contains crucial information about the gravity state) is causally disconnected from the asymptotic boundary. Furthermore, it was shown in \cite{Balasubramanian:2025zey} that sufficiently large sets of either type A or B shell states form a complete basis for the non-perturbative single-boundary gravity Hilbert space. Hence any gravity state  can be expression as superpositions of these shell state. We will therefore first focus on detecting a particular shell state, and extend this to an arbitrary superposition in Sec.~\ref{sec:superposition}.

\begin{figure}[h]
  \begin{subfigure}[c]{0.42\linewidth}
    \centering
    \includegraphics[width=0.7\linewidth]{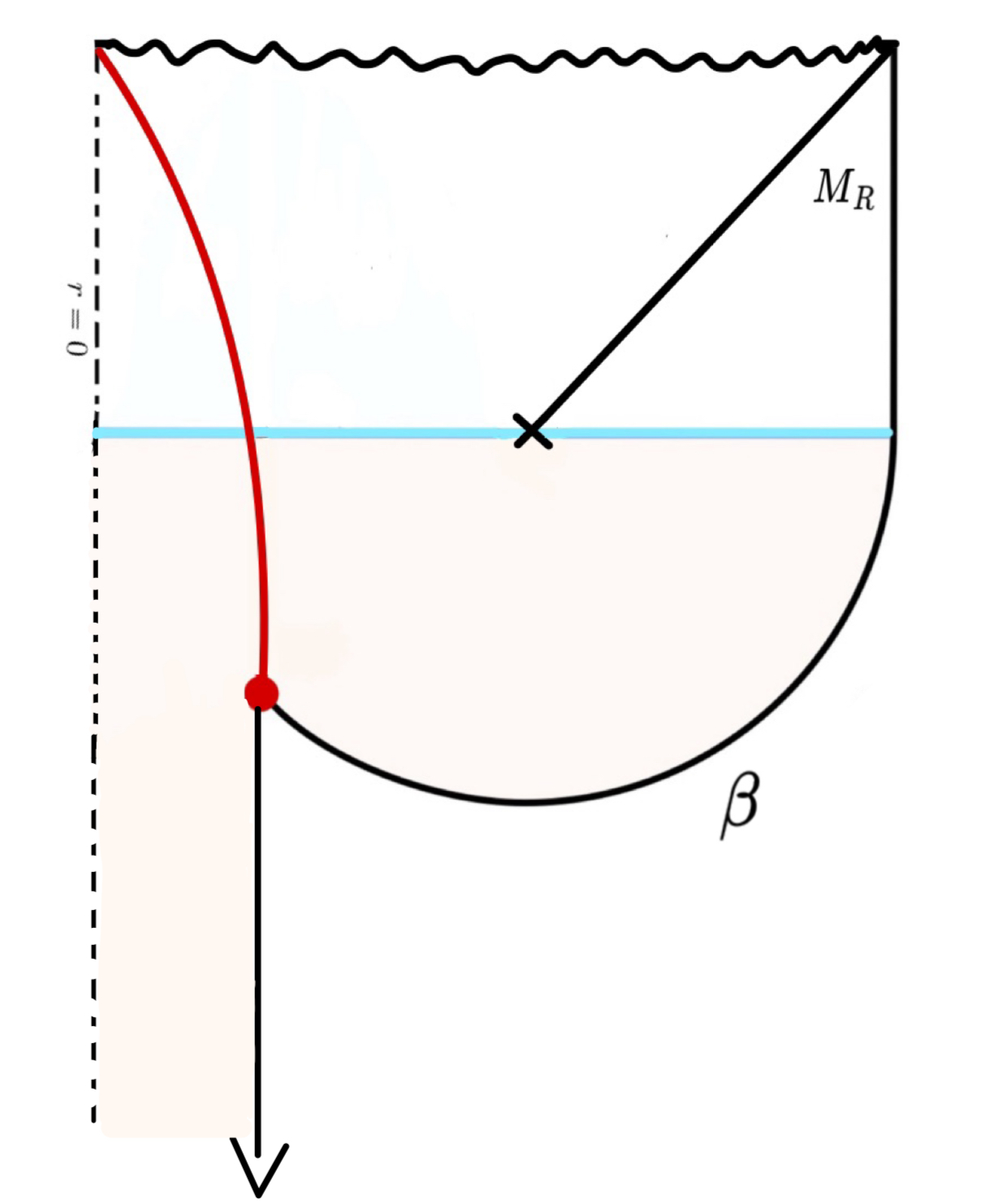}
    \caption{}
    \label{fig:type1}
    \end{subfigure}
    \hfill
 \begin{subfigure}[c]{0.42\linewidth}
    \centering
    \includegraphics[width=0.7\linewidth]{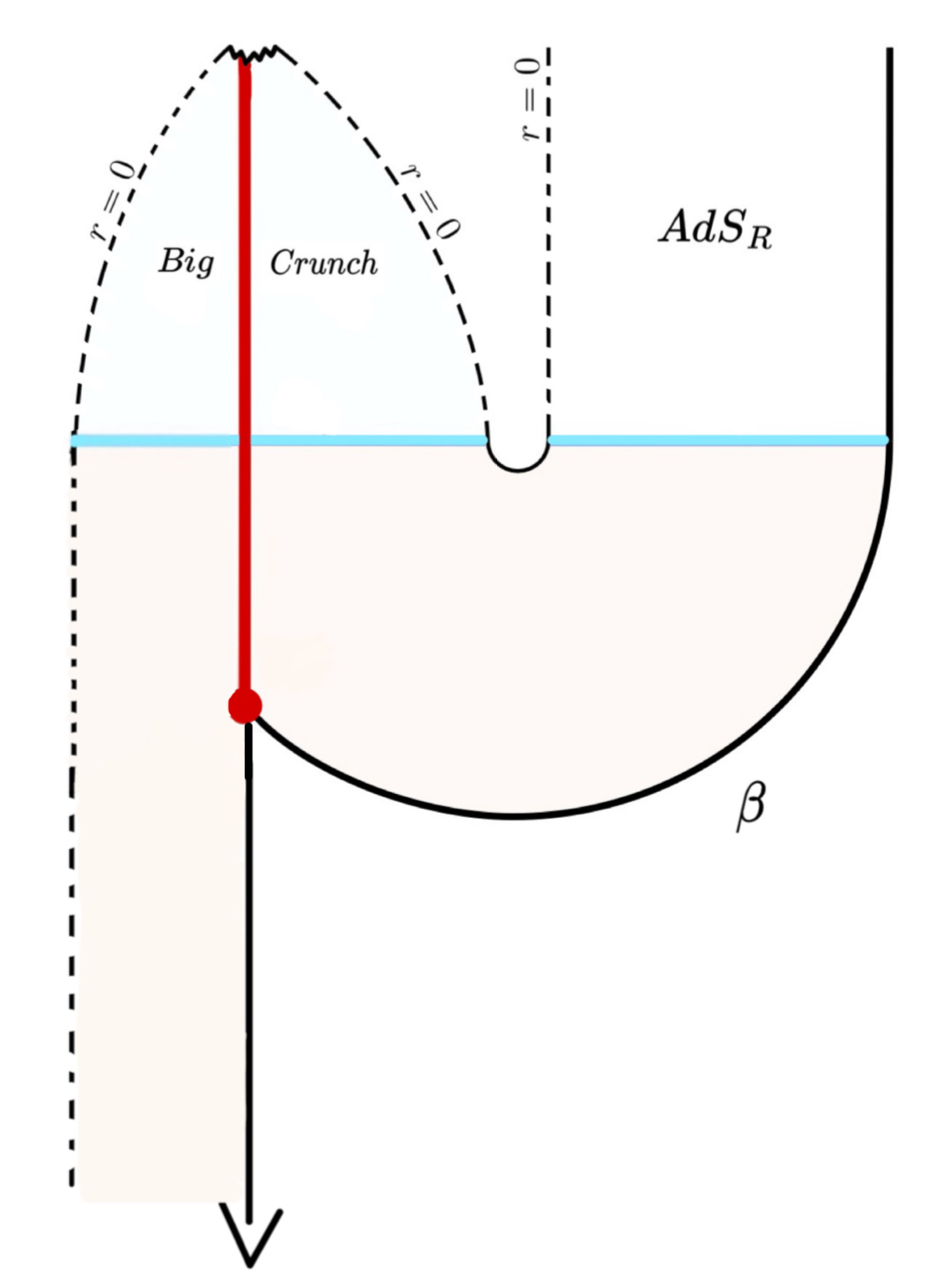}
    \caption{}
    \label{fig:type2}
    \end{subfigure}
\caption{Analytic continuation of the single-sided shell state saddlepoints to  Lorentzian signature, shown here with asymptotically AdS boundary conditions. For asymptotically flat boundary conditions the vertical lines are replaced by diamonds.  ({\bf a}) Type A shell state corresponding to a single-sided black hole. ({\bf b}) Type B shell state consisting of thermal AdS (or thermal flat space)  with and added disconnected compact Big-Crunch AdS cosmology. Images adapted from \cite{Balasubramanian:2025zey}. }
\end{figure}

\subsection{Single-Boundary Lorentzian Probes}\label{sec:singbdrprobes}
Suppose the universe is in the shell basis state $\ket{i}$, prepared by inserting the shell operator $\mathcal{O}_i$ and time evolving with inverse temperature $\beta$ (Fig.~\ref{fig:sinlge_def}). In this section we show that there exist boundary probes in the gravity theory that can test a proposal for which shell operator $\mathcal{O}_i$ was used to make the state. Whenever such probes exist we will say that a boundary observer can \textit{detect} the state. In Sec.~\ref{sec:QMA} we  discuss how a boundary observer could use these probes to actually verify the state of the universe. 

 Remarkably, these probes exist even though the shell worldvolume is not is causal contact with the boundary.  To see how, consider the magnitude squared of a \textit{probe correlator} $\braket{i|\mathcal{O}_P|i}$ given by $\braket{i|\mathcal{O}_P|i}\braket{i|\mathcal{O}_P^{\dagger}|i}=|\braket{i|\mathcal{O}_P|i}|^2$ where the probe $\mathcal{O}_P$ is some shell operator of mass $m_p$. 
 \begin{figure}
 \centering
\includegraphics[width=0.2\linewidth]{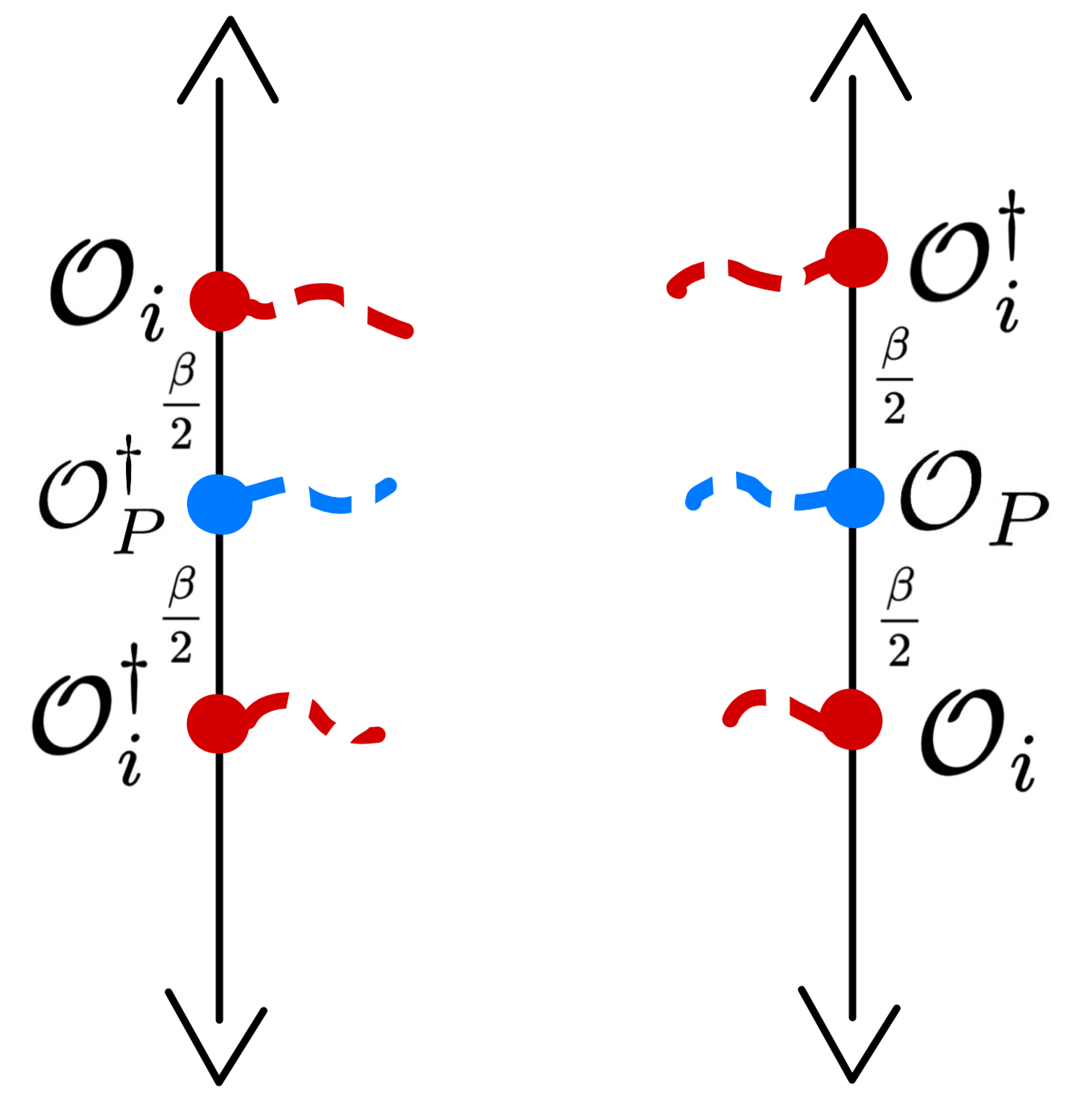}
\caption{ Path integral boundary condition for $\braket{i|\mathcal{O}_P|i}\braket{i|\mathcal{O}_P^{\dagger}|i}$. }
\label{fig:1sLprobecorbc}
\end{figure}
The gravity path integral boundary condition defining the probe correlator $\braket{i|\mathcal{O}_P|i}$ is similar to that computing the norm of the single-sided shell state (Fig.~\ref{fig:1s_shellbdry}) apart from the additional $\mathcal{O}_P$ insertion at $\tau=0$ and similarly for $\braket{i|\mathcal{O}_P^{\dagger}|i}$. By the same reasoning that lead to $\overline{\braket{i|j}}\approx\delta_{ij}$ above we have that $\overline{\braket{i|\mathcal{O}_P|i}} \approx 0$.  However, we again interpret this to be a by-product of the performing the coarse-gained/ensemble averaged nature of the gravity path integral, where erratic phases cause the correlator to average out to zero. Still, the  gravitational path integral can be used to estimate the magnitude of this correlator by computing  $\overline{\braket{i|\mathcal{O}_P|i}\braket{i|\mathcal{O}_P^{\dagger}|i}}$, which does not depend on such phases.  We aim to evaluate this gravity path integral in the saddlepoint approximation. As we will see, the upshot is that if the observer chooses the ``wrong" probe operator $\mathcal{O}_P\neq\mathcal{O}_i$ there is a universal response that jumps up once the ``right" probe operator $\mathcal{O}_P=\mathcal{O}_i$ is selected, allowing the state to be detected.

\begin{figure}
\begin{subfigure}{0.3\linewidth}
    \centering
    \includegraphics[width=0.6\linewidth]{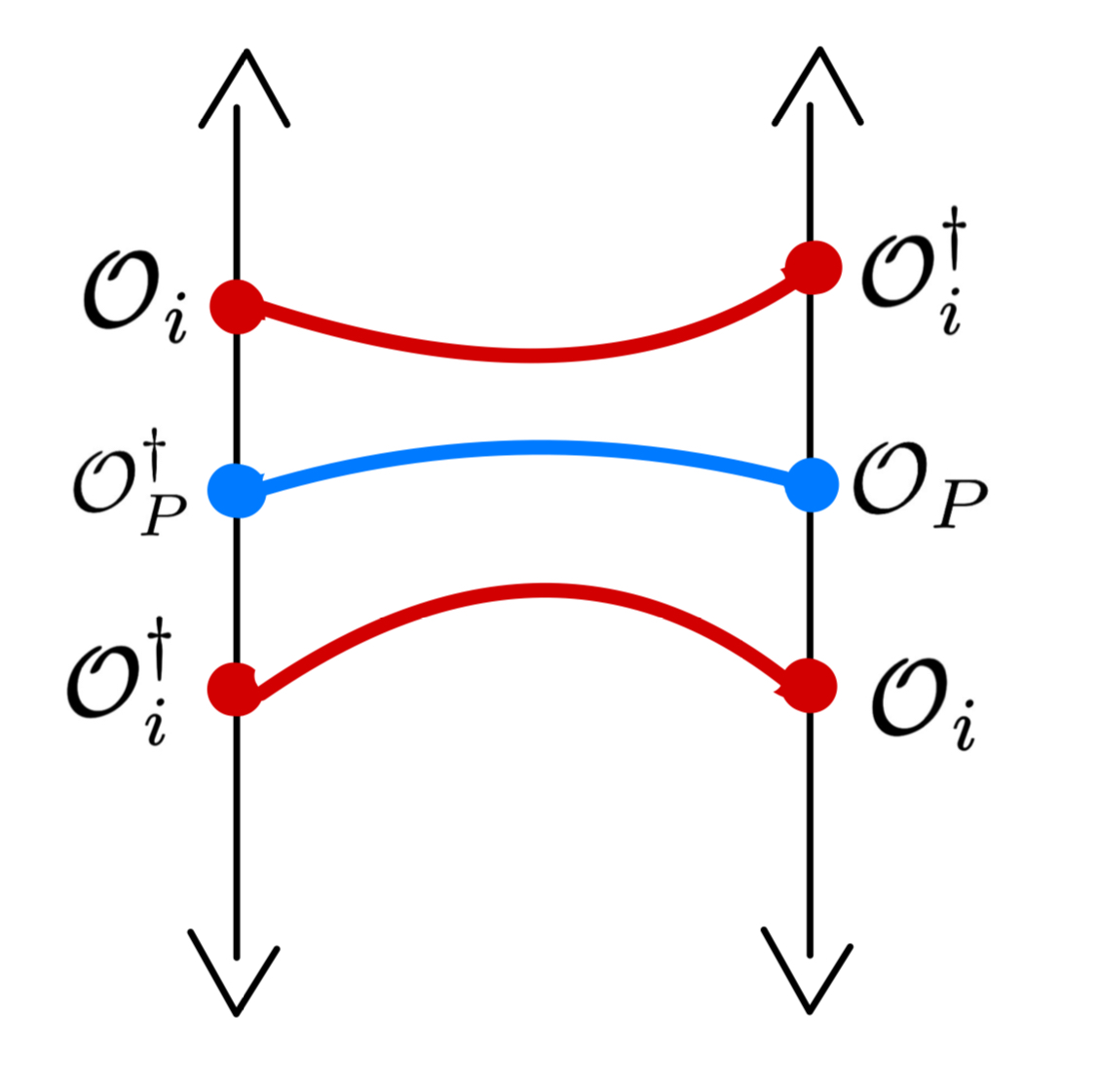}
    \caption{$U_L$}
    \label{fig:U_L}
    \end{subfigure}
\hfill
\begin{subfigure}{0.3\linewidth}
    \centering
    \includegraphics[width=0.6\linewidth]{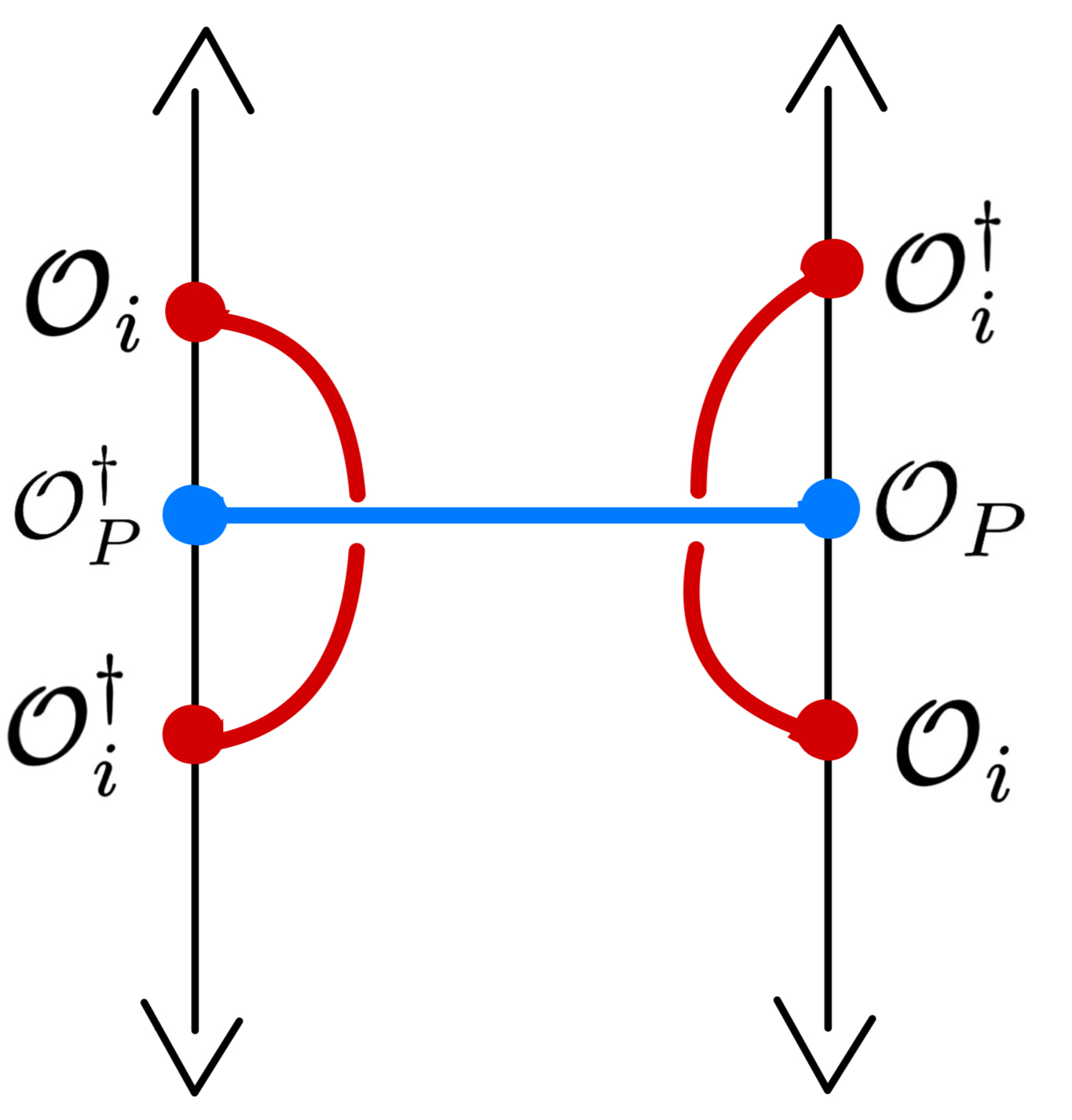}
    \caption{$U_R$}
    \label{fig:U_R}
    \end{subfigure}
\hfill
\begin{subfigure}{0.3\linewidth}
    \centering
    \includegraphics[width=0.6\linewidth]{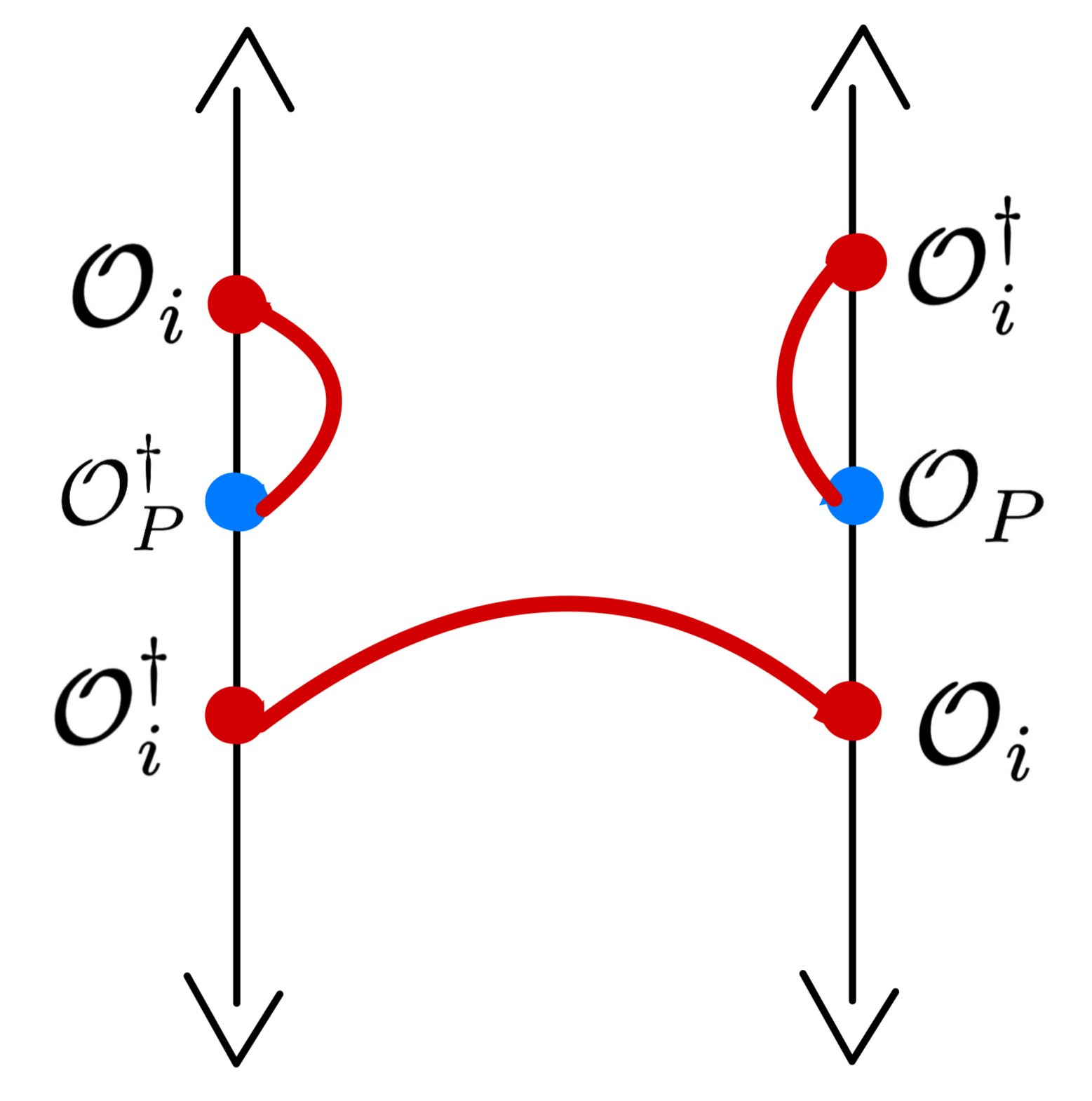}
    \caption{$D^{\uparrow}_L$}
    
    \end{subfigure}
\hfill
\begin{subfigure}{0.3\linewidth}
    \centering
    \includegraphics[width=0.6\linewidth]{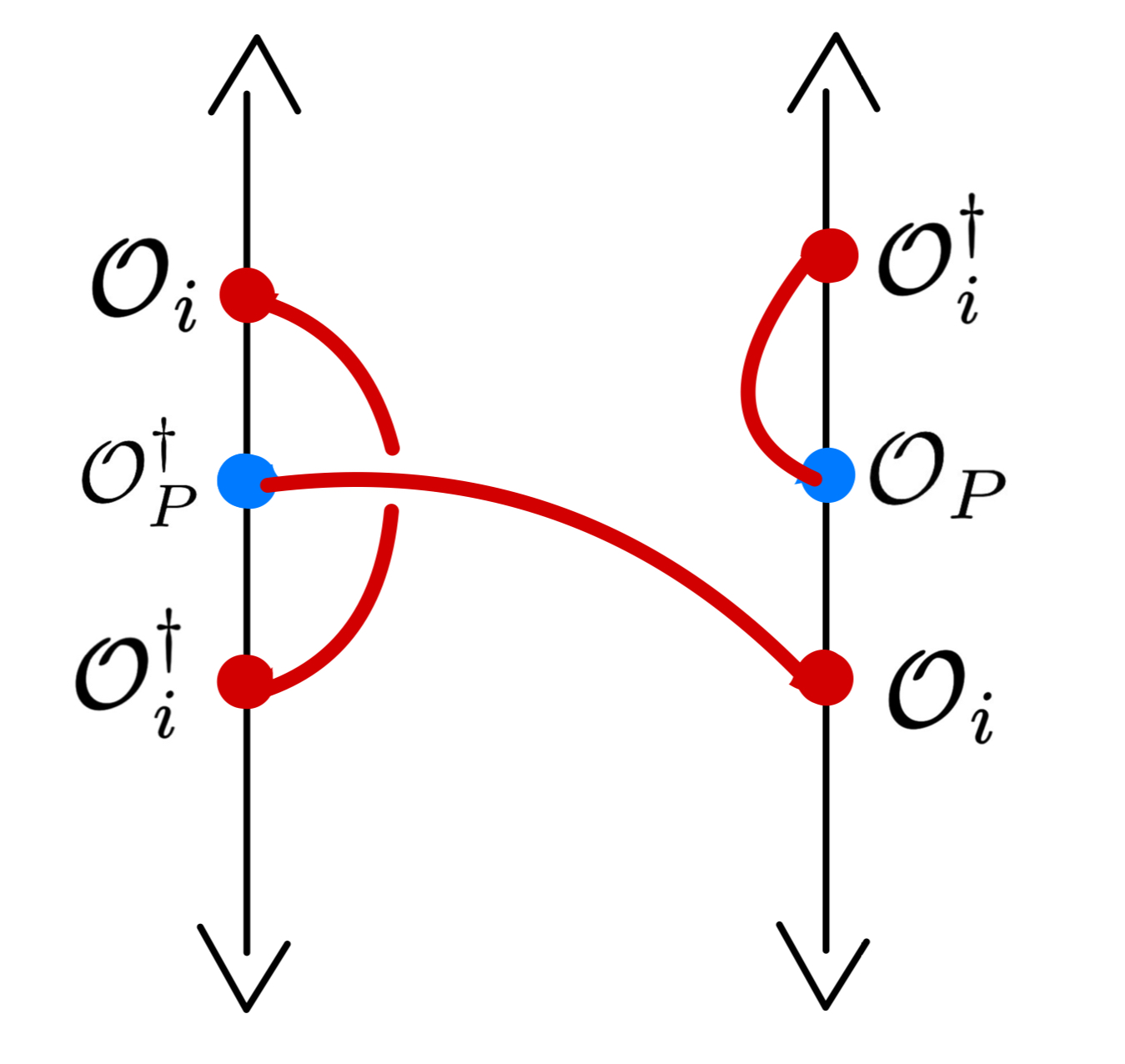}
    \caption{ $D^{\downarrow}_R$}
   
    \end{subfigure}
\hfill
\begin{subfigure}{0.3\linewidth}
    \centering
    \includegraphics[width=0.6\linewidth]{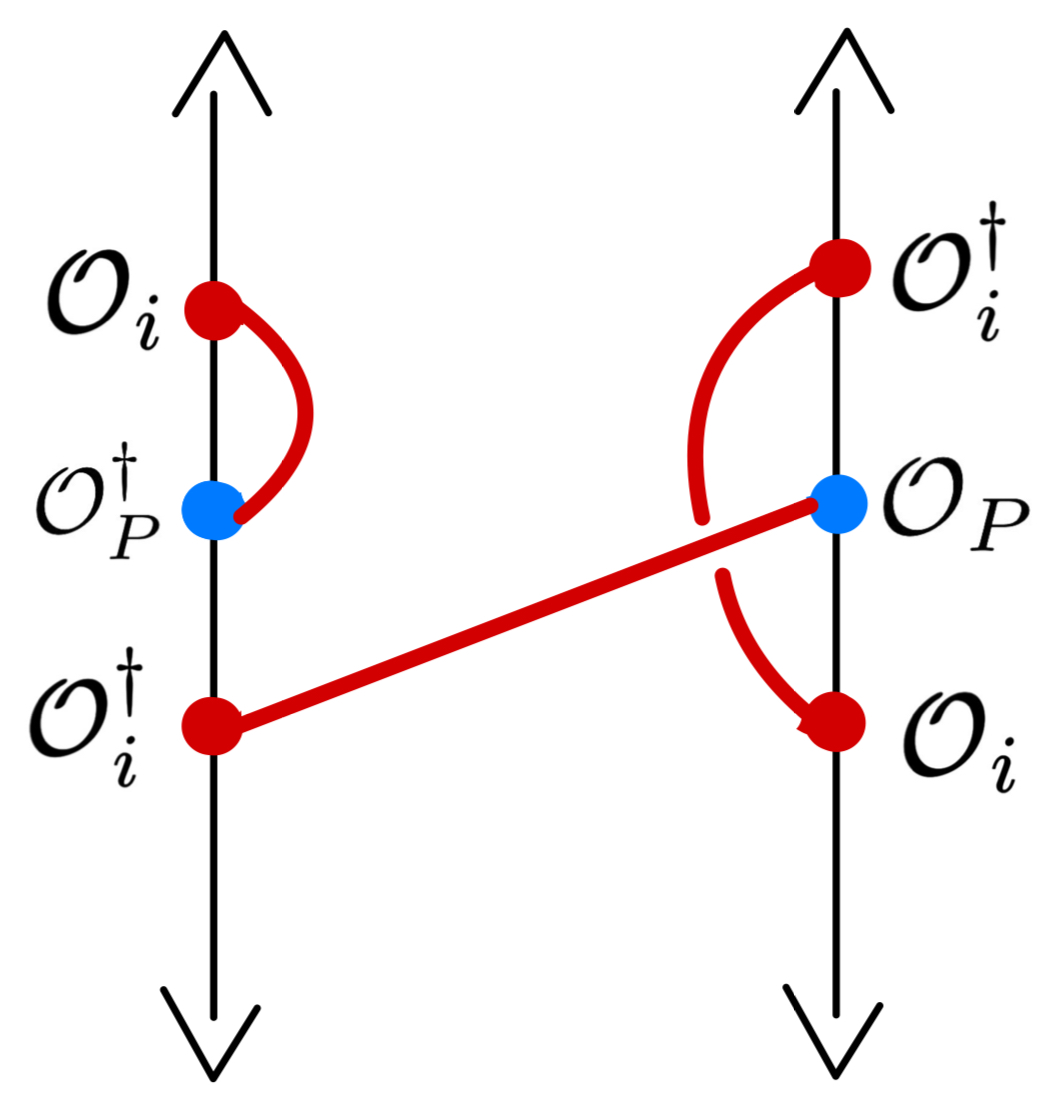}
    \caption{$D^{\downarrow}_L$}
    
    \end{subfigure}
\hfill
\begin{subfigure}{0.3\linewidth}
    \centering
    \includegraphics[width=0.6\linewidth]{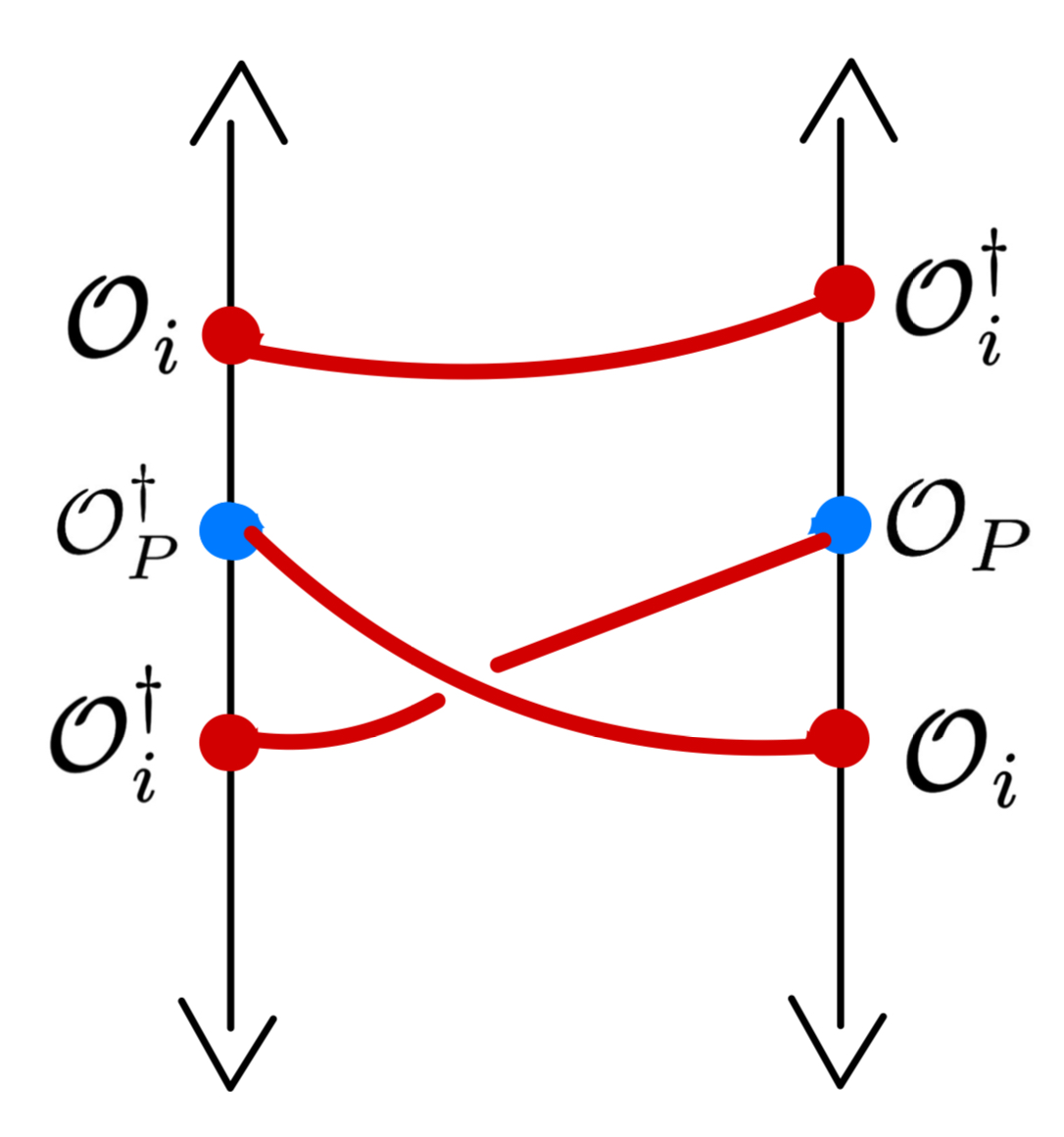}
    \caption{$D^{\uparrow}_R$}
    \label{fig:nonplanarprop}
\end{subfigure}
 \caption{Schematic of the  classes of saddle geometry that contribute to $\braket{i|\mathcal{O}_P|i}\braket{i|\mathcal{O}_P^{\dagger}|i}$. The top row are the universal saddles that always contribute. The lower two rows are the \textit{detection} saddles that only contribute once $\mathcal{O}_P=\mathcal{O}_i$. }
        \label{fig:all_saddle}
\end{figure}

The boundary condition for $\overline{\braket{i|\mathcal{O}_P|i}\braket{i|\mathcal{O}_P^{\dagger}|i}}$ consists of two disconnected asymptotic boundaries (see Fig.~\ref{fig:1sLprobecorbc}). The Euclidean bulk can be filled in either with two disconnected geomtries for each boundaries or by  wormhole geometries connecting the two boundaries. As discussed above, within the saddlepoint approximation $\overline{\braket{i|\mathcal{O}_P|i}}\approx\overline{\braket{i|\mathcal{O}_P^{\dagger}|i}}\approx 0$ as splitting and joining the different shells requires $m_P$ bulk interactions in Planck units and is therefore highly suppressed. Hence the disconnected contribution to $\overline{\braket{i|\mathcal{O}_P|i}\braket{i|\mathcal{O}_P^{\dagger}|i}}$ is negligibly small. If $\mathcal{O}_P\neq \mathcal{O}_i$ there are two classes of wormhole saddles which we collectively refer to as the \textit{universal} saddles, see the top row of Fig.~\ref{fig:all_saddle}. However if $\mathcal{O}_P=\mathcal{O}_i$ there are four \textit{additional} classes of wormhole saddles. As the existence of these additional saddles allows for state detection we refer to them as the \textit{detection} saddles, see the lower two rows of Fig.~\ref{fig:all_saddle}. These universal and detection saddles are simple to construct if we consider the large shell mass limit for both $\mathcal{O}_{i}$ and $\mathcal{O}_{P}$, and we will work in this limit for the remainder of this paper. As these wormhole saddles are constructed  analogously to the ones discussed in  \cite{Balasubramanian:2025zey,Balasubramanian:2025hns}, we will be brief and refer the reader to these works for details.

\begin{figure}
\begin{subfigure}{0.45\linewidth}
    \centering
    \includegraphics[width=\linewidth]{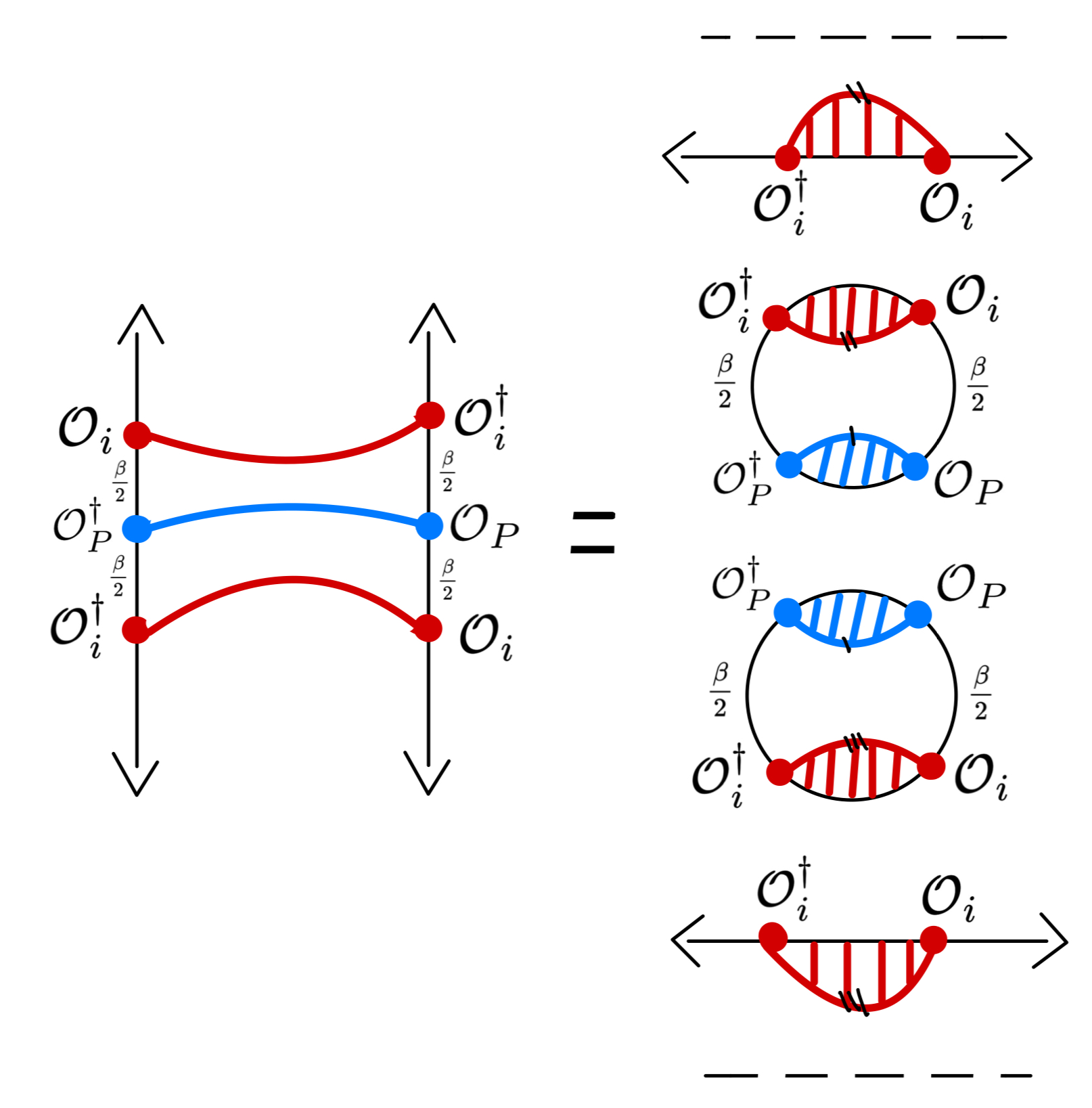}
   \caption{}
    \label{fig:planarprop}
    \end{subfigure}
\hfill
\begin{subfigure}{0.45\linewidth}
    \centering
    \includegraphics[width=\linewidth]{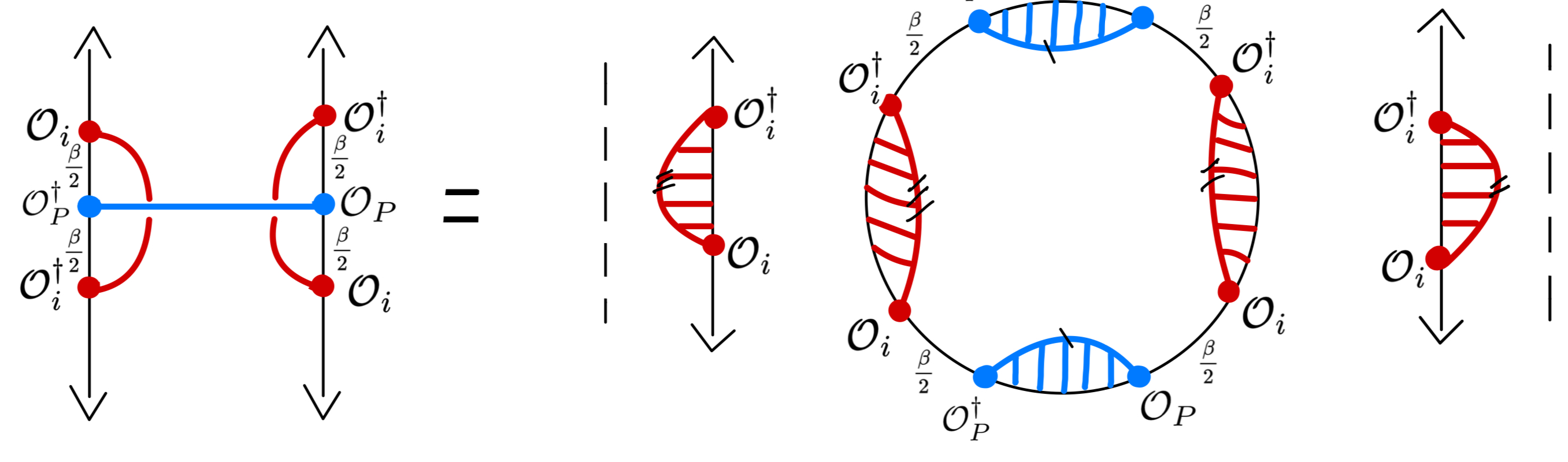}
    \caption{}
    \label{fig:nonplanarprop}
\end{subfigure}
\caption{Construction of the two universal saddles to $\overline{\braket{i|\mathcal{O}_P|i}\braket{i|\mathcal{O}_P^{\dagger}|i}}$ that contribute even when $\mathcal{O}_P\neq \mathcal{O}_i$. (\textbf{a}) The $U_L$ saddles are constructed by gluing two strips into a disk twice and glugin the resulting geometries together along the $\mathcal{O}_P$ shell. (\textbf{b})  The $U_R$ saddles are constructed by gluing two strips into a central disk, and gluing two opposite $\mathcal{O}_P$ shells together on this same disk.  }
\end{figure}

\paragraph{Universal saddles.}
If $\mathcal{O}_i \neq \mathcal{O}_P$ the shell sourced by $\mathcal{O}_P$ must propagate from the asymptotic boundary into the bulk and be absorbed by the $\mathcal{O}^{\dagger}_P$ insertion on the other asymptotic boundary. The $\mathcal{O}_i $ shells can either propagate to the $\mathcal{O}^{\dagger}_i$ insertion on the same asymptotic boundary (Fig.~\ref{fig:U_R}) or propagate to the $\mathcal{O}^{\dagger}_i$ insertion on the second boundary (Fig.~\ref{fig:U_L}), resulting in two classes of saddles. The saddles of the latter kind are depicted in Fig.~\ref{fig:planarprop} and are constructed as follows. Consider a disk saddle  on which a $P$ and $i$ shell propagate from the boundary into the bulk before being re-absorbed at the boundary a time $T_{P}$ and $T_{i,D}$ later respectively. Furthermore consider an $i$ shell inserted on the Euclidean strip propagating into the bulk before being absorbed a boundary time $T_{i,S}$ later. First we discard the $i$ shell homology regions on the disk and the strip and glue the resulting geometries together along the $i$-shell worldvoumes using the junction conditions. The junction conditions dynamically determine $T_{i,S},T_{i,D}$ such that the resulting geometry satisfies the equations of motion. Next, repeat this entire procedure to obtain another copy of that geometry. On each of these copies discard the $P$ shell homology regions and glue the resulting geometries together along the $P$ shell world-volume using the junction conditions, which will determine $T_{P}$. In the large shell mass limit we have $T_{P},T_{i,S},T_{i,D}\to 0 $, the shell homology regions pinch off and the shells contribute universally to the action. Hence in this limit the contribution of these saddles is given by
$\overline{Z}(\beta)^2\times \overline{S}(0)^2\times Z_{m_i}^2 Z_{m_P}$ where $Z_{m_i}, Z_{m_P}$ are the shell matter contributions. 

In the second class of saddles  the $\mathcal{O}_i $ shells connect to the $\mathcal{O}^{\dagger}_i$ insertion on the \textit{same}  boundary are depicted in Fig.~\ref{fig:nonplanarprop} and are constructed as follows. First
consider a disk containing two $P$-shell and two $i$-shell insertions in an alternating fashion, i.e. $P-i-P-i$. This disk is glued into a cylinder by discarding the $P$-shell homology regions and gluing the resulting geometry along the $P$-shell worldvolumes. This procedure is similar to the construction of the so-called ``folded" wormholes in \cite{Balasubramanian:2025jeu}. Next consider two copies of a strip containing an $i$-shell and discard the $i$-shell homology regions on these strips and on the cylinder. Each of the strips is then glued into one of the $i$ shells on the cylinder. In the large shell mass limit all the propagation times again go to zero and the shell homology regions pinch off completely. Hence these saddles contribute a factor $\overline{Z}(2\beta)\times \overline{S}(0)^2\times Z_{m_i}^2 Z_{m_P}$.

Collecting everything together, the contribution to $\overline{\braket{i|\mathcal{O}_P|i}\braket{i|\mathcal{O}_P^{\dagger}|i}}$  from the universal saddles is given by
\beq \label{eq:1spropsaddles}
Z_U= (\overline{Z}(\beta)^2\times \overline{S}(0)^2 + \overline{Z}(2\beta)\times \overline{S}(0)^2) \times Z_{m_i}^2 Z_{m_P}.
\eeq
To obtain the physical value of the correlator we should normalise the shell states $\ket{i}\to \ket{i} / \sqrt{\braket{i|i}}$, which we do by dividing (\ref{eq:1spropsaddles}) by $\overline{\braket{i|i}\braket{i|i}}$.\footnote{Here we follow \cite{Antonini:2023hdh} and assume the normalized quantity is obtained by computing the quantity and the normalization separately using the gravitational path integral and then dividing. However, we will mostly be interested in the ratios between quantities where the normalisation drops out altogether. } This path integral contains  disconnected and wormhole contributions that can be constructed analagously to the saddles above. The disconnected contribution is simply the square of (\ref{eq:overlap}). To see how the wormhole geometries are constructed consider a disk of length $2\beta+2\Delta T_D$ on which two shells propagate for a time $\Delta T_D$, separated from each other on either side by a length $\beta$. Consider also two strips of length $\Delta T_S$ on which shells propagate for a time $\Delta T_S$. These two strips are then glued into disk by discarding the shell homology regions and identifying the corresponding shell worldvolumes. This results in $\overline{Z}(2\beta)\times \overline{S}(0)^2 \times Z_{m_i}^2$. Hence in total $\overline{\braket{i|i}\braket{i|i}}=(\overline{Z}(\beta)^2\times \overline{S}(0)^2 + \overline{Z}(2\beta)\times \overline{S}(0)^2 )\times Z_{m_i}^2$ (see for example \cite{Balasubramanian:2025zey,Balasubramanian:2025hns}). 

Remarkably, the normalised universal contribution simply becomes $Z_U= Z_{m_P} $. This result is universal for any shell state probe that is not equal to $\mathcal{O}_i$, and contains no information about the state whatsoever. This is in line with the expectation from \cite{Balasubramanian:2005mg} that generic probes of the black hole interior should yield universal responses. 

\begin{figure}

\begin{subfigure}{\linewidth}
    \centering
    \includegraphics[width=0.7\linewidth]{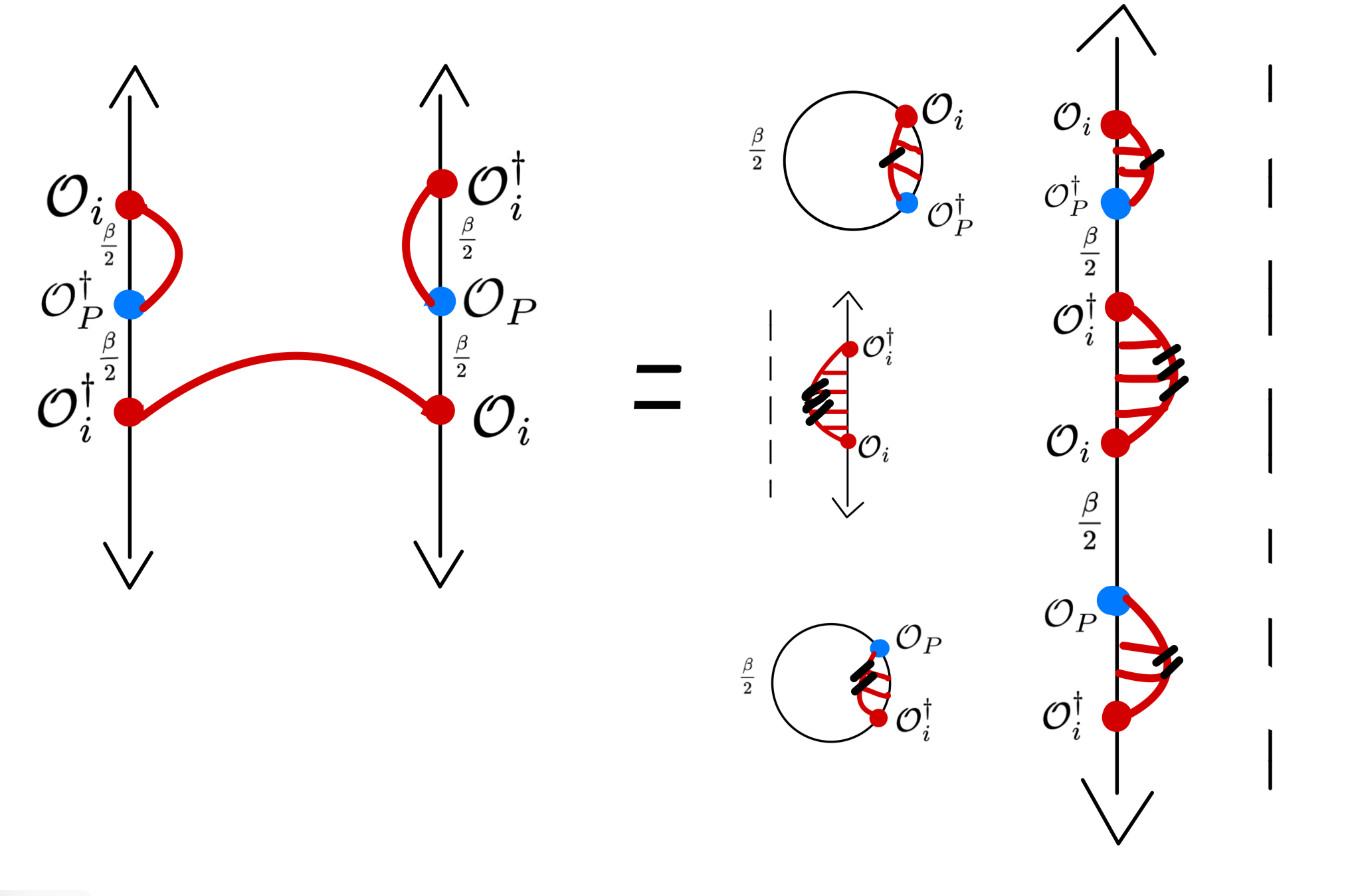}
    \caption{}
    \label{fig:UU_L_const}
    \end{subfigure}
\hfill
\begin{subfigure}{\linewidth}
    \centering
    \includegraphics[width=\linewidth]{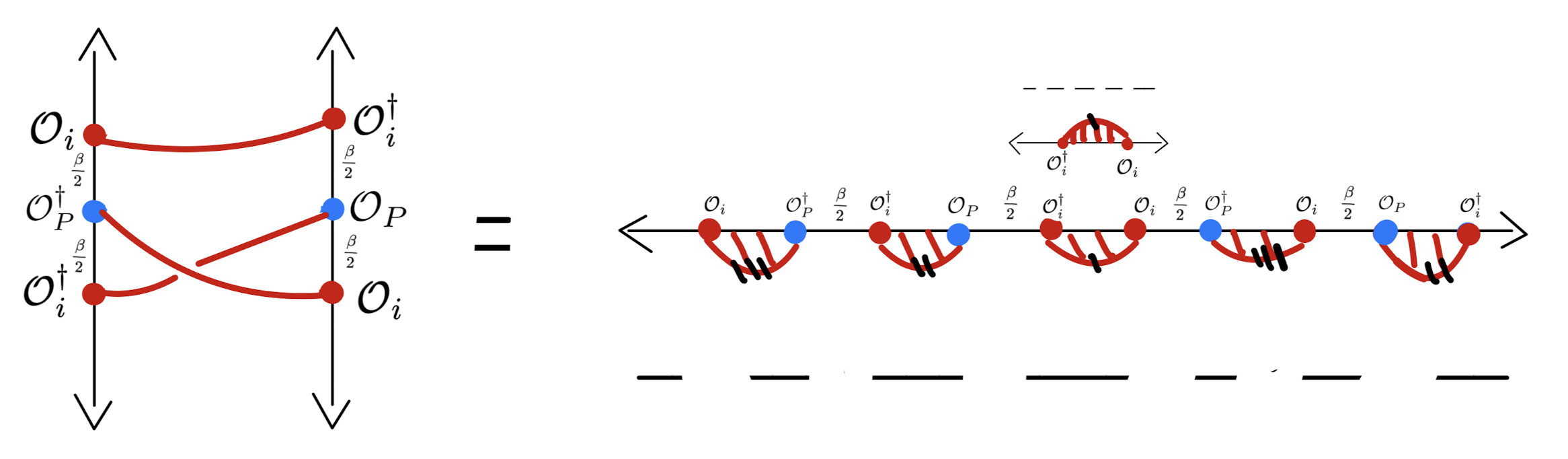}
    \caption{}
    \label{fig:Down_R_const}
\end{subfigure}
\begin{subfigure}{\linewidth}
    \centering
    \includegraphics[width=\linewidth]{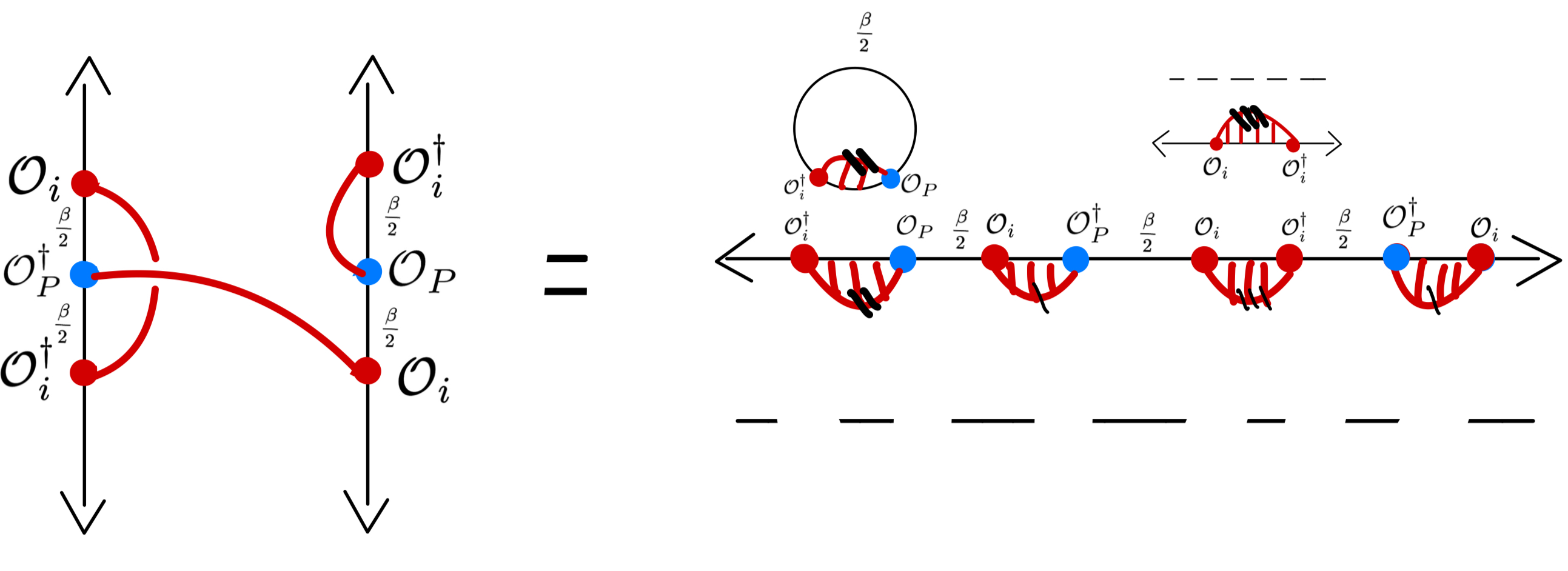}
    \caption{}
    \label{fig:UU_R_const}
\end{subfigure}
\caption{(\textbf{a}) Construction of the $D^{\uparrow}_L$ saddles. (\textbf{b}) Construction of the $D^{\uparrow}_R$ saddles. (\textbf{c}) Construction of the $D^{\downarrow}_R$ saddles.} 
\end{figure}

\paragraph{Detection saddles.}
It was also argued in \cite{Balasubramanian:2005mg} that probing the state with the same operator as the one used to make the state, in this case the heavy shell state $\mathcal{O}_i$, should give a large response, allowing the state to be detected. If  $\mathcal{O}_P=\mathcal{O}_i$ there are $3!=6$ different ways of connecting up the shells, and therefore six classes of saddles. The saddles in which $\mathcal{O}_P$ connects to $ \mathcal{O}^{\dagger}_P$ correspond to the universal saddles above. The additional four classes of saddles are the detection saddles in which  $ \mathcal{O}_P^{(\dagger)}$ are absorbed by one of the $\mathcal{O}_i ^{(\dagger)}$ insertions. To see how these detection saddles are constructed we label the boundary on which  $ \mathcal{O}_P$  is inserted as $B_1$, the one with $ \mathcal{O}^{\dagger}_P$ as $B_2$, and use ($\mathcal{O}_P$, $ \mathcal{O}^{\dagger}_{i,B_1}$) to denote that the $\mathcal{O}_P$ shell is absorbed by the $ \mathcal{O}^{\dagger}_{i}$ insertion on $B1$ etc.  There are then four options to consider:
\begin{enumerate}
    \item ($\mathcal{O}_P$, $ \mathcal{O}^{\dagger}_{i,B_1}$), ($\mathcal{O}^{\dagger}_P$, $ \mathcal{O}_{i,B_2}$), ($ \mathcal{O}_{i,B_1}$, $ \mathcal{O}^{\dagger}_{i,B_2}$): We refer to this group of saddles as $D^{\uparrow}_L$. These saddles are constructed by starting with a strip on which three shells are placed with propagation times $T_{S1},T_{S2},T_{S3}$. We glue disks  of boundary length $\beta/2 +T_{1D}$ and $\beta/2 +T_{3D}$ into each of shell 1 and 3 respectively, where $T_{1,3D}$ are  shell propagation times on the disk. Additionally, a strip containing a shell with propagation time $T_{2S}$ is glued into shell 2  along the worldvolume of this shell, completing the construction of the saddle. In the large shell mass limit all propagation times go to zero and these saddles contribute a factor $\overline{Z}(\frac{\beta}{2})^2\times \overline{S}(\beta)\times \overline{S}(0) \times Z_{m_i}^2 Z_{m_P}$. See Fig.~\ref{fig:UU_L_const}.

     \item  ($\mathcal{O}_P$, $ \mathcal{O}^{\dagger}_{i,B_2}$), ($   \mathcal{O}^{\dagger}_{P},\mathcal{O}_{i,B_1}$)($ \mathcal{O}_{i,B_2}$, $ \mathcal{O}^{\dagger}_{i,B_1}$): We refer to this group of saddles as $D^{\uparrow}_R$. Consider a ``central" shell strip on with five shells propagate, which we number 1-2-3-4-5, each separated from each other by $\beta /2$ time evolution on the boundary and with propagation times $T_{1 \cdots 5}$.  First glue a strip on which a shell propagates for a time $T_S$ into this central strip along the worldvolume of shell 3. Then discard the homology regions of shell 2 and 4 and glue the resulting geometry together along their worldvolumes, and to the same for the 1 and 5 shell. In the large shell mass limit we are left with $\overline{S}(2\beta) \times\overline{S}(0)\times Z_{m_i}^2 Z_{m_P}$.  See Fig.~\ref{fig:UU_R_const}.
     
     \item ($\mathcal{O}_P$, $ \mathcal{O}^{\dagger}_{i,B_1}$), ($\mathcal{O}^{\dagger}_P$, $ \mathcal{O}_{i,B_1}$), ($ \mathcal{O}_{i,B_2}$, $ \mathcal{O}^{\dagger}_{i,B_2}$):  We refer to this group of saddles as $D^{\downarrow}_R$.
     Consider a central shell strip on with four shells propagate, which we number 1-2-3-4 from above to below, each separated from each other by $\beta /2$ time evolution on the boundary and with propagation times $T_{1 \cdots 4}$.   First the central strip is folded in on itself by gluing the 1 shell to the 3 shell by cutting out the homology regions and gluing the resulting geometry along the shell worldvolumes. Then a disk with a boundary length  $\beta/ 2+T_{D}$ is glued into the central strip along the worldvolume of shell 4, and an additional strip on which a shell propagates for a time $T_S$ is glued into the central strip along the worldvolume of shell 2. In the large shell mass limit the propagation times go to zero and the shell homology regions pinch off, giving $\overline{S}(\frac{3\beta}{2}) \times \overline{Z}(\frac{\beta}{2}) \times\overline{S}(0)\times Z_{m_i}^2 Z_{m_P}$.   See Fig.~\ref{fig:Down_R_const}.

     \item ($\mathcal{O}_P$, $ \mathcal{O}^{\dagger}_{i,B_2}$), ($\mathcal{O}^{\dagger}_P$, $ \mathcal{O}_{i,B_2}$), ($ \mathcal{O}_{i,B_1}$, $ \mathcal{O}^{\dagger}_{i,B_1}$) : We refer to this group of saddles as $D^{\downarrow}_L$, and their construction is identical to those of $D^{\downarrow}_R$ outlined above. 
    
\end{enumerate}

Collecting everything together, the  additional contribution to $\overline{\braket{i|\mathcal{O}_P|i}\braket{i|\mathcal{O}_P^{\dagger}|i}}$ once $\mathcal{O}_P=\mathcal{O}_i$   due to the detection saddles is given by:
\beq \label{eq:annhilcontri1s}
Z_D = \left(2\times \overline{Z}(\frac{\beta}{2})\times \overline{S}(\frac{3\beta}{2}) \times\overline{S}(0) + \overline{S}(2\beta)\times \overline{S}(0) + \overline{Z}(\frac{\beta}{2})^2\times \overline{S}(\beta)\times \overline{S}(0)\right) \times Z_{m_i}^2 Z_{m_P} \, .
\eeq
Upon normalising the shell states we obtain:
\beq
Z_D = \frac{\left(2\times \overline{Z}(\frac{\beta}{2})\times \overline{S}(\frac{3\beta}{2}) \times\overline{S}(0) + \overline{S}(2\beta)\times \overline{S}(0) + \overline{Z}(\frac{\beta}{2})^2\times \overline{S}(\beta)\times \overline{S}(0)\right)}{(\overline{Z}(\beta)^2+\overline{Z}(2\beta))\times \overline{S}(0)^2} \times Z_{m_P}.
\eeq
While the strength of the detection signal is independent of the particular state $\ket{i}$ in question, the fact that they exist only when $\mathcal{O}_P=\mathcal{O}_i$ can be used to detect the state. More precisely, the existence of the detection saddles allows one to easily check a hypothesis that the universe is in a particular shell state. In Sec.~\ref{sec:QMA} we discuss wehther this also allows a boundary observer to find the state without prior knowledge. 

\paragraph{Detection-to-universal ratio.} Above we showed the universal saddles are present for any shell probe at all, while the \textit{additional} detection saddles  only exist when $\mathcal{O}_P=\mathcal{O}_i$ . The probe correlator is therefore always strictly larger when detection saddle are present, and to see  by how much we consider the ratio: 
\beq
\frac{Z_D}{Z_U}= \frac{2\times \overline{Z}(\frac{\beta}{2})\times \overline{S}(\frac{3\beta}{2}) \times\overline{S}(0) + \overline{S}(2\beta)\times \overline{S}(0) + \overline{Z}(\frac{\beta}{2})^2\times \overline{S}(\beta)\times \overline{S}(0)}{\overline{Z}(\beta)^2\times \overline{S}(0)^2 + \overline{Z}(2\beta)\times \overline{S}(0)^2 },
\eeq
where the shell dependent factors $Z_{m_i}^2 Z_{m_P}$ have canceled out. Recall that the notation $\overline{S}(\alpha)$ is used to denote an infinite strip boundary condition of length $\lim_{\alpha \to \infty} [-\alpha, \beta + \alpha]$ \cite{Balasubramanian:2025zey}. The ratio $\frac{\overline{S}(\alpha)}{\overline{S}(0)}\equiv \hat{S}(\alpha)$ therefore corresponds to a ``renormalized" strip action given by the volume of the Euclidean AdS strip of boundary length $[0,\alpha]$; i.e. dividing by $\overline{S}(0)$ subtracts off  the infinite half line portions of the strip action. Furthermore, as the Euclidean strip has the same local geometry as thermal AdS but with uncompactified time, we have $\hat{S}(\alpha)= \overline{Z}_{TAdS}(\alpha)$ where $\overline{Z}_{TAdS}(\beta)\equiv e^{-I_{TAdS}(\beta)}$. In this notation the ratio becomes:
 \beq \label{eq:rationormalised}
\frac{Z_D}{Z_U}= \frac{2\times \overline{Z}(\frac{\beta}{2})\times \overline{Z}_{TAdS}(\frac{3\beta}{2})  + \overline{Z}_{TAdS}(2\beta)+ \overline{Z}(\frac{\beta}{2})^2\times \overline{Z}_{TAdS}(\beta)}{\overline{Z}(\beta)^2 + \overline{Z}(2\beta) }.
\eeq

We can estimate the size of this ratio in certain parameter regimes. 
First, note that if $\beta / 2 > \beta_{HP}$  the leading saddle for $\overline{Z}(\frac{\beta}{2})$, $\overline{Z}(\beta)$ and $\overline{Z}(2\beta)$ is the thermal AdS saddle, for which $(\overline{Z}_{TAdS}(x))^n= \overline{Z}_{TAdS}(nx)$. Hence when $\beta /2 > \beta_{HP}$  we have $\frac{Z_D}{Z_U}\approx 2$. As discussed above, in the regime $\beta / 2 > \beta_{HP}$ the shell states are of type B and therefore in these cases the information is hidden in a compact universe. 

For the type A black hole case $\beta < \beta_{HP}$ there are two regimes of interest:
\begin{enumerate}
    \item $\beta < \beta_{HP}$ and $2\beta > \beta_{HP}$: In this case the leading saddle to $\overline{Z}(\beta)$ and $\overline{Z}(\beta /2 )$ are the black hole, while the leading saddle for $ \overline{Z}(2\beta) $ is  thermal AdS (TAdS): 
 \beq
\frac{Z_D}{Z_U} \approx  \frac{2\times \overline{Z}_{BH}(\frac{\beta}{2})\times \overline{Z}_{TAdS}(\frac{3\beta}{2})  + \overline{Z}_{TAdS}(2\beta)+ \overline{Z}_{BH}(\frac{\beta}{2})^2\times \overline{Z}_{TAdS}(\beta)}{\overline{Z}_{BH}(\beta)^2 + \overline{Z}_{TAdS}(2\beta) }.
\eeq

    \item $\beta < \beta_{HP}$ and $2\beta < \beta_{HP}$: In this case the leading saddle for all the partition functions is the black hole saddle: 
   \beq
\frac{Z_D}{Z_U} \approx \frac{2\times \overline{Z}_{BH}(\frac{\beta}{2})\times \overline{Z}_{TAdS}(\frac{3\beta}{2})  + \overline{Z}_{TAdS}(2\beta)+ \overline{Z}_{BH}(\frac{\beta}{2})^2\times \overline{Z}_{TAdS}(\beta)}{\overline{Z}_{BH}(\beta)^2 + \overline{Z}_{BH}(2\beta) }.
\eeq
\end{enumerate}
Either way if $\beta < \beta_{HP}$  the dominant term ratio is approximately equal to $\frac{\overline{Z}_{BH}(\frac{\beta}{2})^2\times \overline{Z}_{TAdS}(\beta)}{\overline{Z}_{BH}(\beta)^2}$.

\paragraph{Microcanonical states.} We can obtain a more illuminating expression for this ratio by considering the detection-to-universal ratio for shell states projected into a micro-canonical window.  Labeling the energy eigenstates of $H$ as $\{\ket{E_n}\}$  we write the shell states in the energy basis as $\ket{i,{\beta}}\equiv e^{-\frac{\beta}{2}H}\ket{\hat{i}}= \sum_n e^{-\frac{\beta}{2}E_n}\braket{E_n|\hat{i}}\ket{E_n}\approx \int_{0}^{\infty}d\tilde{E}  \, \rho(\tilde{E} ) e^{-\frac{\beta}{2}\tilde{E} }\braket{\tilde{E}|\hat{i}}\ket{\tilde{E} }$. The projection of this state into the micro-canonical band $[E-\frac{\Delta E}{2},E+\frac{\Delta E}{2}] $ is given by $\ket{i,\beta,E}=\int_{E-\frac{\Delta E}{2}}^{E+\frac{\Delta E}{2}}d\tilde{E} \, \rho({\tilde{E} }) e^{-\frac{\beta}{2}\tilde{E} }\braket{\tilde{E} |\hat{i}}\ket{\tilde{E} }$ and can be performed using the Laplace transform 
\beq
\ket{i,\beta,E}= \int_{E-\frac{\Delta E}{2}}^{E+\frac{\Delta E}{2}}d\tilde{E} e^{-\frac{\beta}{2}\tilde{E} } \int d\tilde{\beta} e^{\tilde{E} \frac{\tilde{\beta}}{2}} \ket{i,\tilde{\beta}}.
\eeq

The micro-canonical probe correlator $\overline{\braket{i,\beta,E|\mathcal{O}_P|i,\beta,E}\braket{i,\beta,E|\mathcal{O}^{\dagger}_P|i,\beta,E}}$ can then be obtained by Laplace transform of the canonical one:
\beq
 \int_{E-\frac{\Delta E}{2}}^{E+\frac{\Delta E}{2}} \prod_{k=1}^{4} dE_k  e^{-\frac{\beta}{2} \sum_{k=1}^4 E_k } \int \prod_{k=1}^{4} d\beta_k e^{\frac{1}{2} \sum_{k=1}^4 E_k\beta_k} \overline{\braket{\beta_1|\mathcal{O}_P|\beta_2}\braket{\beta_3|\mathcal{O}^{\dagger}_P|\beta_4}}.
\eeq

The detection saddle contributions to $\overline{\braket{\beta_1|\mathcal{O}_P|\beta_2}\braket{\beta_3|\mathcal{O}^{\dagger}_P|\beta_4}}$ are found by generalising (\ref{eq:annhilcontri1s}) above:
\begin{align}
& Z_{D,micro}= (Z(\frac{\beta_1}{2})Z(\frac{\beta_3}{2})Z_{TAdS}(\frac{\beta_2+\beta_4}{2}) + Z_{TAdS}(\frac{\beta_1+\beta_2+\beta_3+\beta_4}{2}) + Z(\frac{\beta_1}{2})Z_{TAdS}(\frac{\beta_2+\beta_3+\beta_4}{2}) \\ \nonumber
&+ Z(\frac{\beta_4}{2})Z_{TAdS}(\frac{\beta_1+\beta_2+\beta_3}{2})) \times Z^2_{m_i}Z_{m_P}S(0)^2
\end{align}
and similarly the universal saddles are given by generalising (\ref{eq:1spropsaddles}):
\beq 
Z_{U,micro}= \left (Z(\frac{\beta_1+\beta_2+\beta_3+\beta_4}{2}) + Z(\frac{\beta_1+\beta_4}{2}) \times Z(\frac{\beta_2+\beta_3}{2})   \right) \times Z^2_{m_i}Z_{m_P}S(0)^2.
\eeq

We will approximate the Laplace transform of the partition functions within the saddlepoint approximation:
\beq \label{eq:Laplace1}
\int dx\, e^{Ex} \,  \overline{Z}(x) \approx \sqrt{\frac{2\pi}{h}}e^{S_{BH}(E)} 
\eeq
where h is the Hessian determinant $\partial^2_x \ln{Z}(x)|_{x=x^*}$ evaluated at the saddlepoint $x^{*}$ and $S_{BH}(E)= (1-\beta\partial_{\beta}) \ln( \overline{Z}(\beta))\,|_{\beta=\beta(x^*)}
$.  Furthermore restricting to windows $E >> \Delta E$  we define \beq \label{eq:Laplace2}
\int^{E +\frac{\Delta E}{2}}_{E -\frac{\Delta E}{2}} d\tilde{E} \sqrt{\frac{2\pi}{h}}e^{S_{BH}(\tilde{E})}  \approx \Delta E \sqrt{\frac{2\pi}{h}}e^{S_{BH}(E)}\equiv e^{\mathbf{S}(E)}
\eeq
and 
\beq \label{eq:Laplace3}
\int^{E +\frac{\Delta E}{2}}_{E -\frac{\Delta E}{2}} d\tilde{E} \sqrt{\frac{2\pi}{h}}e^{S_{BH}(\tilde{E})}  e^{-\beta \tilde{E}} \approx  e^{-\beta E  } e^{\mathbf{S}(E)}.
\eeq

Within these approximations the microcanonical version of (\ref{eq:rationormalised}) becomes:
\beq
\frac{Z_D}{Z_U}=\frac{e^{2\mathbf{S}(E)}  +2\times e^{\mathbf{S}(E)} +1}{e^{2\mathbf{S}(E)}+e^{\mathbf{S}(E)}} = 1 + e^{-\mathbf{S}(E)}
\eeq
 Above the black hole threshold $e^{\mathbf{S}(E)}  >> 1 $ and hence $\frac{Z_D}{Z_U} \approx 1$.  So the response is factor of two larger when the right probe ($\mathcal{O}_P=\mathcal{O}_i$)  is selected. Below the black hole threshold we have $e^{\mathbf{S}(E)} =1 $ and therefore $\frac{Z_D}{Z_U}=2$, and hence the response is a factor of three larger when the correct operator is chosen.

\section{Observing a Two-Boundary Universe} \label{sec:two_bdry}
Having established probes to detect the state of a single-boundary universe, we now turn to the case of universes with two asymptotic boundaries. It was shown in \cite{Balasubramanian:2025jeu} that the two-boundary shell states introduced in \cite{Balasubramanian:2022gmo,Balasubramanian:2022lnw,Climent:2024trz} provide a family of bases that span the full non-perturbative two-boundary gravity Hilbert space. In this section we assume the universe is in one of these basis states, and identify a Lorentzian boundary probe that can be used to detect the state. Again, by detection here we mean a probe that allows for the verification of a proposal for the shell operator used to create the state. We will show that this is possible using coordinated probes places on both boundary simultaneously. Remarkably, we show detection is also possible using a probe localized on just one of the asymptotic boundaries.

\begin{figure}[h]
        \begin{subfigure}[b]{.43\linewidth}
            \centering
            \includegraphics[width=0.6\linewidth]{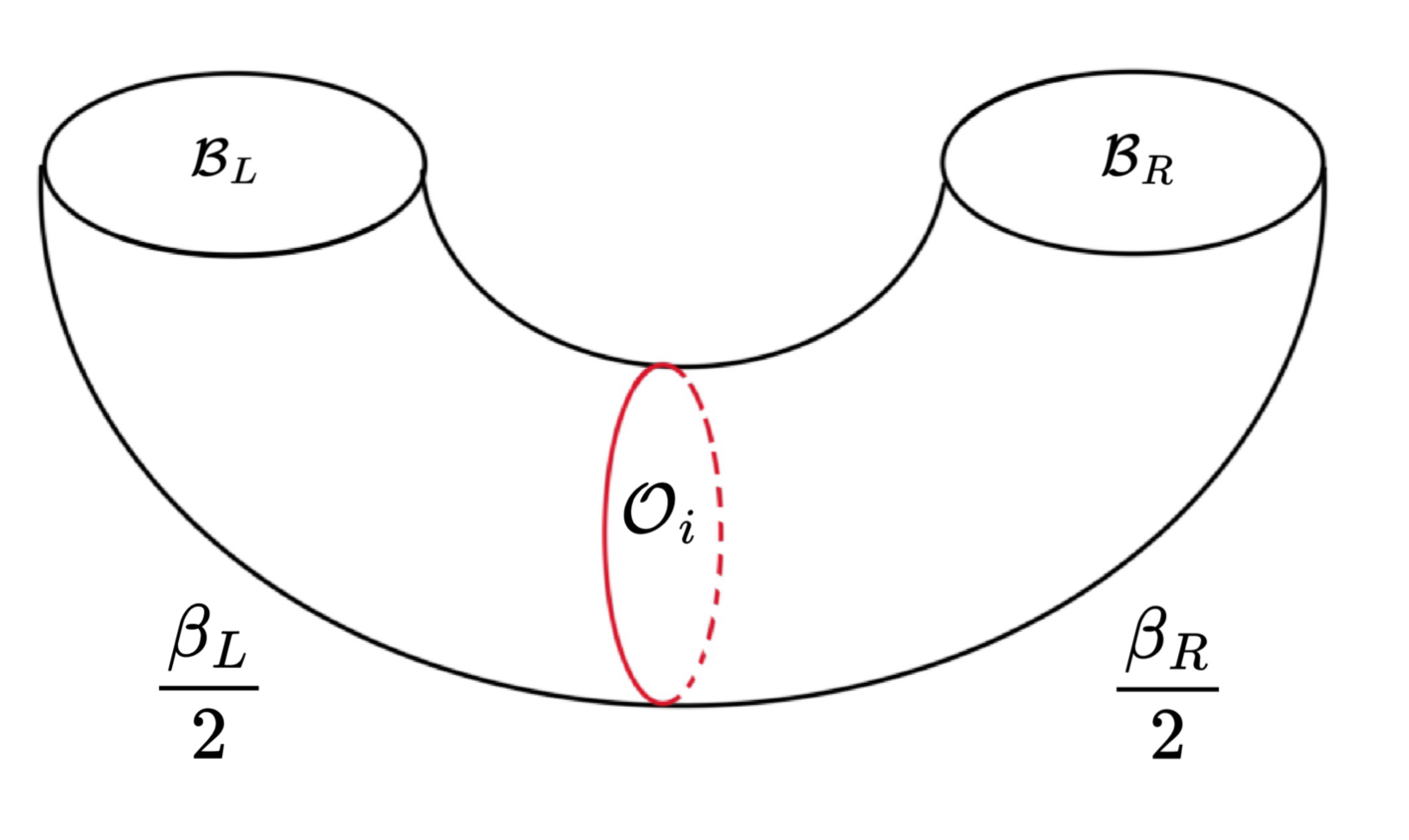}
            \caption{}
       
        \end{subfigure}
 \hfill
        \begin{subfigure}[b]{.5\linewidth}
            \centering  
            \includegraphics[width=0.6\linewidth]{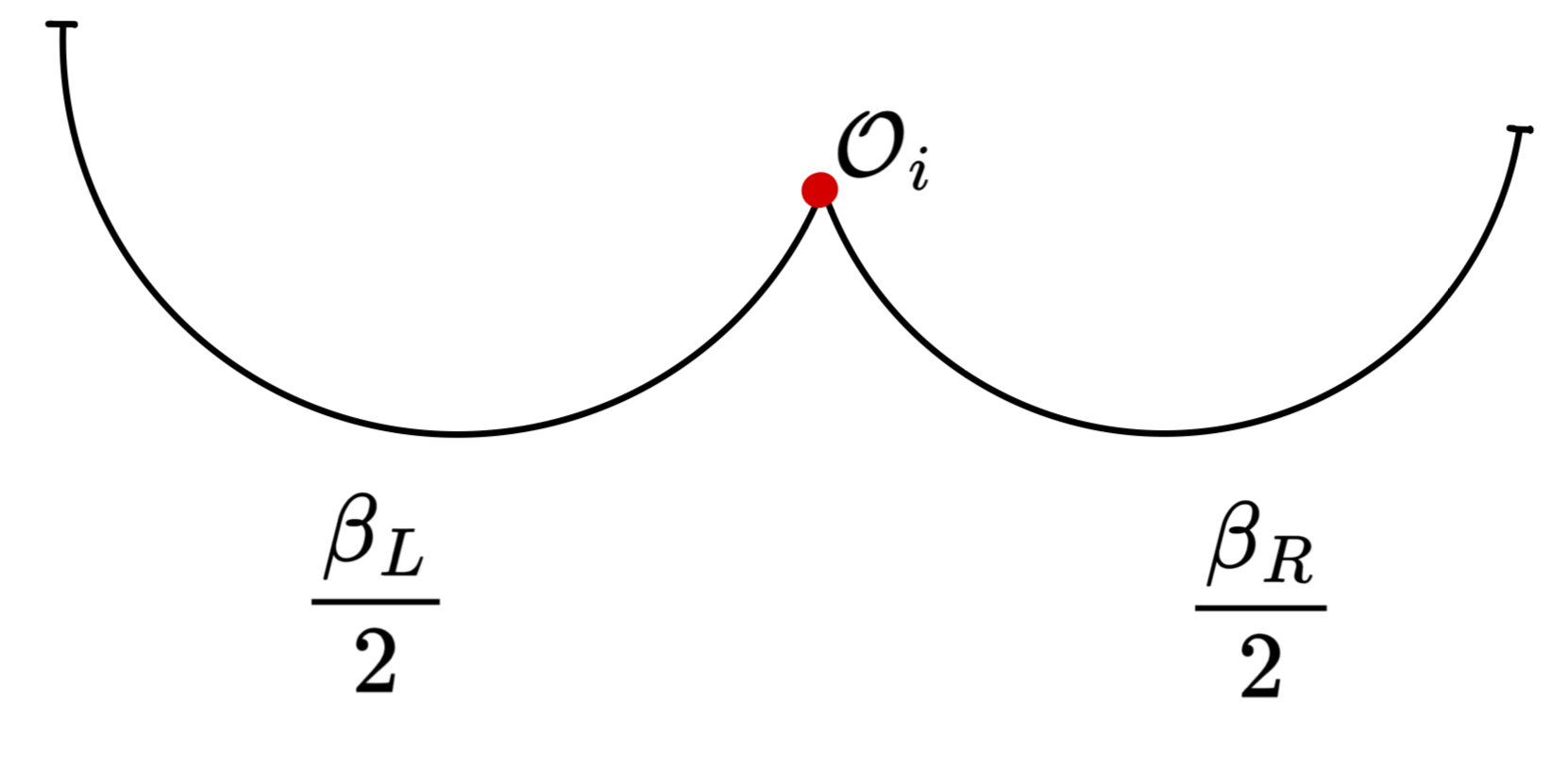}
            \caption{}
       
        \end{subfigure}
    \caption{Asymptotic boundary condition for the gravity path integral defining the shell state.  ({\bf a}) Cut-open Euclidean boundary with topology $\mathbb{I}_{\frac{\beta_L+\beta_R}{2}}\times\mathbb{S}^{d-1}$ for preparation of the shell states. The shell operator $\mathcal{O}_{i}$  is pictured in red. In AdS/CFT we can also perform the path integral in the boundary CFT with insertion of a $\mathbb{S}^{d-1}$ symmetric operator dual to the shell.  ({\bf b}) Euclidean boundary with the $\mathbb{S}^{d-1}$ suppressed. We adopt this convention for the rest of the paper. Here $\beta_{L,R}/2$ are  Euclidean ``preparation times''. Figure adapted from \cite{Balasubramanian:2025jeu}.}
    \label{fig:shell_b.c}
\end{figure}

 First, we briefly review construction of the shell state basis for the non-perturbative two-boundary Hilbert space developed in  \cite{Sasieta:2022ksu,Balasubramanian:2022gmo,Balasubramanian:2022lnw,Antonini:2023hdh,Balasubramanian:2024rek,Balasubramanian:2025jeu} and shown to be a full basis in \cite{Balasubramanian:2025jeu}.\footnote{In the AdS/CFT context, these are related to Partially Entangled Thermal States (PETS) \cite{Goel:2018ubv}.}  These states are defined by cutting open the Euclidean gravity path integral with shell operators inserted on a periodic asymptotic boundary. The cut consists of two disconnected $\mathbb{S}^{d-1}$ components which we call $\mathcal{B}_{L,R}$ (Fig.~\ref{fig:shell_b.c}).  Slitting open the path integral in this way produces two disconnected pieces of topology $\mathbb{I}\times \mathbb{S}^{d-1}$ corresponding to the bra and ket. The shell states are defined by inserting a $\mathbb{S}^{d-1}$ symmetric dust shell operator $\mathcal{O}_{i}$ of mass $m_i \sim \mathit{O}(1/G_{N})$ on the asymptotic boundary, at distances $\frac{\beta_L}{2}$ and $\frac{\beta_R}{2}$ from the $\mathcal{B}_{L,R}$ cuts. Varying $m_i$ results in an infinite family of shell states $\ket{\mathbf{i}}$, which we denote in boldface to distinguish them from the single-sided shell state prepared using the same operator insertion.

\begin{figure}[h]
    \centering
    \includegraphics[width=0.3\linewidth]{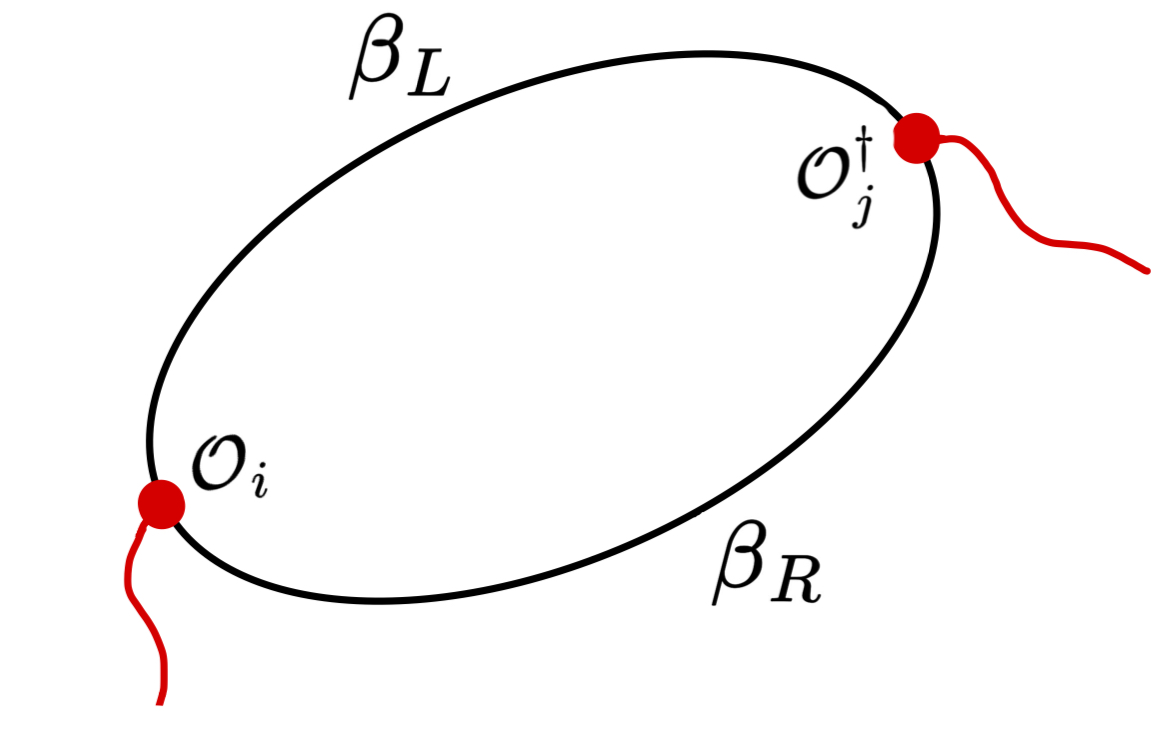}
    \caption{Shell asymptotic boundary condition for $\overline{\langle j|i\rangle} $, consisting of the operator insertions $\mathcal{O}_{i}$ and  $\mathcal{O}^{\dagger}_{j}$ separated by asymptotic time extent $\beta_{L}$ and $\beta_{R}$ respectively. The red lines represent the shells propagating into the bulk. Figure adapted from \cite{Balasubramanian:2025jeu}.}
    \label{fig:shell_bdry}
\end{figure}

\begin{figure}[h]
            \centering
            \includegraphics[width=1\linewidth]{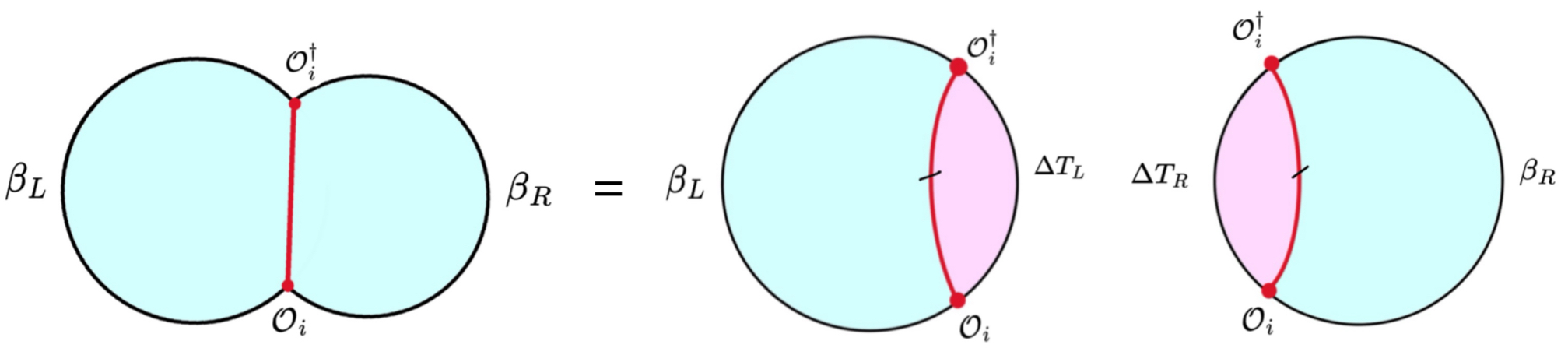}
            \caption{The saddlepoints for the shell norm path integral $\overline{\langle i|i\rangle}$ are constructed by gluing together two disks. In particular the shell homology regions (purple) on each disk are discarded and the resulting geometries glued together along the shell worldvolume. Figure adapted from \cite{Balasubramanian:2025jeu}.}
            \label{fig:shell_norm}
\end{figure}

The path integral boundary conditions computing elements of the Gram matrix $G_{\mathbf{i}\mathbf{j}}\equiv \langle \mathbf{i}|\mathbf{j}\rangle$  are given by sewing together the boundary conditions defining $\bra{\mathbf{i}}$ and $\ket{\mathbf{j}}$ along their cuts. This results in boundary operator insertions $\mathcal{O}_{j}$ and  $\mathcal{O}^{\dagger}_{i}$ separated by asymptotic times  $\beta_{L}$ and $\beta_{R}$ on a closed boundary manifold of periodicity $\beta_{L}+\beta_{R}$ (Figs.~\ref{fig:shell_bdry}, \ref{fig:shell_norm}). The overlaps are then computed by the gravity path integral with this sewn boundary condition. As in the single-boundary case,  if the shell inertial masses $m_{i,j}$ are sufficiently large we have 
\begin{equation}
\overline{\langle \mathbf{i}|\mathbf{j} \rangle} =\delta_{\mathbf{ij}} Z_{1} \,.
\label{eq:defZ1}
\end{equation} 
to leading order, while higher topology  contributions stabilized by the shell matter  modify the overlap to
\begin{equation}
\overline{|\langle \mathbf{i}|\mathbf{j} \rangle|^2}=\overline{\langle \mathbf{i}|\mathbf{j} \rangle\langle \mathbf{j}|\mathbf{i} \rangle} = Z_{2} + \delta_{\mathbf{ij}}Z_{1}^2 \, ,
\end{equation}
where $Z_{2}$ is a wormhole saddlepoint contribution \cite{Balasubramanian:2022gmo,Balasubramanian:2025jeu}.  

\paragraph{Norm.}The path integral for the norm $\overline{\braket{\mathbf{i}|\mathbf{i}}}$ can be computed in the saddlepoint approximation. To find such saddles, the geometry on either side of the shell can be filled in with anything that locally solves the equations of motion; hence by symmetry these are portions of the ``disk" saddles  discussed above. Consider two disk saddles, which we label ($L,R$), of boundary lengths $\beta_{L,R}+ \Delta T_{L,R}$, on which a shell propagates into the bulk before getting absorbed a boundary time $\Delta T_{L,R}$ later. The saddles to  $\overline{\braket{\mathbf{i}|\mathbf{i}}}$ are constructed by taking these two disks, discarding the shell homology regions (purple in Fig.~\ref{fig:shell_norm}), and gluing the resulting geometries together along the shell worlvolumes. The junction conditions dynamically determine $\Delta T_{L,R}$ such that the resulting geometry satisfies the equations of motion. Again, in the large shell mass limit  the  turning points of the shell trajectories approach the asymptotic boundary  of the disk so that the shell homology regions pinch off. In this limit, $\Delta T_{L,R} \to 0$, and each shell contributes simply contributes a factor $Z_{m_i}\sim e^{-2(d-1)\log(\mathrm{G_N} m_i)}$.  This behavior is independent of the disk saddle geometries that are  glued along the shells \cite{Balasubramanian:2022gmo,Balasubramanian:2022lnw,Antonini:2023hdh}. 
There are three disk saddles for asymptotically AdS boundary conditions and any two of these can be glued together in this way to obtain a saddle to $\overline{\braket{\mathbf{i}|\mathbf{i}}}$. There are therefore $3\times 3=9$ such saddles. In the large shell mass limit the contribution to the action from the geometry on either side of the shell is simply the action of that entire disk geometry. Hence using (\ref{eq:partitionfunc}) we can economically write the sum over all saddles to the norm as:   

\beq \label{eq:twobnorm}
\overline{\braket{\mathbf{i}|\mathbf{i}}}= \overline{Z}(\beta_L)\times \overline{Z}(\beta_R)\times Z_{m_i}.
\eeq

\begin{figure}[h]
    \centering
    \begin{subfigure}[c]{0.45\linewidth}
        \centering
        \includegraphics[width=0.7\linewidth]{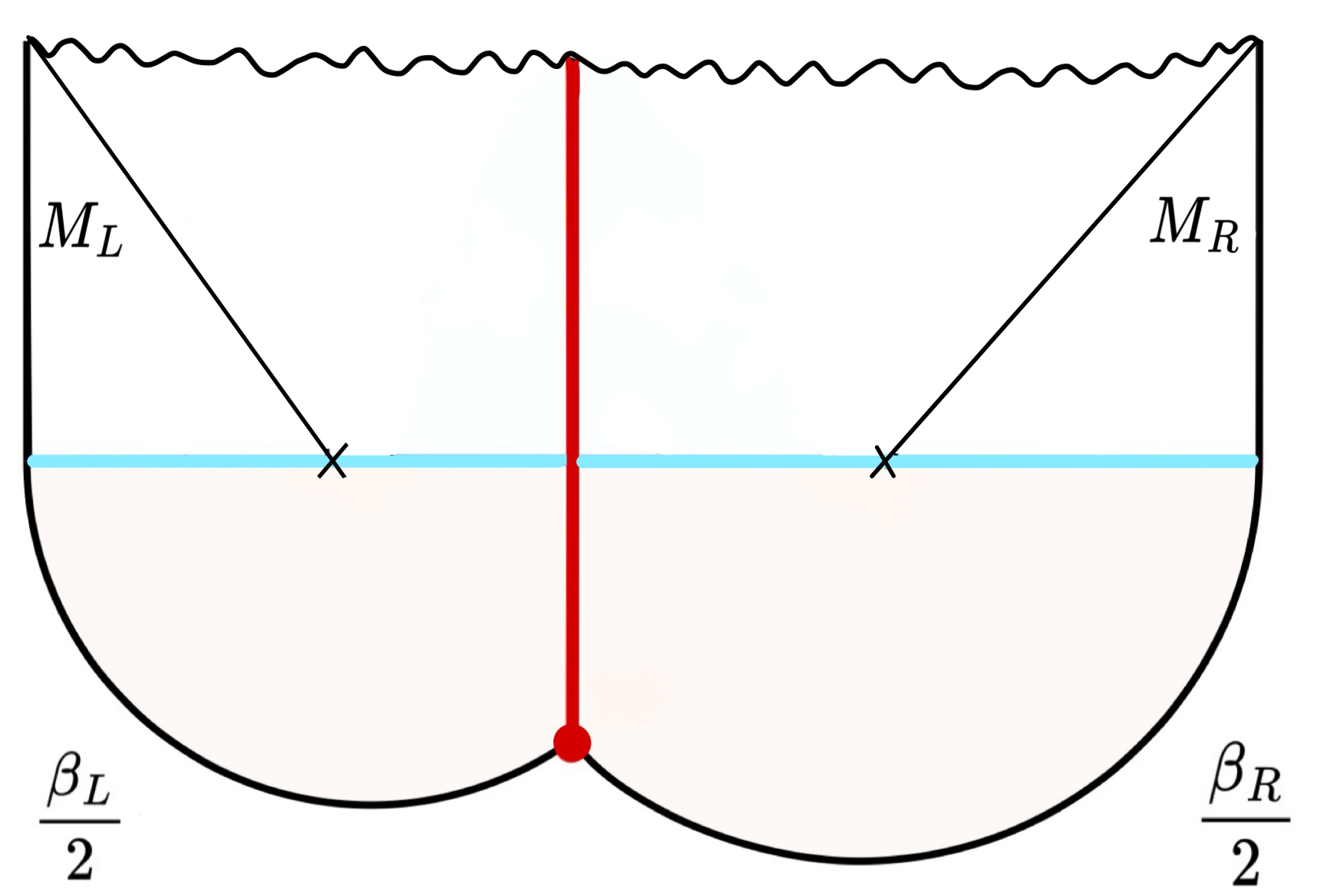}
        \caption{}
        \label{type1}
    \end{subfigure}
    \hfill
    \begin{subfigure}[c]{0.45\linewidth}
        \centering
        \includegraphics[width=0.7\linewidth]{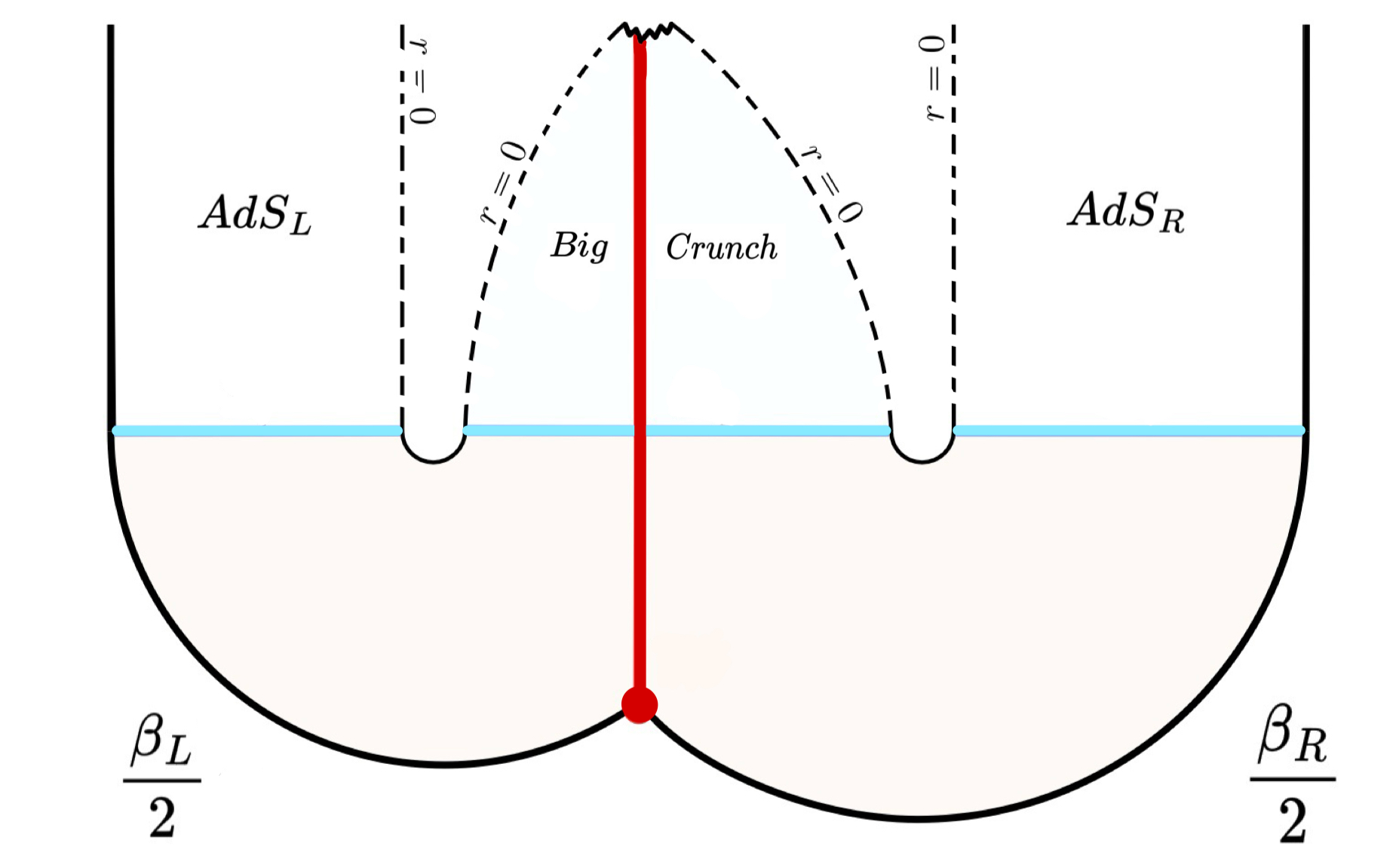}
        \caption{}
        \label{type2}
    \end{subfigure}
    \hfill
    \begin{subfigure}[c]{0.45\linewidth}
        \centering
        \includegraphics[width=0.7\linewidth]{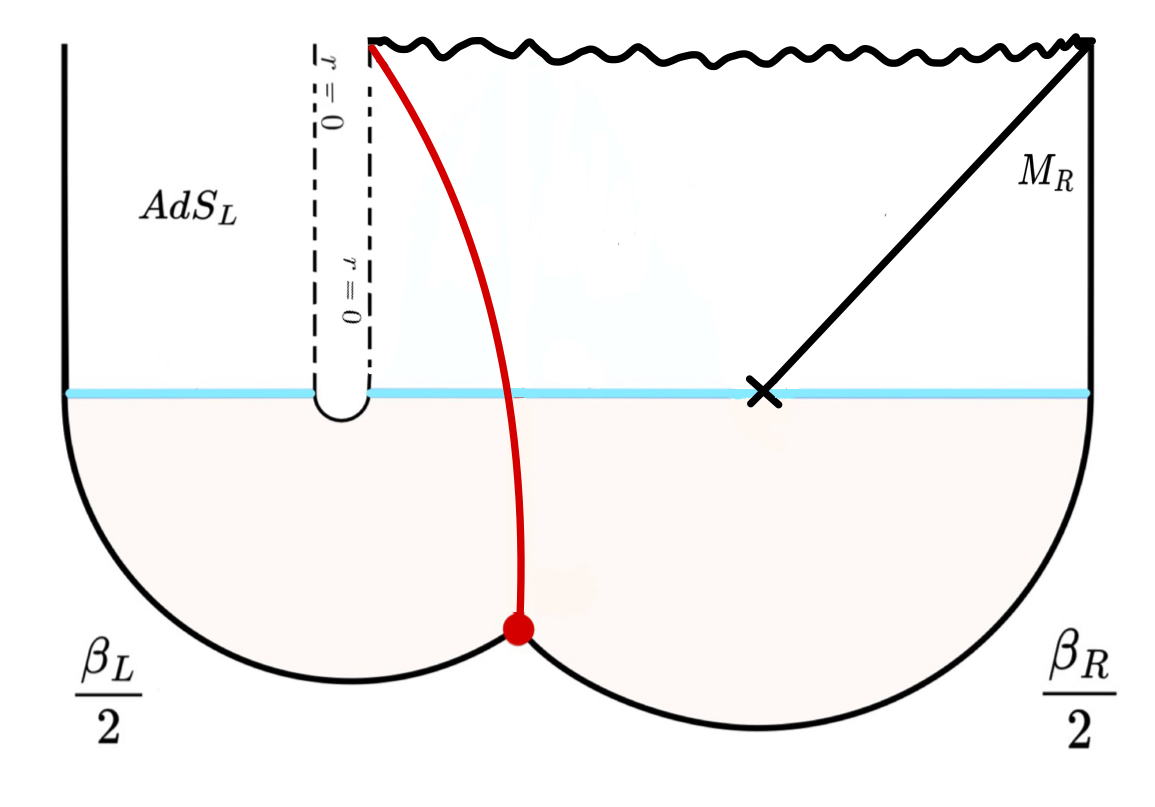}
        \caption{}
        \label{type3}
     \end{subfigure}
\caption{Diagrams showing the analytic continuation of the type 1-3 shell states to  Lorentzian signature.  ({\bf a}) Type 1 shell state with $\beta_L < \beta_{HP}$ and $\beta_R < \beta_{HP}$ corresponding to a two-sided black hole. ({\bf b}) Type 2 shell state with $\beta_L > \beta_{HP}$ and $\beta_R>\beta_{HP}$ consisting of two copies of thermal AdS with the addition of a compact Big-Crunch AdS cosmology. ({\bf c}) Type 3 shell state with $\beta_L > \beta_{HP}$ and $\beta_R < \beta_{HP}$ corresponding to a disconnected geometry with an $R$ single sided BH and an $L$ thermal AdS factor. Figure adapted form \cite{Balasubramanian:2025jeu}. }
\end{figure}

\paragraph{Shell geometry.} \label{sec:shellgeom}
We again associate a geometry to these states by considering the leading saddle to the norm (\ref{eq:twobnorm}) and analytically continuing to Lorentzian signature. 
Focusing on the asymptotically AdS case for concreteness, this leading saddle is determined by wether $\beta_{L,R}$ is above or below $\beta_{HP}$. If $\beta_{L} >\beta_{HP}$ the leading $L$ disk saddle is thermal AdS, whereas for $\beta_{L} < \beta_{HP}$ it is the Euclidean AdS black hole (similarly for $R$).  There are therefore four regimes to consider:

\begin{enumerate}
    \item $\beta_{L}, \beta_{R} < \beta_{HP}$ : The leading  saddles are two-sided Schwarschild-AdS black holes with a long wormhole interior (Fig.~\ref{type1}). These are \textit{Type 1} shell states. These are the states originally considered in \cite{Balasubramanian:2022gmo}. 
   
    \item $\beta_{L}, \beta_{R} > \beta_{HP}$: The leading saddles  are two copies of thermal AdS with the addition of a compact Big-Crunch AdS cosmology (Fig.~\ref{type2}). This construction was studied in detail in \cite{Antonini:2023hdh,Antonini:2025ioh}. These ares \textit{Type 2} shell states. 
    
    \item $\beta_{L}  < \beta_{HP}$ and  $\beta_{R} > \beta_{HP}$  or vice versa : The leading saddles correspond to an $L$ ($R$) single-sided AdS Black hole and a disconnected $R$ ($L$) copy of thermal AdS (Fig.~\ref{type3}). These  are \textit{Type 3} shell states. Such states were studied in \cite{Balasubramanian:2022lnw}, but they did not include the disconnected thermal AdS.

\end{enumerate} 

\subsection{Information Sharing Two-Boundary Probes} \label{sec:infoshare}

\begin{figure}
  \centering
    \includegraphics[width=0.3\linewidth]{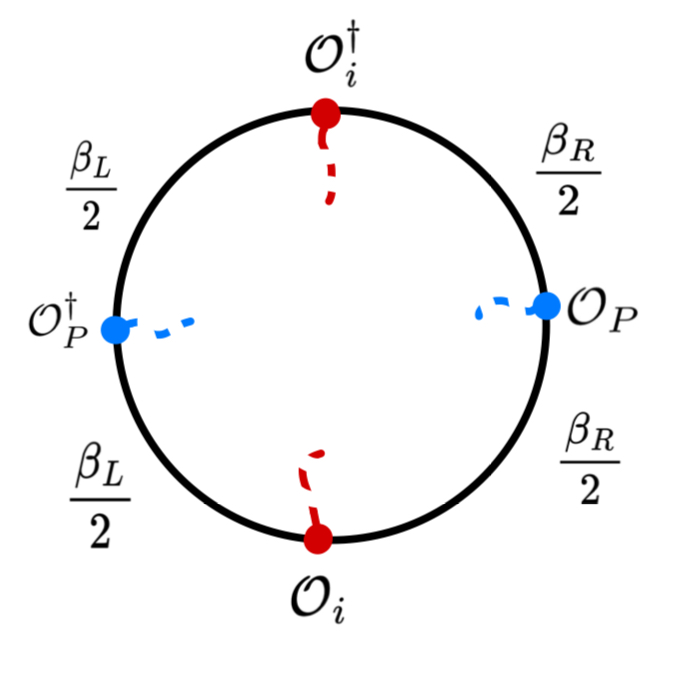}
    \caption{Lorentzian two-sided shell detection.}
    \label{fig:2S_BC}
    \end{figure}

Consider the two-sided shell state $\ket{\mathbf{i}}$ created by insertion of $\mathcal{O}_i$ with $\beta_L, \beta_R$ preparation temperatures. Entanglement wedge reconstruction tells us that the interior of the two-sided black hole, or alternatively the state of the compact universe is guaranteed to be in the entanglement wedge of the union of the boundary theories, but not in both of them separately
\cite{Headrick:2014cta,Jafferis:2015del,Dong:2016eik,Bao:2016skw,Faulkner:2017vdd,Cotler:2017erl,Chen:2019gbt,Kang:2018xqy,Gesteau:2020rtg}. In other words, the semiclassical interior emerges  from the entanglement structure of the state on the two theories jointly, and therefore cannot be reconstructed by linear operators in just one side of the theory. In line with, we will first consider the insertion of a probe operator and its conjugate on the $L,R$ asymptotic boundaries respectively at $\tau=0$, resulting in the two-boundary probe correlator $\braket{\mathbf{i}|\mathcal{O}_{P,L}^{\dagger}\mathcal{O}_{P,R}| \mathbf{i}}$, see Fig.~\ref{fig:2S_BC}. The gravity path integral for this quantity $\overline{\braket{\mathbf{i}|\mathcal{O}_{P,L}^{\dagger}\mathcal{O}_{P,R}| \mathbf{i}}}$  
 again has two classes of saddles, the universal saddles that contribute when $\mathcal{O}_{P}\neq\mathcal{O}_{\mathbf{i}}$ and the additional detection saddles that only contribute when $\mathcal{O}_{P}=\mathcal{O}_{\mathbf{i}}$.

\begin{figure}
   \begin{subfigure}{0.45\linewidth}
   \centering
    \includegraphics[width=\linewidth]{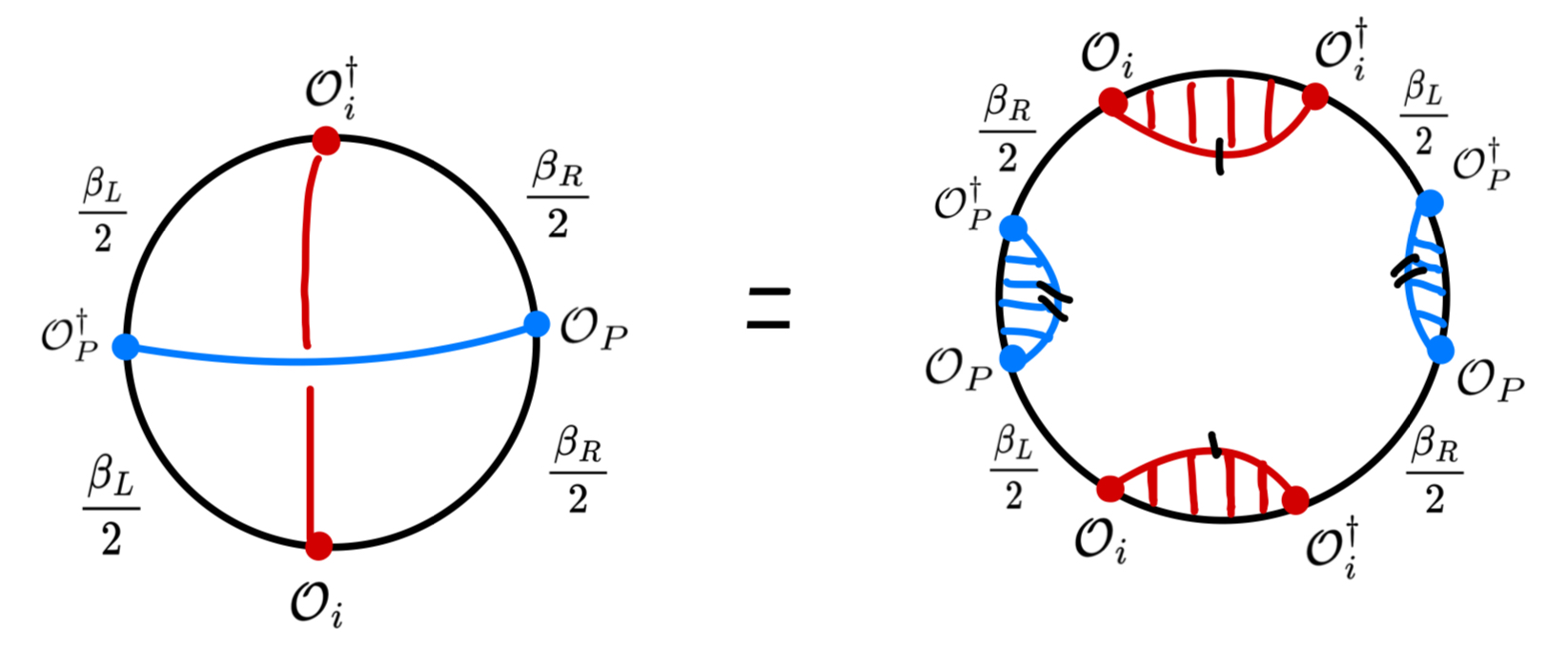}
    \caption{}
    \label{fig:2SU}
    \end{subfigure}
\hfill
\begin{subfigure}{0.45\linewidth}
    \centering
    \includegraphics[width=\linewidth]{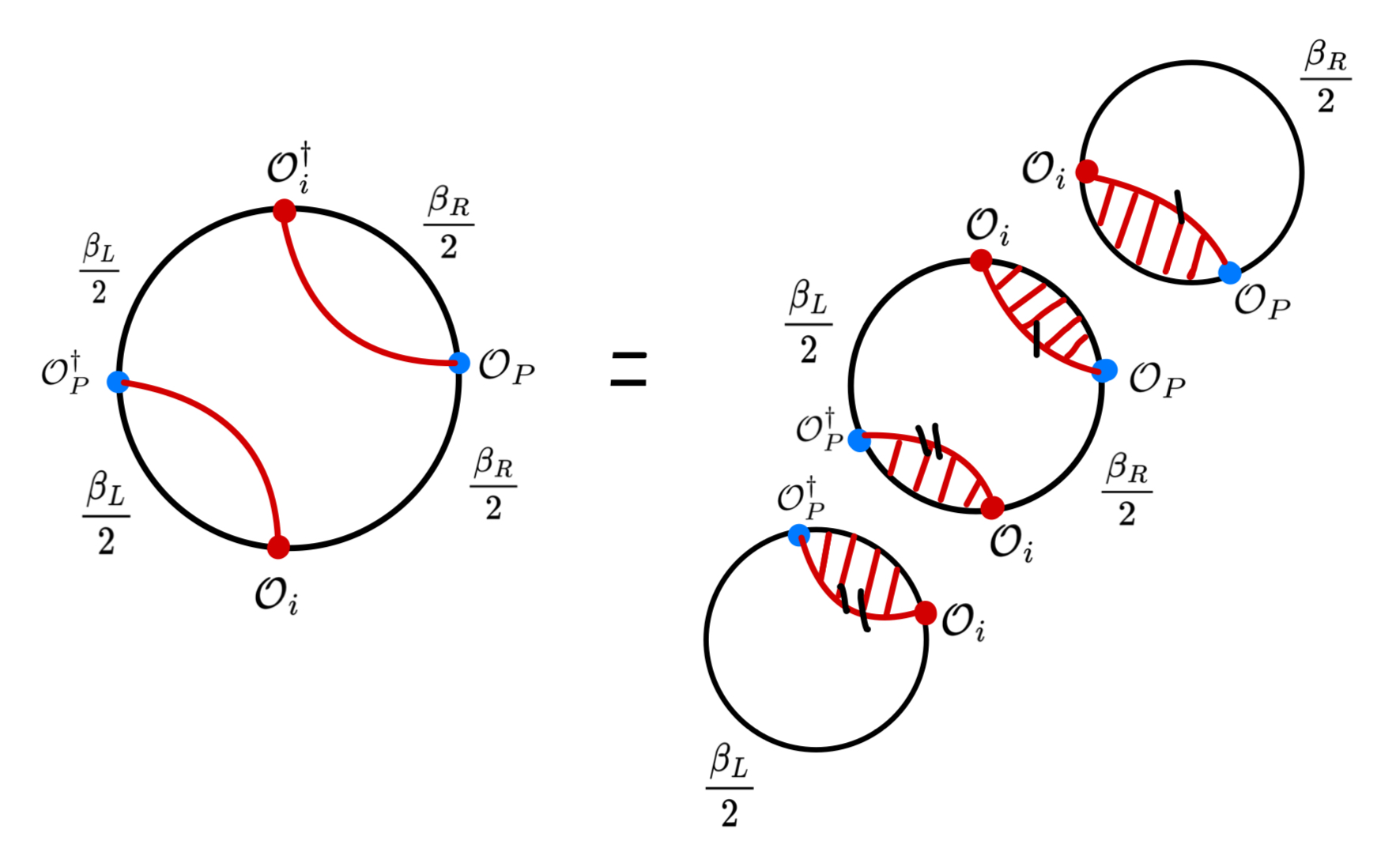}
    \caption{}
    \label{fig:2SD}
    \end{subfigure}
    \caption{ (\textbf{a}) Universal saddle contribution to  $\braket{\mathbf{i}|\mathcal{O}_{P,L}^{\dagger}\mathcal{O}_{P,R}| \mathbf{i}}$. (\textbf{b}) Propagation saddle contribution to $\braket{\mathbf{i}|\mathcal{O}_{P,L}^{\dagger}\mathcal{O}_{P,R}| \mathbf{i}}$. }
    \label{fig:2s_2s}
\end{figure}

\paragraph{Universal saddles.}The  universal saddles are constructed by considering a disk on which four shells labeled $1-2-3-4$ propagate for a boundary time $T_1 \cdots T_4$ and are separated from each other by boundary times $\frac{\beta_L}{2},\frac{\beta_L}{2},\frac{\beta_R}{2},\frac{\beta_R}{2}$ respectively. The shell homology regions are then discarded and opposite shells ($1-3$ and $2-4$)  are glued along the corresponding shell worldvolumes using the junction conditions, see Fig.~\ref{fig:2SU}. In the large shell mass limit the propagation times go to zero, the shell homology regions pinch off,  and the shells contribute universally to the action. These saddles therefore contribute a factor $\overline{Z}(\beta_L + \beta_R) \times Z_{m_i} Z_{m_P}$. Interestingly, the saddle geometry constructed in this manner has the topology of a disk with an additional handle (or equivalently a punctured torus). One of the shells propagates through the handle, while the other passes underneath. The shells therefore do not actually intersect in the bulk, although it appears that way in Fig.~\ref{fig:2SU}. 

\paragraph{Detection saddles.}The detection saddles are constructed by first considering a disk of boundary length $\frac{\beta_L}{2} + T_1 $ on which shell 1 propagates for a time $T_1$ , and a similarly a  disk of length $\frac{\beta_R}{2} + T_2 $ on which shell 2 propagates. Also consider a central disk on which the two shells propagate for a time $\tilde{T}_1, \tilde{T_2}$ respectively and are separated from each other by a boundary times $\frac{\beta_L}{2} , \frac{\beta_R}{2} $. The two aforementioned disks are then glued into this central disk by discarding the shell homology regions and and identifying the corresponding shell worldvolumes, Fig.~\ref{fig:2SD}. In the large shell mass limit the propogation times go to zero and homology regions pinch off. These saddles therefore contribute a factor  $Z(\beta_L  /2)Z(\beta_R  /2) \times Z((\beta_L +\beta_R)  /2)\times Z_{m_i} Z_{m_P}$.

\paragraph{Detection-to-universal ratio.} Putting everything together we obtain:
\beq
\frac{Z_D}{Z_U}= \frac{\overline{Z}(\beta_L  /2) \times \overline{Z}(\beta_R  /2) \times \overline{Z}((\beta_L +\beta_R)  /2)}{\overline{Z}(\beta_L + \beta_R)}.
\eeq
Note that if $\beta_L + \beta_R < \beta_{HP}$ we have $\frac{Z_D}{Z_U} >> 1$ and if $\beta_L/2 ,\beta_R  /2>\beta_{HP}$, then $\frac{Z_D}{Z_U} =1$.

Much like in the single-boundary case, we can project the two-boundary shell states into $L,R$ microcanonical windows $[E_{L,R}-\frac{\Delta E_{L,R}}{2},E_{L,R}+\frac{\Delta E_{L,R}}{2}] $ by the double Laplace transform:
\beq
\ket{i,\beta_{L,R},E_{L,R}}= \int_{E_L-\frac{\Delta E_L}{2}}^{E_L+\frac{\Delta E_L}{2}} \int_{E_R-\frac{\Delta E_R}{2}}^{E_R+\frac{\Delta E_R}{2}} d\tilde{E}^{L,R}  \, e^{-\frac{\beta_{L,R} }{2}\tilde{E}^{L,R}  } \int d\tilde{\beta}_{L,R}  e^{\tilde{E}^{L,R}  \frac{\tilde{\beta}_{L,R} }{2}} \ket{i,\tilde{\beta}_{L,R}}.
\eeq
The micro-canonical probe correlator $\overline{\braket{i,\beta_{L,R},E_{L,R}|\mathcal{O}_{P,L}^{\dagger}\mathcal{O}_{P,R}|i,\beta_{L,R},E_{L,R}}}$ can then be obtained by Laplace transform of the canonical one:
\beq
\int_{E_L-\frac{\Delta E_L}{2}}^{E_L+\frac{\Delta E_L}{2}} dE_{1} \, dE_{3}e^{-\frac{\beta_L}{2}(E_1+E_3)}\int_{E_R-\frac{\Delta E_R}{2}}^{E_R+\frac{\Delta E_R}{2}} dE_{2} \, dE_4 e^{ -\frac{\beta_R}{2}(E_2+E_4) } \int \prod_{k=1}^{4} d\beta_k e^{\frac{1}{2} \sum_{k=1}^4 E_k\beta_k} \overline{\braket{\beta_1,\beta_2|\mathcal{O}_{P,L}^{\dagger}\mathcal{O}_{P,R}|\beta_3,\beta_4}}.
\eeq

where the ration of the detection and universal contributions to $\overline{\braket{\beta_1,\beta_2|\mathcal{O}_{P,L}^{\dagger}\mathcal{O}_{P,R}|\beta_3,\beta_4}}$ is given by 
\beq \label{eq:microcanrat}
\frac{Z_D}{Z_U} =  \frac{\overline{Z}(\beta_2  /2)\overline{Z}(\beta_3  /2) \overline{Z}((\beta_1 +\beta_4)  /2)}{\overline{Z}((\beta_1 + \beta_2 + \beta_3 + \beta_4 )/2) }.
\eeq
Upon evaluating the Laplace transforms in the saddlepoint approximation and using (\ref{eq:Laplace1}), (\ref{eq:Laplace2}) and (\ref{eq:Laplace3}) we obtain:
\beq
{Z_D}=  Z_{m_i} Z_{m_P} \times e^{-\frac{\beta_L}{2}E_L+ \frac{-\beta_R}{2}E_R }\times e^{\mathbf{S}(E_L)+\mathbf{S}(E_R)} \times \Phi(E_L,\Delta_L,E_R,\Delta_R,\beta_L,\beta_R)
\eeq
and 
\beq
Z_U=  Z_{m_i} Z_{m_P}\times\Phi(E_L,\Delta_L,E_R,\Delta_R,2\beta_L,2\beta_R),
\eeq
where 
\beq
\Phi(E_L,\Delta_L,E_R,\Delta_R)= \int_{E_L-\frac{\Delta E_L}{2}}^{E_L+\frac{\Delta E_L}{2}} dE_1 e^{-\frac{\beta_L}{2}E_1} \int_{E_R-\frac{\Delta E_R}{2}}^{E_R+\frac{\Delta E_R}{2}}  dE_4 e^{-\frac{\beta_R}{2}E_4} \delta(E_4-E_1) e^{S((E_1+E_4)/2)}.
\eeq

Interestingly, the mixing of $L,R$ preparation temperatures in the partition function arguments gives rise to the smearing function $\Phi$, which vanishes unless the $L,R$ energy windows to overlap.  As long as this overlap is nonzero, if either of the energies $E_{L,R}$ is above the black hole threshold, then $e^{\mathbf{S}(E_{L,R})}>>1$ and the detection to universal ratio is exponentially large. If both are below threshold $e^{\mathbf{S}(E)}=1$ and the ratio is $\mathcal{O}(1)$. 

\subsection{A Single-Boundary Probe Can Detect a Two-Boundary State}
\begin{figure}
    \centering
    \includegraphics[width=0.5\linewidth]{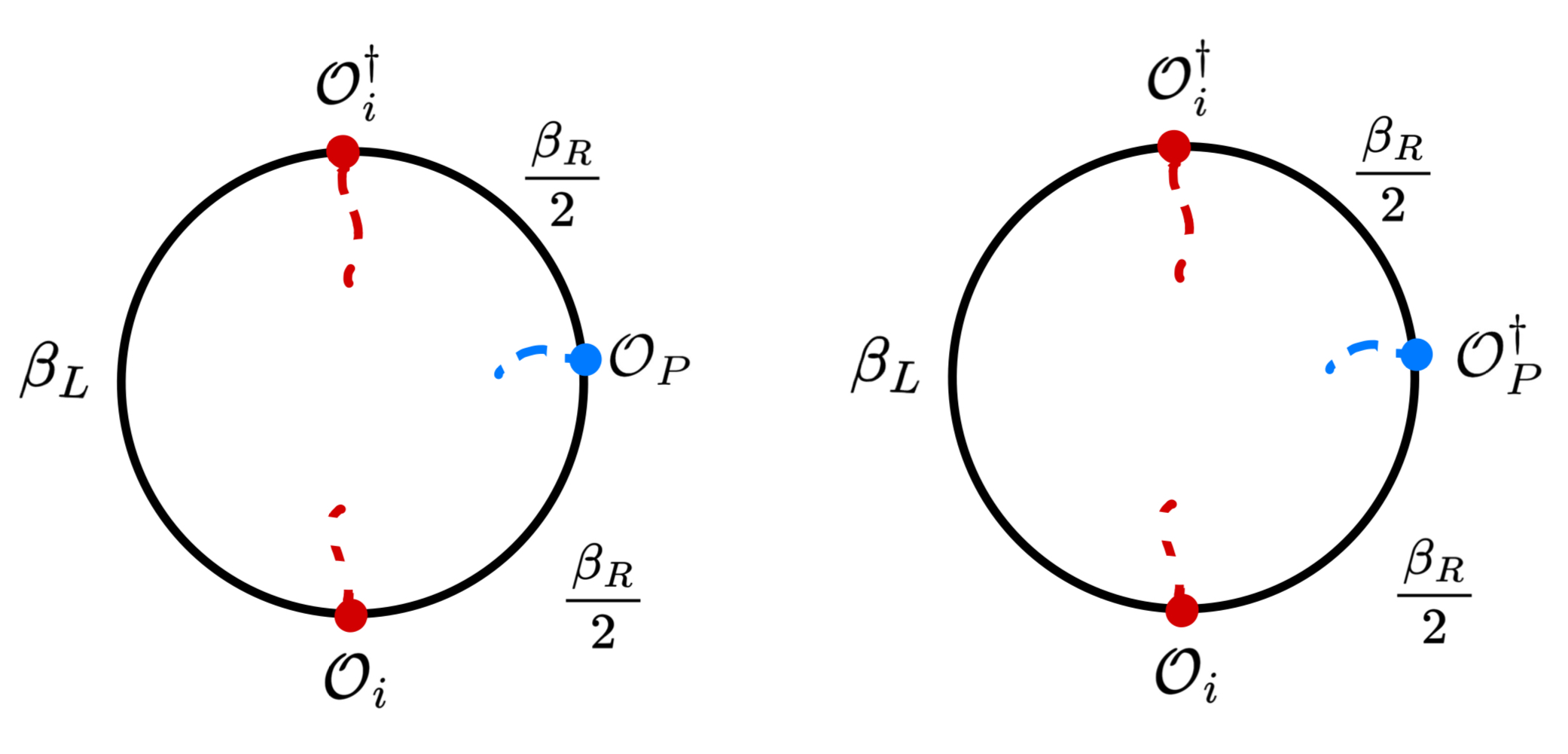}
    \caption{Boundary condition for detecting a two-boundary state using a single-boundary probe.}
    \label{fig:2s1bP}
\end{figure}

We now show that the two-sided gravity states can also be detected with access to just one of the asymptotic boundaries. In particular, we can adapt the single-boundary probe correlator of Sec.~\ref{sec:singbdrprobes} to the two-boundary case. In particular, we use the probe operator $\mathcal{O}_R$ inserted on the $\tau=0$ slice of the right boundary,\footnote{ We could equally have inserted the operator on the  left boundary instead.} and compute $\braket{\mathbf{i}|\mathcal{O}_{R}|\mathbf{i}}\braket{\mathbf{i}|\mathcal{O}_R^{\dagger}|\mathbf{i}}=|\braket{\mathbf{i}|\mathcal{O}_R|\mathbf{i}}|^2$, see Fig.~\ref{fig:2s1bP}. For the same reasons as the single-boundary case, the gravity path integral predicts $\overline{\braket{\mathbf{i}|\mathcal{O}_R|\mathbf{i}}} \approx 0 $ while the magnitude squared $\overline{|\braket{\mathbf{i}|\mathcal{O}_R|\mathbf{i}}|^2}$ has two classes of universal  and four classes of detection wormhole saddle contributions.

\begin{figure}
   \begin{subfigure}{0.45\linewidth}
   \centering
    \includegraphics[width=0.8\linewidth]{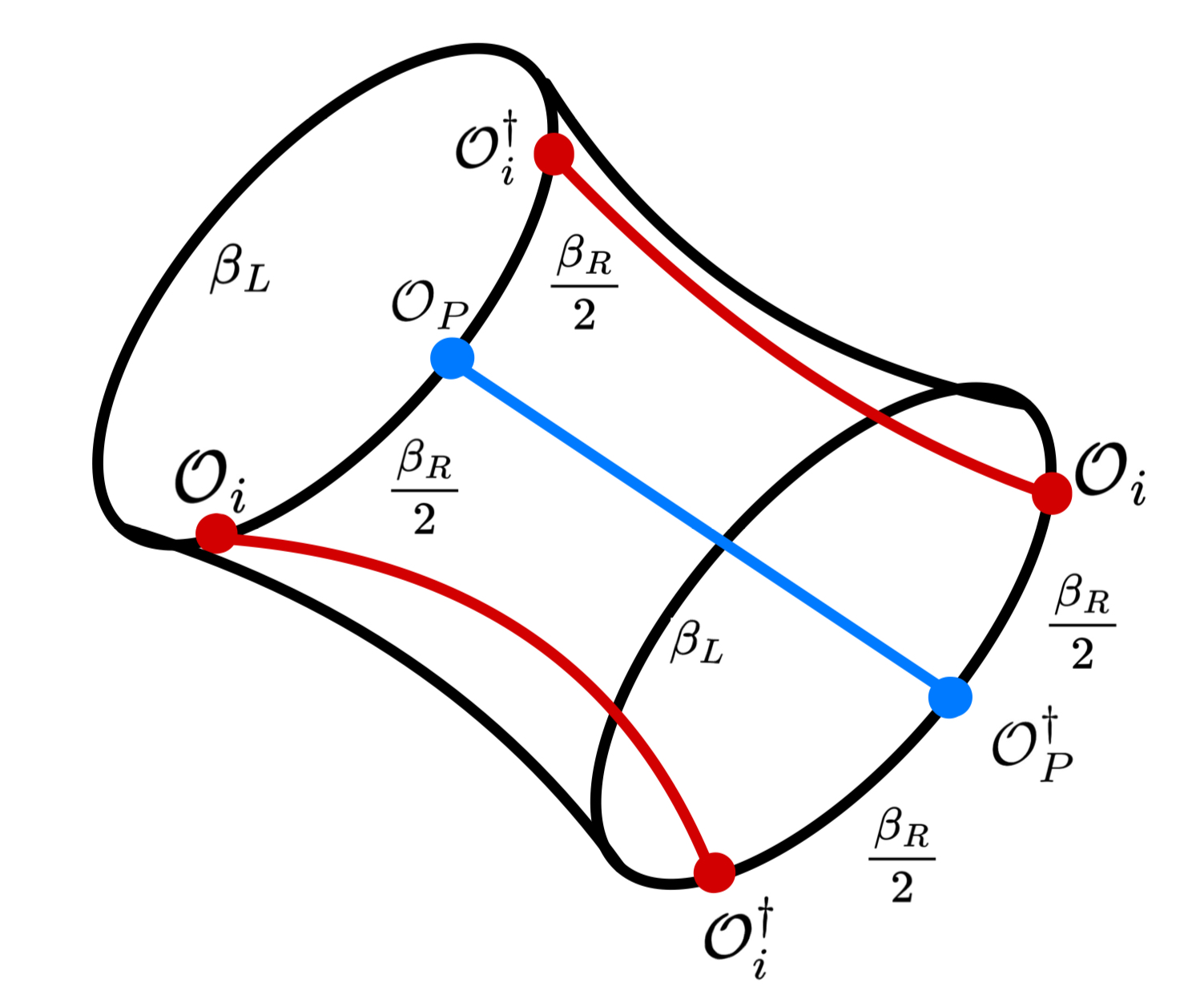}
    \caption{}
    \label{fig:2S_1S_prop1}
    \end{subfigure}
\hfill
\begin{subfigure}{0.45\linewidth}
    \centering
    \includegraphics[width=0.8\linewidth]{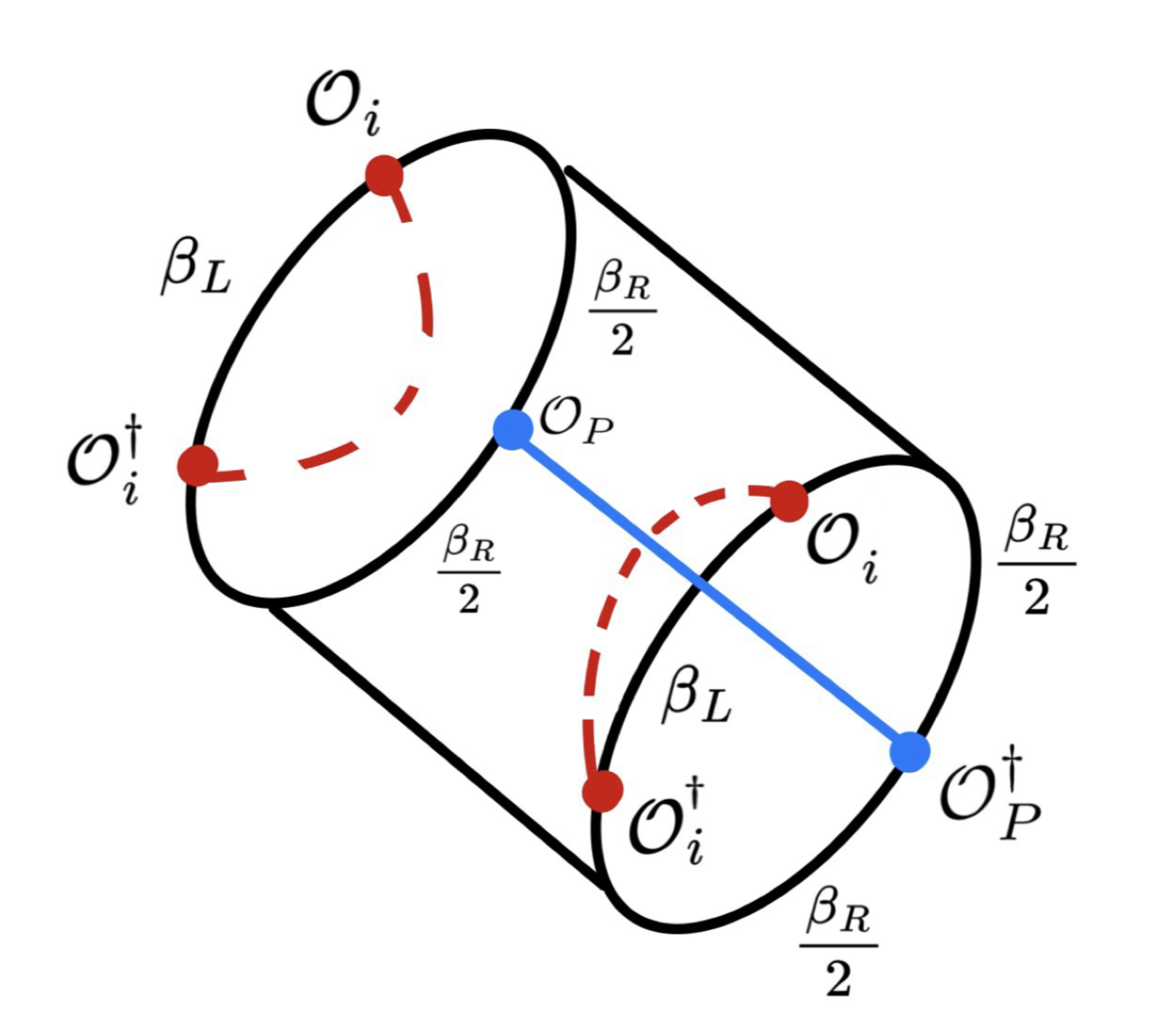}
    \caption{}
    \label{fig:2S_1S_prop2}
    \end{subfigure}
    \caption{Construction of the universal saddle contributions to $\overline{\braket{\mathbf{i}|\mathcal{O}_{R}|\mathbf{i}}\braket{\mathbf{i}|\mathcal{O}_R^{\dagger}|\mathbf{i}}}$.}
\end{figure}

  \paragraph{Universal saddles.}
  The two classes universal saddles contributing to $\overline{\braket{\mathbf{i}|\mathcal{O}_{R}|\mathbf{i}}\braket{\mathbf{i}|\mathcal{O}_R^{\dagger}|\mathbf{i}}}$ are of cylinder topology. In the first class of universal saddle the three shells all propagate from one boundary of the cylinder to the other (Fig.~\ref{fig:2S_1S_prop1}), while in the second class only the probe shell propagates between the two boundaries while the other two propagate to and from the same boundary (Fig.~\ref{fig:2S_1S_prop2}).   As explained in Appendix B \cite{Balasubramanian:2024rek}, the saddle geometries for these kind of quantities must satisfy an additional bulk constraint. In particular, the flow of Euclidean boundary time induces a natural orientation to the boundary condition computing overlaps such $\braket{\mathbf{i}|\mathbf{j}}$ due to operator ordering. The boundary condition for $\braket{\mathbf{j}|\mathbf{i}}$ is then given by replacing operators with their conjugate while keeping orientation of this flow fixed. Only the saddles in which this orientation is smoothly continued into the bulk contribute. We have not been explicit about this requirement in the preceding sections as it does not lead to subtleties there. 
  
 The first class of saddles is constructed as follows. Consider cutting the cylinder open along the shell worldvolumes, resulting in three pieces (called \textit{sheet diagrams} in \cite{Balasubramanian:2025jeu}). Each of these pieces can be obtained as a portion  of a disk with two shell insertions, and in the large shell mass limit the propagation times  of the shells on the disk go to zero, such that the asymptotic boundary length of this disk is equal to the boundary section of sheet. The length of these disk are $2\beta_L$, $\beta_R$ and $\beta_R$, see Fig.~\ref{fig:2S_1S_prop1}. Hence these saddles contribute a factor $\overline{Z}(2\beta_L)\times \overline{Z}(\beta_R)^2 \times Z_{m_i}^2 Z_{m_P}$.  To see how the second class of saddles  is constructed first consider a disk containing two $P$-shell and two $i$-shell insertions in an alternating fashion, i.e., $P-i-P-i$. This disk is glued into a cylinder by discarding the $P$-shell homology regions and gluing the resulting geometry along the $P$-shell worldvolumes.  Next consider two copies of a disk of boundary length $\beta_L + \Delta_T$ containing an $\mathbf{i}$-shell, discard the $\mathbf{i}$-shell homology regions on these disk and the cylinder and glue each of the disk into one side of the cylinder along the respective shell world-volume, see Fig.~\ref{fig:2S_1S_prop2}. In the large shell mass limit all the propagation times  again go to zero and the shell homology regions pinch off completely. Hence these saddles contribute a factor $ \overline{Z}(2\beta_R)\times \overline{Z}(\beta_L)^2\times Z_{m_i}^2 Z_{m_P}$. The universal response is therefore

\beq
Z_U =  \left(\overline{Z}(2\beta_R)\times \overline{Z}(\beta_L)^2 + \overline{Z}(2\beta_L)\times \overline{Z}(\beta_R)^2 \right) \times Z_{m_i}^2 Z_{m_P}
\eeq

\begin{figure}
   \centering
    \includegraphics[width=0.7\linewidth]{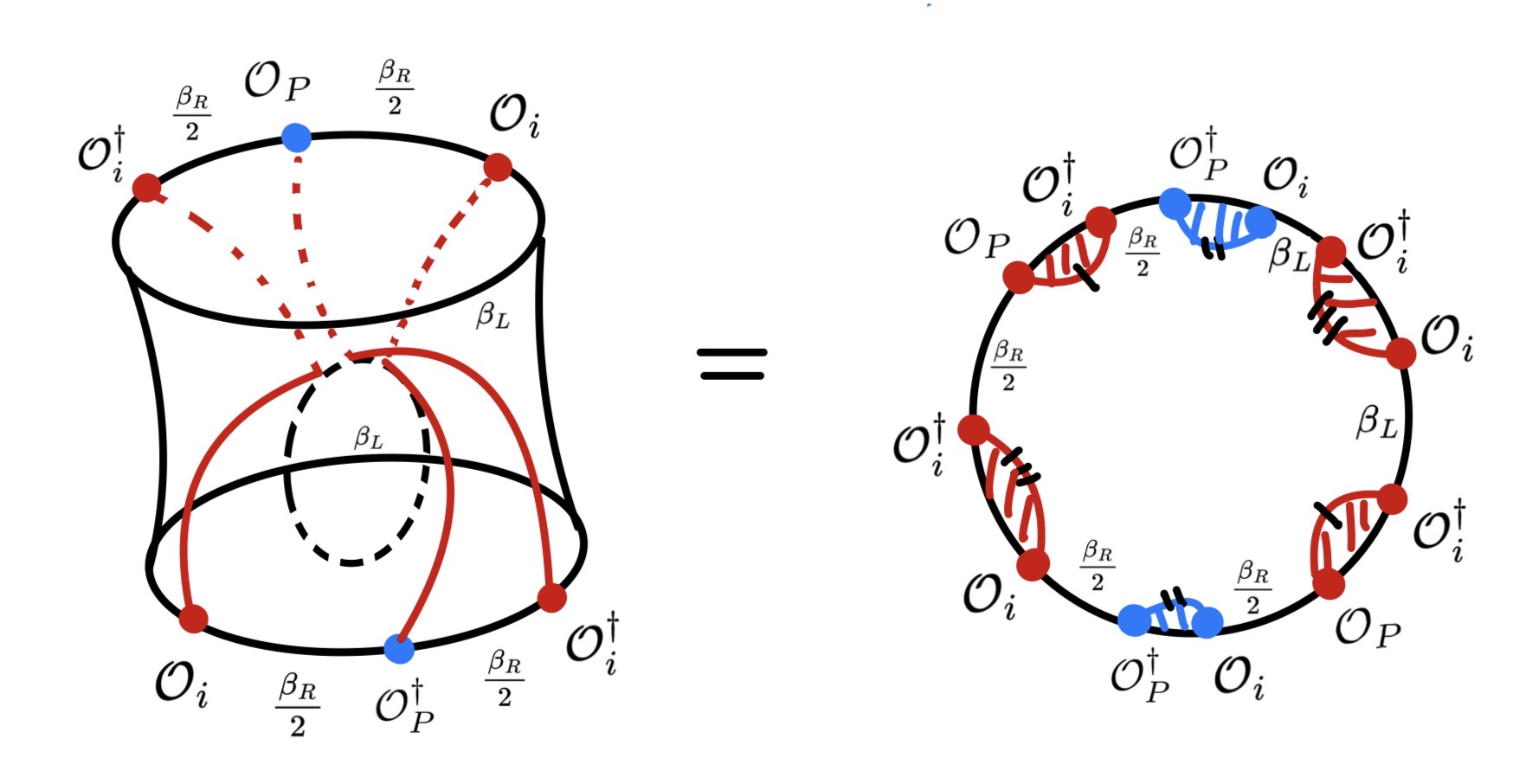}
    \caption{Fourth classes of detection saddle contributions to $\overline{\braket{\mathbf{i}|\mathcal{O}_{R}|\mathbf{i}}\braket{\mathbf{i}|\mathcal{O}_R^{\dagger}|\mathbf{i}}}$. The bulk topology is that of a twice punctured sphere.}
   \label{fig:2s_from_1s_ann3}
\end{figure}

 \paragraph{Detection  saddles.}
The four classes of detection saddles contributing to $\overline{\braket{\mathbf{i}|\mathcal{O}_{R}|\mathbf{i}}\braket{\mathbf{i}|\mathcal{O}_R^{\dagger}|\mathbf{i}}}$  are depicted in (Fig.~\ref{fig:2s_1s_D}). The first three of these are variations of the two universal saddles discussed above. The construction is therefore identical to that in the above paragraph once the correct lengths of the respective disks used to glue together wormholes are taken into account. However, the last saddle is novel. This saddle has the topology of a twice punctured torus such that each shell can propagate from one boundary to the other without intersecting, whilst allowing the orientation of Euclidean time flow form the boundary to continue smoothly into the bulk. This saddle geometry can be obtained by gluing shells propagating on a disk saddle in the manner depicted in Fig.\ref{fig:2s_from_1s_ann3}. In the large shell mass limit this term contributes a factor $ \overline{Z}(2\beta_L+2\beta_R)\times Z_{m_i}^2 Z_{m_P}$. Putting everything together we obtain:
\begin{figure}
\begin{subfigure}{0.24\linewidth}
   \centering
    \includegraphics[width=\linewidth]{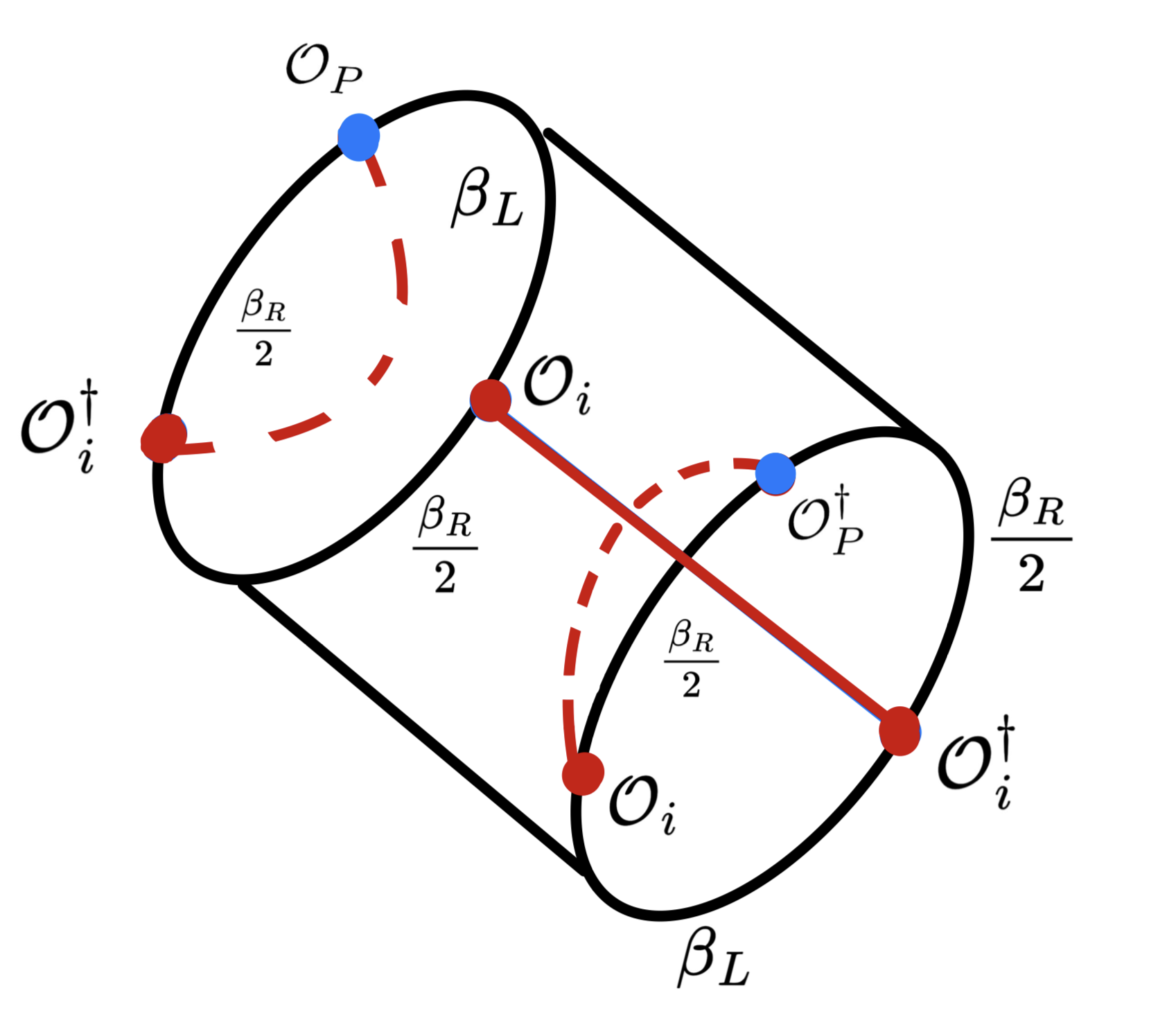}
    \caption{}
    \label{fig:2s_from_1s_ann}
    \end{subfigure}
\hfill
\begin{subfigure}{0.24\linewidth}
    \centering
    \includegraphics[width=\linewidth]{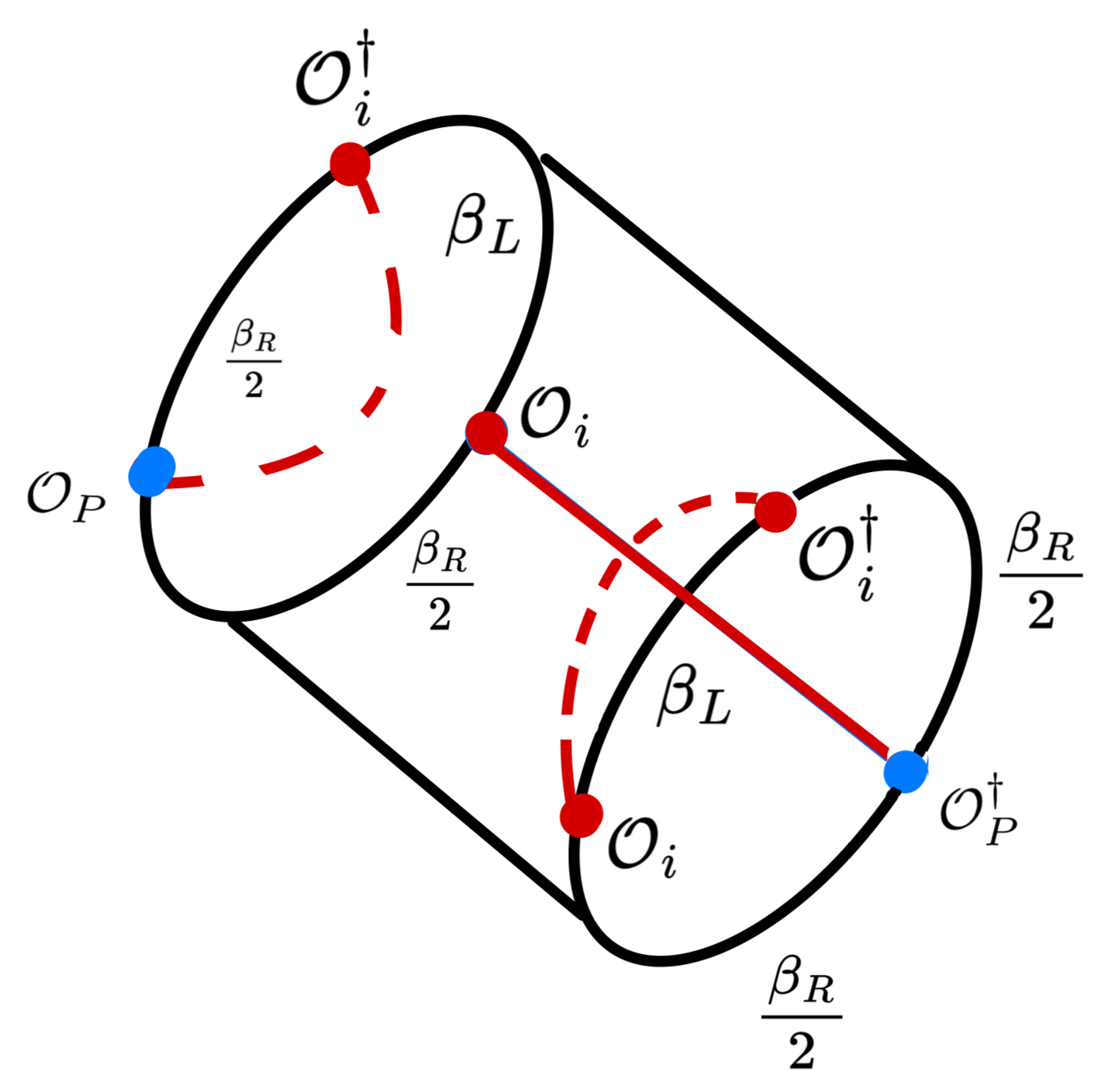}
    \caption{}
    \label{fig:2s_from_1s_ann2}
    \end{subfigure}
\hfill
\begin{subfigure}{0.24\linewidth}
    \centering
    \includegraphics[width=\linewidth]{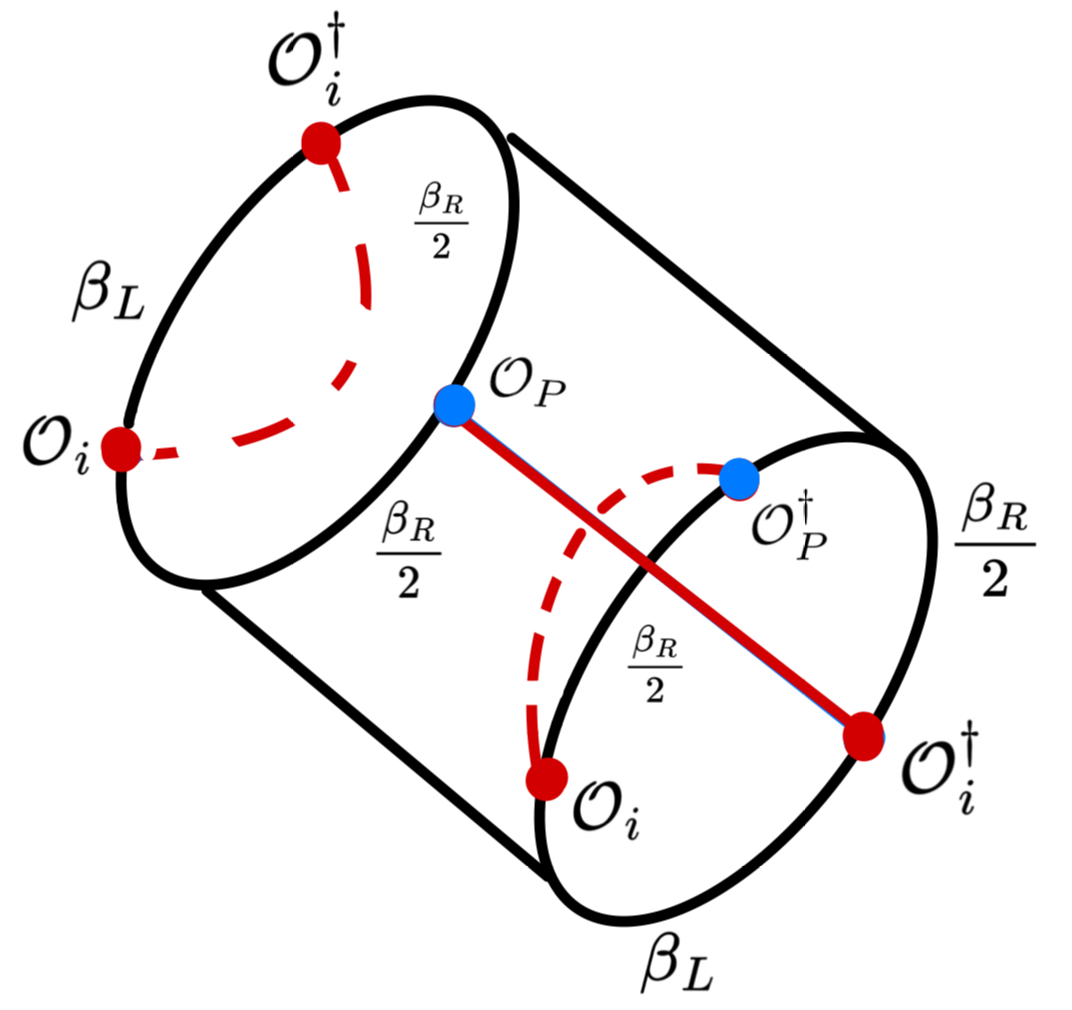}
    \caption{}
    \label{fig:2s_from_1s_ann4}
    \end{subfigure}
\hfill
\caption{First three classes of detection saddle contributions to $\overline{\braket{\mathbf{i}|\mathcal{O}_{R}|\mathbf{i}}\braket{\mathbf{i}|\mathcal{O}_R^{\dagger}|\mathbf{i}}}$.  }
\label{fig:2s_1s_D}
\end{figure}
 \beq
 Z_D = \left ( \overline{Z}(\frac{\beta_R}{2})^2 \times \overline{Z}(\beta_R + 2\beta_L) + 2\times \overline{Z}(\frac{\beta_R}{2})\times \overline{Z}({\beta_L})\times \overline{Z}({\beta_L}+ \frac{3 \beta_R}{2}) + \overline{Z}({\beta_R})\times \overline{Z}(2\beta_L+2\beta_R) \right) \times Z_{m_i}^2 Z_{m_P}
\eeq

\paragraph{Detection-to-universal ratio.}
The size of detection-to-universal ratio,
\beq \label{R-W_2b-1s}
\frac{Z_D}{Z_U}=\frac{\overline{Z}(\frac{\beta_R}{2})^2 \times \overline{Z}(\beta_R + 2\beta_L) + 2\times \overline{Z}(\frac{\beta_R}{2})\times \overline{Z}({\beta_L})\times \overline{Z}({\beta_L}+ \frac{3 \beta_R}{2}) + \overline{Z}(2\beta_L+2\beta_R)}{\overline{Z}(2\beta_R)\times \overline{Z}(\beta_L)^2 + \overline{Z}(2\beta_L)\times \overline{Z}(\beta_R)^2},
\eeq
depends on the relative size of $\beta_L,\beta_R$ in a complicated manner. To obtain a more illuniating ratio we project the two-boundary shell states into the micro-canonical window $[E_{L,R}-\frac{\Delta E_{L,R}}{2},E_{L,R}+\frac{\Delta E_{L,R}}{2}]$. The  Laplace transforms of (\ref{R-W_2b-1s}) can be worked out similarly to those in Sec.~\ref{sec:infoshare} above. Leaving aside the details, we assume for ease of calculation that $E_L=E_R\equiv E$ and $\Delta_L=\Delta_R\equiv \Delta$.\footnote{Note that if we do not choose $E_L=E_R\equiv E$ and $\Delta_L,\Delta_R$ the windows $[E_{L,R}-\frac{\Delta E_{L,R}}{2},E_{L,R}+\frac{\Delta E_{L,R}}{2}]$ must have at least some overlap to yield a nonzero result. As explained above this is due to the ``mixing terms" where the argument of $\overline{Z}$ depends on both $\beta_L$ and $\beta_R$.} In this simplified setting every class of saddle contributes a factor $e^{-2E(\beta_L+\beta_R)+ \mathbf{S}(E)}$, and hence to leading order $\frac{Z_D}{Z_U}=\frac{3}{2}$. 

We again see that when probed with the correct operator, the detection saddles result in an $\mathcal{O}(1)$ increase in signal strength, allowing the state to be verified through wormhole effects. In Sec.~\ref{sec:infoshare} we also showed that a two-boundary state can be detected \textit{without} relying on these wormhole saddles by inserting a tuned probe on both asymptotic boundaries simultaneously. This resulted in an exponentially large response. In a sense, it is therefore easier to detect the state in that way. However, in both detection methods  the left and right microcanonical windows of the state must have some overlap in order for detection to be possible. This arises due to the the ``mixing" of $L,R$ preparation temperatures in arrangements of the partition functions $\overline{Z}(\cdots)$ appearing in the detection saddles.

\section{Detecting the state of the universe and QMA}\label{sec:QMA}
In the above sections we have shown in various settings that probing a shell state $\ket{i}$ with a different heavy shell operator $\mathcal{O}_P\neq\mathcal{O}_i$ gives universal results that do not contain information about the state in question. Furthermore, generic light probes will also not be able to access information regarding the state. Take for example the single-boundary Lorentzian correlator $\overline{\braket{i|\mathcal{O}_P|i}\braket{i|\mathcal{O}_P^{\dagger}|i}}$ considered in Sec.~\ref{sec:singbdrprobes} for a probe $\mathcal{O}_P$ that is light enough to not back-react significantly, but heavy enough to be well described by the geodesic approximation. Such a probe is well described by a geodesic on the wormhole geometry contributing to $\overline{\braket{i|i}\braket{i|i}}$. As explained in the \textbf{universal} paragraph of Sec.~\ref{sec:singbdrprobes}  this wormhole geometry is obtained by taking a disk containing two $i$ shells and gluing in two strips containing an $i$ shell along the shell worldvolumes. The geodesic describing the shell trajectory is therefore simply a geodesic on disk geometry, and contains no information about the shell operator used to create the state.

We also showed that if $\mathcal{O}_P=\mathcal{O}_i$ the result of the probe measurement is generically larger, because there are additional detection saddles. If this detection contribution is large enough for the observer to discern experimentally, an asymptotic boundary observer can use these probe correlators to verify a hypothesis that the universe is in a certain shell state. To see this, suppose the boundary observer wants to verify whether the universe is in a shell state $\ket{i}$ prepared by the  operator $\mathcal{O}_i$  with inverse preparation temperature $\beta$.  The preparation state could include  additional quantum matter fields which are described in the semiclassical limit by a  matter QFT on the type A or B background geometry. In heavy shell mass limit the state of these fields restricted to the region in casual contact with the boundary will look almost exactly thermal at inverse temperature $\beta$.\footnote{See Sec.~3.4 of \cite{Balasubramanian:2022gmo}.} The observer can measure this temperature and determine if it is above or below the Hawking-Page temperature $T_{HP}$.\footnote{The Hawking-Page temperature of the spacetime depends on the AdS radius $\ell$: $T_{HP}=\frac{3}{2 \pi \ell}$. The observer can determine $\ell$ by measuring the curvature.}
This allows the observer to determine wether the state in question is of type A ($\beta > \beta_{HP}$) or type B ($\beta < \beta_{HP}$). Next, the observer needs verify the proposal for the shell operator that created the state ($\mathcal{O}_i$). While the probe operator itself is not Hermitian and therefore not a physical observable, a boundary observer can measure the expectation value of the Hermitian observables $A=\frac{\mathcal{O}_P + \mathcal{O}_P^{\dagger}}{2}$, $B=\frac{\mathcal{O}_P - \mathcal{O}_P^{\dagger}}{2i}$, for which $\braket{i|A|i}^2+\braket{i|B|i}^2=|\braket{i|\mathcal{O}_P|i}|^2$. With access to sufficiently many copies of the state, the observer can use these observables to compute the probe correlator and establish wether it is large enough to contain universal saddle contributions. If so, the proposal is verified, otherwise it is rejected.

\paragraph{Detecting a superposition.} \label{sec:superposition}
The above conclusions can be generalised to any superpositon of shell states. To see this consider the correlator $\overline{\braket{i|O_j|k}\braket{m|O^{\dagger}_l|p}}$. The universal contributions (i.e., the ones where $O_j, O^{\dagger}_l$ annilihate each other) are given by $\delta_{jl}(\delta_{ik}\delta_{mp}U_{R} +\delta_{ip}\delta_{km}U_{L})  $, whereas the annilihation contribtions are given by $\delta_{ji}(\delta_{km}\delta_{lp}D_{L}^{\uparrow} +\delta_{lk}\delta_{mp}D_{R}^{\uparrow}) + \delta_{jm}(\delta_{ik}\delta_{lp}D_{L}^{\downarrow} +\delta_{ip}\delta_{kl}D_{R}^{\downarrow})$. Here the $U_{R},U_{L} \cdots D_{R}^{\downarrow}$ refer to the labeling of universal and detection saddles  in Fig.~\ref{fig:all_saddle}. If we now consider a superposition of $N$ normalized shell states $\ket{\psi}=\sum^{N}_{i=1} \alpha_i \ket{i}$ probed with a linear combination of  corresponding shell operators $\mathcal{O}_P=\sum^{N}_{i=1}\gamma_i \mathcal{O}_i$, the probe correlator gives:
\beq\label{eq:superposition}
\overline{\braket{\psi|\mathcal{O}_P|\psi}\braket{\psi|O^{\dagger}_P|\psi}}= \braket{\vec{\alpha}|\vec{\alpha}}\braket{\vec{\alpha}|\vec{\alpha}}\braket{\vec{\gamma}|\vec{\gamma}}(U_{R} +U_{L}) + \braket{\vec{\alpha}|\vec{\alpha}}\braket{\vec{\alpha}|\vec{\gamma}}\braket{\vec{\gamma}|\vec{\alpha}}(D_{L}^{\uparrow}+D_{R}^{\uparrow}+D_{L}^{\downarrow}+D_{R}^{\downarrow}),
\eeq
where we have defined $(\vec{a})_i\equiv \alpha_i$, $(\vec{\gamma})_i\equiv \gamma_i Z_{m_i}$ (where $Z_{m_i}$ is the universal probe shell mass contribution to the action)  and the inner products between these vectors appearing above is the standard inner product on $\mathbb{C}^N$.  We  normalise $\ket{\psi}$ and $\mathcal{O}_P$ such that $\braket{\vec{\alpha}|\vec{\alpha}}=\braket{\vec{\gamma}|\vec{\gamma}}=1$, although note this does not imply $\braket{\psi|\psi}=1$ because the shell basis states are not orthonormal. We can estimate the response for generic superpositions probed by a generic linear combination by considering $\vec{\alpha},\vec{\gamma}$ to be independently drawn random vectors. Such vectors satisfy $|\braket{\vec{\gamma}|\vec{\alpha}}|\sim 1/ \sqrt{N}$, and hence for large N we get a universal universal response of $\overline{\braket{\psi|\mathcal{O}_P|\psi}\braket{\psi|O^{\dagger}_P|\psi}}\approx P_{R} +P_{L}$, which contains no information regarding the state. For some particular draws of $\vec{\alpha},\vec{\gamma}$ the anniliation contribution is nonzero, but it is clearly maximized when $\vec{\alpha}=\vec{\gamma}$ as this saturates the Cauchy–Schwarz inequality for the inner product on $\mathbb{C}^N$. 

In short, a random superposition state probed by a random linear combination of shell operators will give a universal response. However, when this state is probed by a fine-tuned  linear combination of shell operators, the signal jumps by the same factor as discussed in the above section, due to contributions from the detection saddles.  Hence a guess for the superposition of shell states can also be verified. Moreover, it was shown in \cite{Balasubramanian:2025jeu} and \cite{Balasubramanian:2025zey} that sufficiently large sets of shell states provide a spanning basis for the full non-perturbative gravity Hilbert space.  So being able to verify arbitrary superpositions amounts to an ability to verify any gravity state at all.

\paragraph{Finding the state.}
While a boundary observer can verify a hypothesis for the state of the universes, it is still difficult for an observer to \textit{find} which state their universe is in. 
Consider, for example, the case where observer wants to correctly identify the universe's state correctly out of a list of $N$ proposals. For simplicity we let these states be shell states created the operators $\{O_1,O_2 \cdots O_N\}$. The observer can find the state by measuring the probe correctors for each possible operator and seeing if it is above the detection saddle threshold. To recognize this threshold the observer first establishes a baseline by measuring the probe corelation for a few random choices of $O_s$.\footnote{Of course if one of these random calibration choices gives a significantly larger result than the others the correct operator has already been identified and no search is needed.} Finding the state in this way  naively requires $\mathcal{O}(N)$ steps. Although Grover search \cite{Grover:1996rk} can improve this to at most $\mathcal{O}(\sqrt{N})$, for sufficiently large sets (for example $N\sim e^{\mathbf{S}(E)}$) it is difficult to find the state this way. In principle the state might be found by 
performing quantum state tomography\footnote{See, for example, \cite{Nielsen:2012yss}. } in the shell basis. However, determining the state through tomography requires access to the fine-grained details of the overlaps between individual shell states, which the saddlepoint approximations of the gravitational path integral at the level used here do not have access to. Recall that at the saddlepoint level the overlap between shell states satisfies $\overline{\braket{m|n}}=\delta_{mn} Z_1$ while the magnitude squared of this overlap is nonzero do to the universal wormhole contribution $\overline{\braket{m|n}\braket{n|m}} =\delta_{mn} Z_1^2 + Z_{WH}$. As discussed near (\ref{eq:WHoverlap}), this suggest
 the gravitational path integral computes coarse-grained averages over the fine-grained microscopic data, resulting in a vanishing overlap due to erratic phases averaging to zero. Hence the saddle-point analysis does not have access to this phase information.

Nonetheless, a proposed solution for the state finding problem can be efficiently verified, putting this problem in the Quantum Merlin Arthur (QMA) complexity class (see for example \cite{Watrous:2008any}).

\section{Summary and Discussion} 
\label{sec:diss}
Black hole states in quantum gravity are expected to be highly complex. As a result, most probes of the state should give universal responses and be unable to access the information encoded in the state \cite{Balasubramanian:2005kk,Balasubramanian:2007qv,Brown:2019rox,Akers:2022qdl}. Indeed, semiclassically these states correspond to geometries with an event horizon or baby universe hiding the information about the interior from outside observers. Nonetheless, in this paper we used the gravitational path integral to show that there exist probes that an asymptotic observer can use to detect the state of a black hole or baby universe. In particular, we considered states  constructed by slicing open the Euclidean gravity path integral with heavy dust shell operator insertions. In the appropriate parameter regimes these heavy shell states either correspond to black holes with the shell behind the horizon or empty space entangled with a compact, big crunch baby universe containing the shell. We have shown that probing these states with a light operator or a heavy dust shell operator different than the one used to prepare the state gives universal responses that do not contain enough information to detect the state. If, however, the state is probed with the same heavy dust shell operator as the one preparing the state, there is an additional contribution on top of the universal part, allowing the state to be determined. This is precisely the scenario suggested in \cite{Balasubramanian:2005kk}. The fact that such observables exist is remarkable because for both the black hole and the baby universe the spacetime accessible to the asymptotic observer, as well as the perturbative quantum fields within it, are semi-classically nearly indistinguishable from the thermal state, containing only  fine quantum hair. These findings are in line with the idea that  highly complex operations performed on the late-time Hawking radiation of a black hole can detect the interior, which is believed to be encoded in this radiation after the Page time (see, for example, \cite{Penington:2019npb,Akers:2022qdl, Balasubramanian:2022fiy,Penington:2019kki,Almheiri:2019qdq}). 

\subsection{Detection Through Euclidean Wormholes}
We have considered three types of probe correlators throughout this work. Correlators of the form $\overline{\braket{i|\mathcal{O}_P|i}\braket{i|\mathcal{O}_P^{\dagger}|i}}$ allow single- and two-boundary gravity states to be detected using probes localized on just one of the boundaries, with a signal that is bigger than the baseline result for a random probe by a factor of $O(1)$. 

This method of detection is possible because of non-perturbative wormhole contributions to the gravity path integral. These wormhole contributions extract the magnitude of $\braket{i|\mathcal{O}_P|i}$ which itself averages to zero after coarse-graining in the gravitational path integral because of the presence of erratic phases. In a sense, these non-perturbative effects making detection possible can be thought off as sensitive the associated quantum hair. For the two-boundary case we have also shown that a correlator involving the insertion of a fine-tuned probe on both of the boundaries $\overline{\braket{\mathbf{i}|\mathcal{O}_{P,L}^{\dagger}\mathcal{O}_{P,R}| \mathbf{i}}}$ allows the state to be detected without non-perturbative wormhole effects, resulting in a signal that scales with the size of the accessible Hilbert space. 

As these two types of correlators involve no Euclidean time evolution once the state is prepared, they can be measured by a Lorentzian boundary observer. This suggest that single boundary probes that detect the state without relying on such wormhole saddles the detection response could be much larger. This can be achieved using Euclidean probes, and will be reported on in \cite{Eucliddec}.

\subsection{Seeing Beyond the Entanglement Wedge}The fact that  non-perturbative gravitational effects can be used to detect the state of a two-boundary black hole from just one of the boundaries is striking. In AdS/CFT the two-sided black holes, which have a very long wormhole interiors, correspond to partially-entangled-states (PETS) on two copies of the CFT $\mathcal{H}_{CFT,L}\otimes\mathcal{H}_{CFT,R}$ \cite{Goel:2018ubv}. Entanglement wedge reconstruction \cite{Headrick:2014cta,Jafferis:2015del,Dong:2016eik,Bao:2016skw,Faulkner:2017vdd,Cotler:2017erl,Chen:2019gbt,Kang:2018xqy,Gesteau:2020rtg} suggest that the interior is in the entanglement wedges union of the boundaries, not in the entanglement wedge of either one.  In other words, the interior of this black hole is not encoded in any one of the boundary theories but is encoded instead in the entanglement between the two. Our results suggest that including non-perturbative gravitational effects makes some part of the interior accessible from just one of the boundary theories. Crucially, it did not matter which of the boundaries we chose to put the probe on, even though the shell can at most be in the entanglement wedge of a particular one of the two boundaries. Of course, being able to detect the interior state from one boundary is not the same as being able to reconstruct low energy EFT excitations in interior on just one of the boundary theories.  It would be interesting to explore the connection between the two further.

\subsection{Complexity of the Baby Universe} 
The universal response to simple operators and non-universal response to a fine-tuned probe found in this paper are characteristic of complex states \cite{Balasubramanian:2005mg,Engelhardt:2024hpe}. 
Remarkably, our conclusions apply equally to the black hole and the baby universe case. Indeed,  the analogy between the two cases suggests that the fact that the  baby universe is disconnected from the rest of the spacetime is also a semiclassical manifestation of the underlying complexity of the gravity state. Note that the non-perturbative Hilbert space of closed universes appears to be one dimensional when counted naively from the boundary \cite{Penington:2019kki,Usatyuk:2024mzs,Marolf:2020xie,Wang:2024itz}. Within our framework, this could  be seen as a by-product of the state being very complex, and thereby hard to probe, forming a perfect black box. However, our findings suggest that while probing the baby universe from the boundary is hard, it is not impossible. This is consistent  with  results from \cite{Antonini:2023hdh,Antonini:2025ioh} that recovers a  nontrivial baby universe Hilbert space by entangling it with a reference; the "observer", in our case this system is the disconnected AdS factor. 

This apparent baby universe complexity is surprising as the ADM energy of the baby universe state vanishes, and it is therefore below the black hole threshold at which the gravity theory is expected to become highly chaotic \cite{Shenker:2013pqa,Cotler:2016fpe,Saad:2018bqo,Schlenker:2022dyo}. Furthermore, as noted in \cite{Balasubramanian:2022gmo}, the shell operator itself is also  ``simple", being just a gas of EFT particles. It is therefore not clear what aspect of the time evolution leads to the complex characteristics displayed by the state, and it would be interesting to explore this further from the tensor network perspective of the baby universe  proposed in \cite{Antonini:2023hdh,Antonini:2025ioh}.

Relatedly, \cite{Antonini:2024mci} proposed  in the context of the AdS/CFT correspondence that the PETS states dual to the baby universe are states of $\mathcal{O}(1)$ energy and entropy when suitably truncated, and should therefore be well described the extrapolate dictionary of AdS/CFT. However, applying this dictionary results in a  dual bulk state of consisting of an entangled gas of light operators acting on the AdS vacuum, without any baby universe. The same CFT state therefore seemingly has two possible bulk duals with distinct semiclassical geometry, see \cite{Engelhardt:2025vsp,Engelhardt:2025azi,Higginbotham:2025dvf,Antonini:2025ioh} for discussion, with the latter seeming  ``simple". This puzzle is in close analogy with the one above, and it would interesting to see if the resolution proposed in \cite{Antonini:2025ioh} helps resolve both.

For concreteness, the discussion in this paper was centered around asymptotically AdS quantum gravity in any dimension.  However, as shell states and their overlaps/correlators can also be constructed in asymptotically flat quantum gravity, any result in this paper that does not rely on the Hawking-Page transition carries over to asymptotically flat boundary conditions also.  Indeed, we expect our results to hold for any consistent asymptotically flat or AdS theory of quantum gravity that is well described at low energies by General Relativity.

\paragraph{Acknowledgments.}
We would like to thank William Chan and Chitraang Murdia for useful discussion.  VB was supported in part by the DOE through DE-SC0013528 and QuantISED grant DE-SC0020360, and in part by the Eastman Professorship at Balliol College, University of Oxford. TY thanks the Peter Davies Scholarship for continued support.

\section*{}
\bibliographystyle{jhep}
\bibliography{main}




\end{document}